\documentclass[a4paper,fleqn,usenatbib]{mnras}
\usepackage{newtxtext,newtxmath}
\usepackage[T1]{fontenc}
\usepackage{ae,aecompl}
\usepackage{graphicx}	% Including figure files
\usepackage{amsmath}	% Advanced maths commands
\usepackage{amssymb}% Extra maths symbols
\usepackage{verbatim}
\usepackage{relsize}

\def\bhar{$\rm \overline{BHAR}$}
\def\dbhar{$\Delta \rm \overline{BHAR}$}
\def\mstar{$M_\star$}
\def\sigmae{$\Sigma_{\rm e}$}
\def\sigmaone{$\Sigma_{1}$}
\def\cone{$C_{1}$}

\def\re{$r_{\rm e}$}
\def\agnf{$f_{\rm AGN}$}
\def\dagnf{$\Delta f_{\rm AGN}$}
\def\sersic{$\rm S{\acute e}rsic$}

\usepackage{xcolor, soul}
\definecolor{OliveGreen}{rgb}{0,0.6,0}

\title[Does black-hole growth depend fundamentally on host-galaxy compactness?]
{Does black-hole growth depend fundamentally on host-galaxy compactness?} 

\author[Ni et al.]{Q. Ni,$^{1,2}$\thanks{E-mail: qxn1@psu.edu}
G. Yang$,^{1,2}$\thanks{E-mail: gyang206265@gmail.com}
W. N. Brandt,$^{1,2,3}$
D. M. Alexander,$^{4}$
C.-T. J. Chen,$^{5}$\newauthor
B. Luo,$^{6,7,8}$
F. Vito,$^{9,10}$
and Y. Q. Xue$^{11,12}$
\\
$^{1}$Department of Astronomy and Astrophysics, 525 Davey Lab, The Pennsylvania State University, University Park, PA 16802, USA\\
$^{2}$Institute for Gravitation and the Cosmos, The Pennsylvania State University, University Park, PA 16802, USA\\
$^{3}$Department of Physics, 104 Davey Laboratory, The Pennsylvania State University, University Park, PA 16802, USA\\
$^{4}$Centre for Extragalactic Astronomy, Department of Physics, Durham University, South Road, Durham DH1 3LE, UK\\
$^{5}$Marshall Space Flight Center, Huntsville, AL 35811, USA\\
$^{6}$School of Astronomy and Space Science, Nanjing University, Nanjing 210093, China\\
$^{7}$Key Laboratory of Modern Astronomy and Astrophysics (Nanjing University), Ministry of Education, Nanjing 210093, China\\
$^{8}$Collaborative Innovation Center of Modern Astronomy and Space Exploration, Nanjing, 210093, China\\
$^{9}$Instituto de Astrof{\'{\i}}sica and Centro de Astroingenier{\'{\i}}a, Facultad de F{\'{i}}sica, Pontificia Universidad Cat{\'{o}}lica de Chile, Casilla 306, Santiago 22, Chile\\
$^{10}$Chinese Academy of Sciences South America Center for Astronomy, National Astronomical Observatories, CAS, Beijing 100012, China\\
$^{11}$CAS Key Laboratory for Research in Galaxies and Cosmology, Department of Astronomy, University of Science and Technology of China,\\
~~ Hefei 230026, China\\
$^{12}$School of Astronomy and Space Science, University of Science and Technology of China, Hefei 230026, China
}

\date{Accepted XXX. Received YYY; in original form ZZZ}
\pubyear{2019}

\begin{document}
\label{firstpage}
\pagerange{\pageref{firstpage}--\pageref{lastpage}}
\maketitle

% Abstract of the paper
\begin{abstract}
Possible connections between central black-hole (BH) growth and host-galaxy compactness have been found observationally, which may provide insight into BH-galaxy coevolution: compact galaxies might have large amounts of gas in their centers due to their high mass-to-size ratios, and simulations predict that high central gas density can boost BH accretion. However, it is not yet clear if BH growth is fundamentally related to the compactness of the host galaxy, due to observational degeneracies between compactness, stellar mass (\mstar), and star formation rate (SFR).
To break these degeneracies, we carry out systematic partial-correlation studies to investigate the dependence of sample-averaged BH accretion rate (\bhar) on the compactness of host galaxies, represented by the surface-mass density, \sigmae, or the projected central surface-mass density within 1~kpc, \sigmaone. We utilize 8842 galaxies with $H < 24.5$ in the five CANDELS fields at $z = 0.5$--3.
We find that \bhar~does not significantly depend on compactness when controlling for SFR or \mstar\ among bulge-dominated galaxies and galaxies that are not dominated by bulges, respectively. However, when testing is confined to star-forming galaxies at $z = 0.5$--1.5, we find that the \bhar-\sigmaone\ relation is not simply a secondary manifestation of a primary \bhar-\mstar\ relation, which may indicate a link between BH growth and the gas density within the central 1~kpc of galaxies.
\end{abstract}

 
%%%%%%%%%%%%%%%%%%%%%%%%%%%%%%%%%%%%%%%%%%%%%%%%%%
%%%%%%%%%%%%%%%%% BODY OF PAPER %%%%%%%%%%%%%%%%%%

\section{Introduction} \label{sec-intro}
Understanding the connections between supermassive black holes (BHs) and their host galaxies has been an essential problem for the past two decades. 
It is well established that BH mass ($M\rm _{BH}$) is correlated with the stellar mass, luminosity, and velocity dispersion of the host-galaxy bulge in the local universe \citep[e.g.][]{Magorrian1998,KH2013}, suggesting the co-evolution of BHs and their host galaxies. 
However, the fundamental link between BH accretion and galaxy growth is still not well understood, and remains one of the most debated issues in astrophysics \citep[e.g.][]{Mullaney2011,Chen2013,Hickox2014,Yang2017,Yang2018a,Yang2018b,Yang2019,Aird2018}.
Researchers have investigated the relations between different galaxy properties and BH growth to determine what drives BH-galaxy co-evolution, and both star formation rate (SFR; which partly traces the total amount of cold gas available) and stellar mass ($M_\star$; which indicates the potential wells of galaxies) have been found to relate to BH growth.

To identify the fundamental link in BH-galaxy co-evolution, one promising avenue is to investigate the relation between BH growth and host-galaxy \textit{compactness}, which, nevertheless, has not been conducted in detail. 
Compactness can be represented by the surface-mass density, \sigmae; $\Sigma_{\rm e} = 0.5M_\star/\pi r_{\rm e}^2$, where $r_{\rm e}$ is the effective radius of the galaxy within which half of the total light is emitted (e.g. \citealt{Barro2017,Kocevski2017}). This widely adopted measurement of compactness measures the mass-to-size ratio in the central 50\% of a galaxy by its definition, thus assessing the compactness globally.  Alternatively, compactness can be represented by the central surface-mass density within 1 kpc, \sigmaone; \sigmaone\ $ = M_\star(< \rm 1~kpc) / \pi (1~kpc)^2$, where $M_\star(< \rm 1~kpc)$ is the stellar mass enclosed in the central 1 kpc of a galaxy. It has been suggested that the central stellar density within 1~kpc is more effective at connecting galaxy morphology and star formation activity when compared with surface-mass density \citep[e.g.][]{Cheung2012,vanDokkum2014,Whitaker2017,Lee2018}. Thus, \sigmaone\ might also be a more effective parameter connecting galaxy morphology and BH growth compared with \sigmae.
It is plausible that large amounts of gas are located within the nuclear regions of some compact galaxies (particularly, those that are actively star forming) due to their high mass-to-size ratios, and simulations predict that high central gas density can boost BH accretion \citep[e.g.][]{Wellons2015,Habouzit2019}.
Recent galaxy evolution simulations and models also predict a dissipative-contraction process (i.e., wet compaction event; e.g. \citealt{Dekel2014,Zolotov2015,Tacchella2016a, Tacchella2016b}) that triggers a compact starburst, which can also trigger concurrent growth of the central BH.
In this paper, we will sometimes speak of compactness, $\Sigma$, generally, where $\Sigma$ could mean either \sigmae\ or \sigmaone\ (i.e. ``$\Sigma$'' should be interpreted as ``\sigmae/\sigmaone'').

In the local universe, several overmassive black-hole ``monsters'' have been found in notably compact galaxies, which have $M\rm _{BH}$ values significantly larger than those expected from the relation with bulge mass \citep[e.g.][]{KH2013,IF2017}. Adding $r_{\rm e}$ as an additional parameter can indeed tighten the local relation between $M\rm _{BH}$ and the stellar mass/velocity dispersion of the host-galaxy bulge \citep[e.g.][]{MH2003,Beifiori2012}.

Possible connections between compactness and BH growth have been found with the great depth and high angular resolution of the \textit{HST} CANDELS survey \citep[e.g.][]{Grogin2011,Koekemoer2011}.
\citet{Kocevski2017} found that the AGN fraction among massive compact star-forming galaxies is significantly higher when compared with mass-matched extended star-forming galaxies at $1.4 < z < 3$. \citet{Rangel2014} suggested that absorption-corrected AGN X-ray luminosities correlate with the host-galaxy compactness at $M_\star > 10^{10.5} M_\odot$.
While those studies provided important clues about the role of compactness in BH-galaxy coevolution, neither of them could answer the question: is BH growth \textit{fundamentally} linked with the compactness of its host galaxy?

Compactness is correlated with stellar mass by construction, raising questions about which of these quantities is most fundamentally linked to BH growth. 
Could the observed correlation in \citet{Rangel2014} between compactness and AGN X-ray luminosity simply be a secondary manifestation of a primary correlation between stellar mass and BH growth \citep[e.g.][]{Yang2017,Yang2018a}?
Also, bulge-dominated galaxies are generally more compact. Could the observed high AGN fraction among compact star-forming galaxies in \cite{Kocevski2017} be a natural consequence of a large amount of BH growth expected among star-forming bulges \citep[e.g.][]{Silverman2008,Yang2019}?
Or, if compactness is indeed a critical property linked with BH growth, perhaps serving as an indicator of central gas density, could the relation between BH growth and \mstar\ found in \citet{Yang2017,Yang2018a} simply be reflecting this linkage? Could the relation between BH growth and SFR among bulge-dominated galaxies presented in \citet{Yang2019} simply be a manifestation of the predicted compact starburst with concurrent BH growth?

In this paper, we aim to break such observational degeneracies and probe if BH growth is fundamentally related to host-galaxy compactness, by carrying out a systematic partial-correlation (PCOR) study for a large galaxy sample. This systematic investigation will contribute to the overall understanding of BH-galaxy co-evolution.
This paper is structured as follows. In Section \ref{sec-ds}, we describe the data-assembly process for this work and define our samples. In Section \ref{sec-ar}, we perform data analyses and present the results. We discuss our results in Section \ref{sec-discuss}. We summarize our work and discuss future prospects in Section \ref{sec-sum}.

Throughout this paper, we assume a cosmology with $H_0=70$~km~s$^{-1}$~Mpc$^{-1}$, $\Omega_M=0.3$, and $\Omega_{\Lambda}=0.7$.
A Chabrier initial mass function \citep{Chabrier2003} is adopted.
$M_\star$\ is in units of $M_\odot$.
SFR and black-hole accretion rate are in units of $M_\odot$~yr$^{-1}$.
\sigmae\ and \sigmaone\ are in units of $M_\odot / {\rm kpc}^2$.
$L_X$ indicates X-ray luminosity at rest-frame 2--10 keV in units of erg s$^{-1}$ that has been systematically corrected for absorption (see Section~\ref{ssec-bhar} for further discussion). 
Quoted uncertainties are at the $1\sigma$\ (68\%) confidence level, unless otherwise stated. 
We consider two quantities to be significantly different if the significance level of their difference is greater than 3$\sigma$ ($p$-value $=$ 0.0027), more stringent than the \hbox{``$p$-value $<$ 0.05''} hypothesis testing which can result in a high rate of false positives \citep[e.g.][]{benjamin2018}. When multiple independent hypothesis tests are being conducted simultaneously, we use the Bonferroni correction \citep{bonferroni1936} to adjust the required significance level corresponding to $p$-value = \hbox{0.0027/$n$}, where $n$ is the number of tests. 
We consider a partial correlation to be significant if its test statistic from the PCOR analyses has a $p$-value $<$ 0.0027, which corresponds to a significance level $>3\sigma$.
Significant results throughout the paper are marked in bold in the tables.

\begin{table*}
 \begin{center}
 \caption{Summary of sample properties. (1) CANDELS field name. (2) Field area in arcmin$^2$. (3) Number of galaxies in an $M_\star$-complete sample. The numbers of
galaxies in the $z = 0.5$--1.5/$z = 1.5$--3 range are quoted in parentheses. (4) Number of spec-$z$/photo-$z$ sources. (5) Reference for CANDELS galaxy catalog. (6) Number of X-ray detected galaxies in the sample. The numbers of X-ray detected galaxies in the $z = 0.5$--1.5/$z = 1.5$--3 range are quoted in parentheses. (7) X-ray depth in terms of exposure time. (8) Reference for \textit{Chandra} X-ray catalog.}
  \begin{tabular}{ccccccccc}
  \hline\hline
Field & Area &  Number of  & Number of & Galaxy      & Number of                      &X-ray& X-ray \\
&  (arcmin$^2$) &  Galaxies &  Spec-$z$/Photo-$z$ & Reference  & X-ray Detections &Depth & Reference \\
(1)    &  (2)    &  (3)            &  (4)             &   (5)  &  (6)   &  (7)  &  (8)  \\
\hline
GOODS-S & 170 & 1274 (907/367)   &643/631& \citet{Santini2015} & 284 (182/102) &7 Ms  &  \citet{Luo2017} \\
GOODS-N & 170 & 1645 (1216/429) &355/1290& \citet{Barro2019} & 203 (133/70) &2 Ms & \citet{Xue2016} \\
EGS           & 200 & 2065 (1361/704) &194/1871& \citet{Stefanon2017} &121 (64/57) &800 ks & \citet{Nandra2015} \\
UDS           & 200 & 1863 (1267/596) &227/1636& \citet{Santini2015} & 97 (53/44) &600 ks& \citet{Kocevski2018} \\
COSMOS   & 220 & 1995 (1496/499) &9*/1986& \citet{Nayyeri2017}& 48 (29/19) &160 ks& \citet{Civano2016} \\
\hline
Total           & 960 & 8842 (6247/2595) &1428/7414 &        -                         & 753 (461/292) & - & -\\
\hline
  \end{tabular}
     \newline
     \raggedright *The latest version of the CANDELS/COSMOS catalog is mostly based on photo-$z$.
  \end{center}
  \label{table1}
\end{table*}

\section{Data and Sample Selection} \label{sec-ds}
We perform analyses based on a sample of 8842 galaxies at $0.5 \leqslant z < 3$ in the five CANDELS fields, i.e., GOODS-S, GOODS-N, EGS, UDS, and COSMOS \citep{Grogin2011,Koekemoer2011}.
All of these CANDELS fields have deep multiwavelength observations from \textit{HST}, \textit{Spitzer}, \textit{Herschel}, and ground-based telescopes such as Keck, Subaru, and VLT, enabling high-quality measurements of galaxy morphology (see Section \ref{ssec-morph}), $M_\star$, and SFR (see Section \ref{ssec-msfr}). At the same time, all these fields have deep X-ray observations from \textit{Chandra}, enabling estimation of BH growth utilizing X-ray data (see Section \ref{ssec-bhar}). We define our sample in Section~\ref{ssec-sampleselect}, and the sample properties are summarized in Table~1.

\subsection{Structural and morphology measurements} \label{ssec-morph}
We adopt the structural measurements in \citet{vanderwel2012}\footnote{\citet{vanderwel2012} carry out structural measurements based on CANDELS images processed by the CANDELS team, and \citet{vanderwel2014} perform structural measurements based on CANDELS images processed by the 3D-HST team. For the purpose of consistency, we utilize the results in \citet{vanderwel2012}. Note that for objects in our sample, structural measurements from \citet{vanderwel2012} and \citet{vanderwel2014} agree well.} for CANDELS \textit{HST} $H_{\rm F160W}$-selected objects derived utilizing \texttt{GALFIT} \citep{Peng2002}. 
With background estimated from \texttt{GALAPAGOS} \citep{Barden2012} and point-spread functions constructed using the \texttt{TinyTim} package \citep{Krist1995}, \citet{vanderwel2012} measured structural properties including total magnitude, effective radius (\re), S$\rm \acute{e}$rsic index ($n$), axis ratio, and position angle for all galaxies identified in the CANDELS $H$-band mosaics from single-component S$\rm \acute{e}$rsic model fits, and quantified the systematic and statistical uncertainties utilizing simulated mosaics \citep{Haussler2007}. The detailed assessments of the uncertainty of structural properties including \re\ and $n$ are given in Table~3 of \citet{vanderwel2012}.
The CANDELS \hbox{$J/H$-band} images reach \hbox{$J/H \sim 27$--28}. Thus, even for galaxies with $H \sim 24$--24.5 (which is the magnitude range for the faintest galaxies selected in our sample; see Section~\ref{ssec-sampleselect}), the median signal-to-noise ratio is $\approx 40$. 
For objects with $0.5 \leqslant z < 1.5$, we adopt structural measurements from the \textit{HST} $J$-band (1.25~$\mu \rm{m}$); for objects with $1.5 \leqslant z < 3$, we adopt structural measurements from the \textit{HST} $H$-band (1.6~$\mu \rm{m}$), thus minimizing the effects of the ``morphological $k$-correction'' with all structural measurements being made in the rest-frame optical consistently. 

We utilize the machine-learning-based $H$-band morphology measurements in \citet{HC2015} for CANDELS galaxies with $H < 24.5$ to distinguish bulge-dominated galaxies from galaxies that are not dominated by bulges. Since we only utilize these morphological measurements for a basic selection, and the morphological $k$-correction is weak in the optical/NIR wavelength range \citep[e.g.][]{TM2007}, our results should not be affected qualitatively by the morphological $k$-correction (see Section~3.4 of \citealt{Yang2019} for details).
In this catalog, probabilities that a hypothetical classifier would have voted for a galaxy having a spheroid ($f_{\rm sph}$), a disk ($f_{\rm disk}$), and some irregularities ($f_{\rm irr}$), being point-like ($f_{\rm pt}$) and unclassifiable ($f_{\rm unc}$) are presented. 

We note that the UV-to-near-IR spectral energy distributions (SEDs) of most ($\gtrsim$ 90\%) \hbox{X-ray} AGNs in these fields are dominated by host-galaxy starlight, and thus their morphological measurements should be reliable \citep[e.g.][]{Luo2010,Xue2010, Kocevski2017,Li2019}.

\subsection{Redshift, stellar mass, and star formation rate} \label{ssec-msfr}
The redshift, stellar mass (\mstar), and star formation rate (SFR) used in this paper are identical to those used in \citet{Yang2019}.
We obtain redshift measurements from the CANDELS catalogs (see Table~1). Spectroscopic redshifts (spec-$z$) are adopted when available, and photometric redshifts (photo-$z$) are taken for the rest of galaxies (see Table~1).
Photo-$z$ values for the CANDELS catalogs are of very high quality: they have $\sigma_{\rm NMAD} = 0.018$ and an outlier fraction of 2\% compared with spec-$z$.\footnote{$\sigma_{\rm NMAD}$ is defined as 
$1.48 \times \mathrm{median}(\frac{|\Delta z- \mathrm{median}(\Delta z)|}
{1+z_\mathrm{spec}})$, where $\Delta z$ is the difference between spec-$z$ and photo-$z$. Outliers are those sources with ${|\Delta z|/(1+z_{\rm spec})>0.15}$.}
The CANDELS catalogs also provide $M_\star$ and SFR measurements from independent teams based on SED-fitting utilizing UV-to-NIR photometric bands. The $M_\star$ and SFR used in this work are the median $M_\star$ and SFR values from the five available teams (2$a_\tau$, 6$a_\tau$, 11$a_\tau$, 13$a_\tau$, and 14a).\footnote{For GOODS-N, only three teams are available (2$a_\tau$, 6$a_\tau$, and 14a).} The \mstar~values obtained from SED-fitting are generally robust and insensitive to different parameterizations of the star formation history \citep[e.g.][]{Santini2015}, and there is an overall agreement between different teams. 
While SED-based SFR values are also generally reliable (see Figure~3 of \citealt{Yang2017} for a comparison between SED-based SFR values and SFR values derived from \textit{Herschel} photometry), it has been suggested that the SED-based SFR estimation may underestimate SFR in the high-SFR regime \citep[e.g.][]{Wuyts2011sfr,Yang2017}, where FIR detections are typically expected.
Thus, when robust \textit{Herschel} detections with $\rm S/N > 3$ are available ($\approx 27\%$; \citealt{Lutz2011,Oliver2012,Magnelli2013}), we calculate SFR from FIR photometry to alleviate this issue (using the reddest available \textit{Herschel} band to avoid possible AGN emission). 
For galaxies with $z > 1.5$, we discard all 100 $\mu$m detections to avoid the contamination of hot-dust emission linked with AGN activity at rest-frame $< 40~\mu$m.
For galaxies in the sample we define in Section~\ref{sssec-ss}, the median rest-frame wavelength of utilized \textit{Herschel} detections is $\approx$ 130~$\mu$m, where the AGN emission has limited contribution to the overall emission (that is dominated by galactic emission; e.g. \citealt{Stalevski2016,Zou2019}).
The procedures for calculating SFR from FIR flux are detailed in \citet{Yang2017,Yang2019}. Basically, we utilize star-forming galaxy templates in \citet{Kirkpatrick2012} to derive the total infrared luminosity from the FIR flux, and then convert it to SFR with the equation:
\begin{equation}\label{eq:sfr}
\begin{split}
\frac{\mathrm{SFR}}{M_\odot / {\rm yr}} = 1.09 \times 10^{-10} 
\frac{L_{\rm IR}}{L_\odot}.
\end{split}
\end{equation}
We note that our results do not change qualitatively when using SED-based SFR solely, or perturbing adopted SFR values randomly by $0-0.5$ dex (the typical scatter between FIR-based SFR and SED-based SFR; \citealt{Yang2017}).

\subsection{Black-hole accretion rate}  \label{ssec-bhar}
We calculate sample-averaged BH accretion rate ($\rm \overline{BHAR}$) contributed by both X-ray detected and undetected sources to cover all BH accretion, thus estimating \textit{long-term average BH growth}.
BH accretion has large variability \citep[e.g.][]{Sartori2018,Yuan2018} on the relevant BH-growth time scales ($\sim 10^{6-8}$ yr) that may hide any BH-galaxy connection within individual objects, making \bhar~an ideal estimator for our study.
The inclusion of X-ray undetected sources also enables us to analyze all sources in different CANDELS fields seamlessly with different \hbox{X-ray} depths (see Table~1). 

For each X-ray detected source, we calculate $L_X$ from the \hbox{X-ray} flux reported in the corresponding \hbox{X-ray} catalog assuming a photon index of $\Gamma = 1.7$ \citep[e.g.][]{Yang2016,Liu2017}. Following \citet{Yang2018b}, we choose, in order of priority, hard-band (observed-frame \hbox{2--7}~keV), full-band (observed-frame \hbox{0.5--7~keV}), or soft-band (observed-frame 0.5--2 keV) flux to minimize X-ray obscuration effects.
At $z = 0.5$--3, the hard band can probe rest-frame \hbox{X-ray} flux up to $10.5$--28 keV, enabling good estimation of $L_X$ until the column density reaches $N_{\rm H} \sim 10^{23} \rm~cm^{-2}$. 
For X-ray detected galaxies in the sample defined in Section~\ref{sssec-ss}, $\approx$ 62\% of them have hard-band detections; full-band detections are utilized for $\approx$ 31\% of them; soft-band detections are utilized for $\approx$ 7\% of them.
Utilizing bright X-ray sources in the CDF-S, \citet{Yang2018b} compare the X-ray flux obtained via this scheme of band choice with the absorption-corrected X-ray flux in \citet{Luo2017}, and show that the underestimation of X-ray flux due to obscuration in this scheme is typically small ($\approx 20\%$).
Following \citet{Yang2019}, we increase the X-ray fluxes of our X-ray sources by 20\% to account for the systematic effects of obscuration.\footnote{Since \citet{Yang2018b} utilized CDF-S X-ray sources above the COSMOS flux limits to assess obscuration, the derived obscuration correction factor should be applicable to bright X-ray sources in all the survey fields in this paper, which contribute most of the accretion power. We have also verified that X-ray detected galaxies in different survey fields utilized in the paper do not have significant differences in the average hardness ratio, demonstrating similar levels of X-ray obscuration.}
For X-ray undetected sources, we employ the stacking results from \citet{Yang2019} to estimate their X-ray emission.

With $L_{\rm X}$ for each individual X-ray detected source and the average X-ray luminosity for any group of X-ray undetected sources obtained via stacking ($\overline{L_{\rm X,stack}}$),
the average AGN bolometric luminosity for a sample of sources can be calculated as \citep{Yang2019}: 

\begin{equation}\label{equ:lx}
 \overline{L_{\rm bol}}  = \frac{ \bigg[{\mathlarger{\sum\limits_{n=0}^{N_{\rm det}}}} (L_{\rm X} -L_{\rm X,XRB}) k_{\rm bol}\bigg]+
 		(\overline{L_{\rm X, stack}} -  \overline{L_{\rm X, XRB}} )N_{\rm non}  \overline{k_{\rm bol}}}
		{N_{\rm det}+N_{\rm non}}
\end{equation}
Here, $N_{\rm det}$ and $N_{\rm non}$ represent the numbers of X-ray detected and undetected sources in the sample, respectively.
The summation in the first term of the numerator is over all X-ray detected galaxies.
Note that when deriving $\overline{L_{\rm X,stack}}$, some X-ray undetected galaxies are too close to X-ray sources to be stacked ($\approx 12\%$). However, they are still included when counting $N_{\rm non}$, and thus are appropriately accounted for statistically.
$L_{\rm X, XRB}$ is the expected luminosity from X-ray binaries (XRBs) for each individual X-ray detected source, and $\overline{L_{\rm X,XRB}}$ is the average expected XRB luminosity for the stacked sources.
$L_{\rm X, XRB}$ and $\overline{L_{\rm X,XRB}}$ are obtained from model 269 of \citet{Fragos2013}, which describes XRB X-ray luminosity as a redshift-dependent linear function of $M_\star$ and SFR, utilizing observations at $z = 0$--7 by \citet{Lehmer2016}. XRBs typically contribute $\approx 10-25\%$ of the total X-ray luminosity in the sample, and thus our analyses should not be affected materially by the uncertainties related to the XRB modeling.
$k_{\rm bol}$ and $\overline{k_{\rm bol}}$ are the $L_{\rm X}$-dependent bolometric corrections at $L_{\rm X} - L_{\rm X, XRB}$ and $\overline{L_{\rm X, stack}} - \overline{L_{\rm X, XRB}}$, respectively, calculated from the model in \cite{Hopkins2007} and then multiplied by a factor of 0.7 to reconcile the overestimation due to the double counting of IR reprocessed emission (see Footnote 4 of \citealt{MH2013}). 

Assuming a constant radiative efficiency of $\epsilon = 0.1$ \citep[e.g.][]{Brandt2015,Yang2019}, we can convert $\overline{L_{\rm bol}}$ to $\rm \overline{BHAR}$ as:
\begin{equation}\label{equ:bhar}
\begin{split}
\overline{\mathrm{BHAR}} &= \frac{(1-\epsilon)
			\overline{L_{\rm bol}}}{\epsilon c^2} \\
	 &= \frac{1.58 \overline{L_{\rm bol}}}{10^{46}\ \rm{erg~s^{-1}}}
	    M_{\sun}\ \mathrm{yr}^{-1}
\end{split}
\end{equation}
\noindent
The uncertainties on $\rm \overline{BHAR}$ are obtained by bootstrapping the sample 1000 times.

\subsection{Sample construction} \label{ssec-sampleselect}
\subsubsection{Sample selection} \label{sssec-ss}
First, we select all $H <$ 24.5 galaxies from the CANDELS \textit{HST} $H$-band selected catalogs \citep{Santini2015,Nayyeri2017,Stefanon2017,Barro2019}.
We note that all $H < 24.5$ galaxies have structural and morphological measurements from \citet{vanderwel2012} and \citet{HC2015}, and thus we should not have any biases due to systematic incompleteness issues when performing sample construction.
Then, following \citet{Yang2019}, we exclude $\approx$ 8\% of sources that have $f_{\rm unc}$ or $f_{\rm pt}$ greater than any of $f_{\rm sph}$, $f_{\rm disk}$, and $f_{\rm irr}$, to exclude stars, broad-line (BL) AGNs, and spurious detections. 
We also discard the 79 spectroscopic BL AGNs reported in the literature (\citealt{Barger2003,Silverman2010,Cooper2012,Newman2013,Marchesi2016}; Suh et al., in prep.).
BL AGNs are excluded since their host-galaxy starlight measurements are typically contaminated by AGN light, significantly affecting the \mstar, SFR, and morphology measurements. Assuming the unified model (e.g. \hbox{\citealt{Antonucci1993,Netzer2015}}), we note that the exclusion of BL AGNs will not qualitatively change our results: if BL AGNs are purely AGNs observed at certain orientations (not intercepting the torus), a group of BL AGNs sharing similar host properties should have average X-ray luminosity close to that of a group of type 2 AGNs with the same host properties, and the relative fraction of BL AGNs among all AGNs should not change significantly with host properties. Evidence for the validity of these assumptions to first order is given in \citet{Merloni2014} and \citet{Zou2019}. Thus, excluding BL AGNs only decreases a similar fraction of \bhar~for bins and subsamples utilized in Section \ref{sec-ar}, which should not affect the existence of trends between \bhar~and host properties.

\begin{figure}
\begin{center}
\includegraphics[scale=0.6]{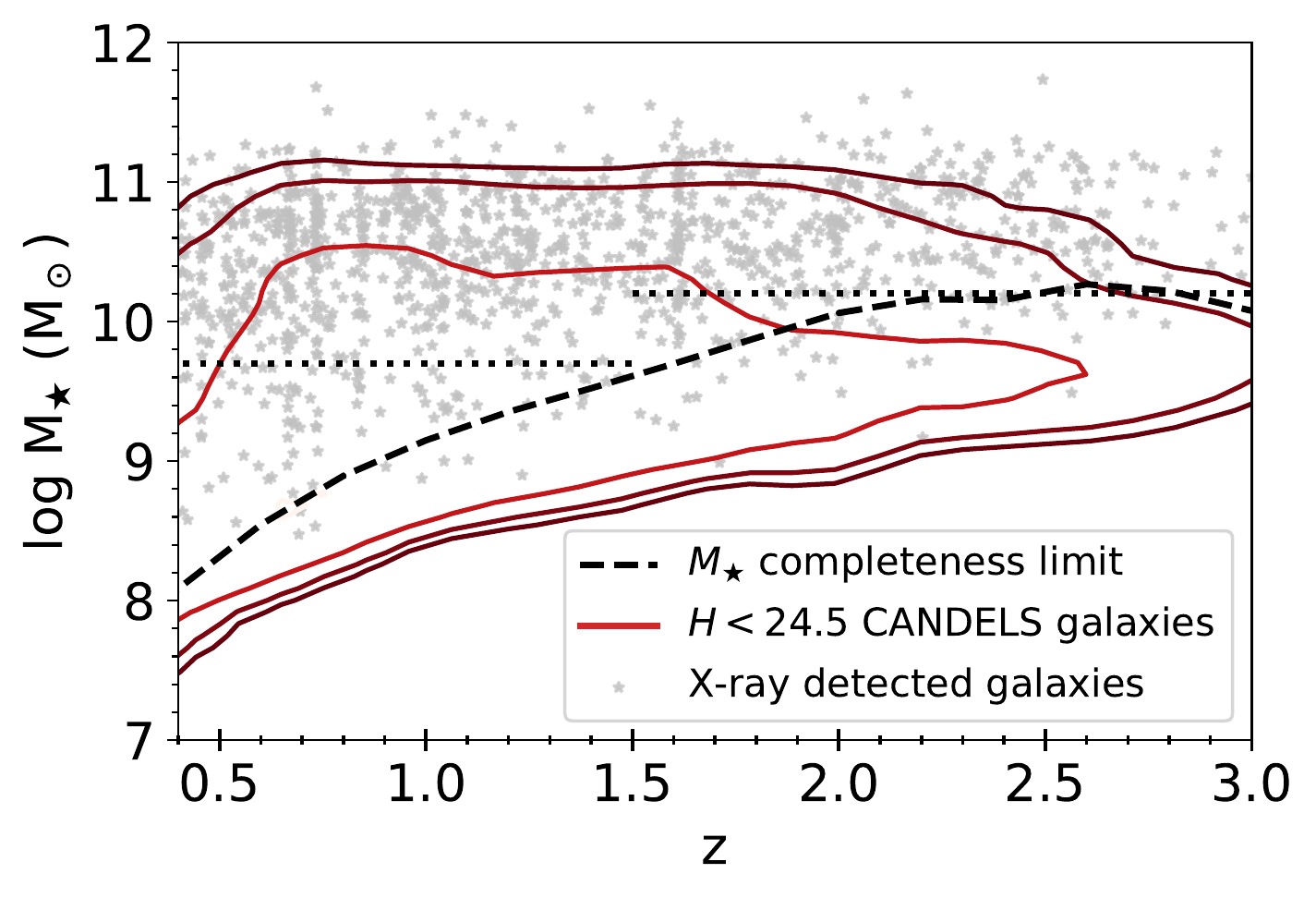}
\caption{$M_\star$ as a function of redshift. The contours encircle 68\%, 90\%, and 95\% of all $H <$ 24.5 galaxies. The gray stars represent X-ray detected sources. The dashed curve indicates the $M_\star$ completeness limit (see Section~\ref{sssec-ss}). The horizontal dotted lines represent our $M_\star$-completeness cuts for the $z = 0.5-1.5/1.5-3$ samples selected in Section~\ref{sssec-ss}.}
\label{mz}
\end{center}
\end{figure}

We limit our analyses to an $M_\star$-complete (corresponding to $H <$ 24.5) sample.
The limiting $M_\star$ ($M_{\rm lim}$) for $H <$ 24.5 is displayed in Figure \ref{mz}.
 The $M_{\rm lim}$-redshift curve is derived based on an empirical method \citep[e.g.][]{Ilbert2013}. 
We first divide our sources into narrow redshift bins with width of $\Delta z = 0.2$. For each redshift bin,
we calculate ${\rm log} M^{\rm ind} _{\rm lim} = {\rm log} M_\star + 0.4 \times (H - 24.5)$ for individual galaxies in the bin.
We then adopt $M_{\rm lim}$ as the 90th percentile of the $M^{\rm ind}_{\rm lim}$ distribution for the redshift bin.

For the studies in Section \ref{sec-ar}, we divide the objects into two redshift bins: $0.5 \leqslant z < 1.5$ and $1.5 \leqslant z < 3$, to probe if the relation between BH growth and host-galaxy compactness changes over cosmic time, and alleviate the influence of the cosmic evolution of compactness \citep[e.g.][]{Barro2017} in our study.
Since the limiting $M_\star$ at $z=1.5$ and $z=3.0$ are ${\rm log} M_{\star} \approx 9.7$, and ${\rm log} M_{\star} \approx 10.2$ ($M_{\star}$ in units of $M_\odot$), respectively, 
we limit our analyses to ${\rm log} M_{\star} > 9.7$ and ${\rm log} M_{\star} > 10.2$ galaxies for the low-redshift and high-redshift bins, respectively. 
The relatively broad redshift bins are necessary to provide sufficiently large samples for our statistical analyses. We also require \texttt{GALFIT\_flag = 0} for the selected galaxies, which includes $\approx$ 86\% of sources in the \mstar-complete sample. The sample properties are shown in Table~1. Here, \texttt{GALFIT\_flag = 0} indicates good quality of the structural parameters. Sources with \texttt{GALFIT\_flag = 1} (9\% of the sample) are less certain: they are not necessarily bad fits, but their magnitudes do not fall within the 3$\sigma$ confidence intervals of the magnitude integrated from the light profile measured with \texttt{GALFIT}. We do not include them in the sample to avoid large systematic uncertainties induced by those uncertain measurements, but our results do not vary qualitatively when adding those uncertain sources (see Appendix \ref{append-flag1}). \texttt{GALFIT\_flag = 2} indicates sources with one or more parameters reaching the constraint set in \texttt{GALFIT}, which means that the derived structural parameters are not meaningful.  \texttt{GALFIT\_flag = 3} indicates non-existing results. Thus, we do not consider a flag value of 2 or 3 (5\%) for the purpose of this work.

\subsubsection{Sample division} \label{ssec-sd}
For sources in our samples, we classify them as bulge-dominated (\hbox{$\approx25\%$}; 2212 galaxies) if they have $f_{\rm sph} \geqslant 2/3$, $f_{\rm disk} < 2/3$, and $f_{\rm irr} < 1/10$, and those that do not satisfy the criteria (that are not dominated by bulges) are classified into the non-bulge sample (6630 galaxies).
This classification approach is supported by visual inspection of the galaxies (see \citealt{Yang2019} for more details).
Hereafter, we will call the bulge-dominated sample the ``BD sample'', and the sample of galaxies that are not dominated by bulges the ``Non-BD sample'' in short.

\begin{figure*}
\begin{center}
\includegraphics[scale=0.6]{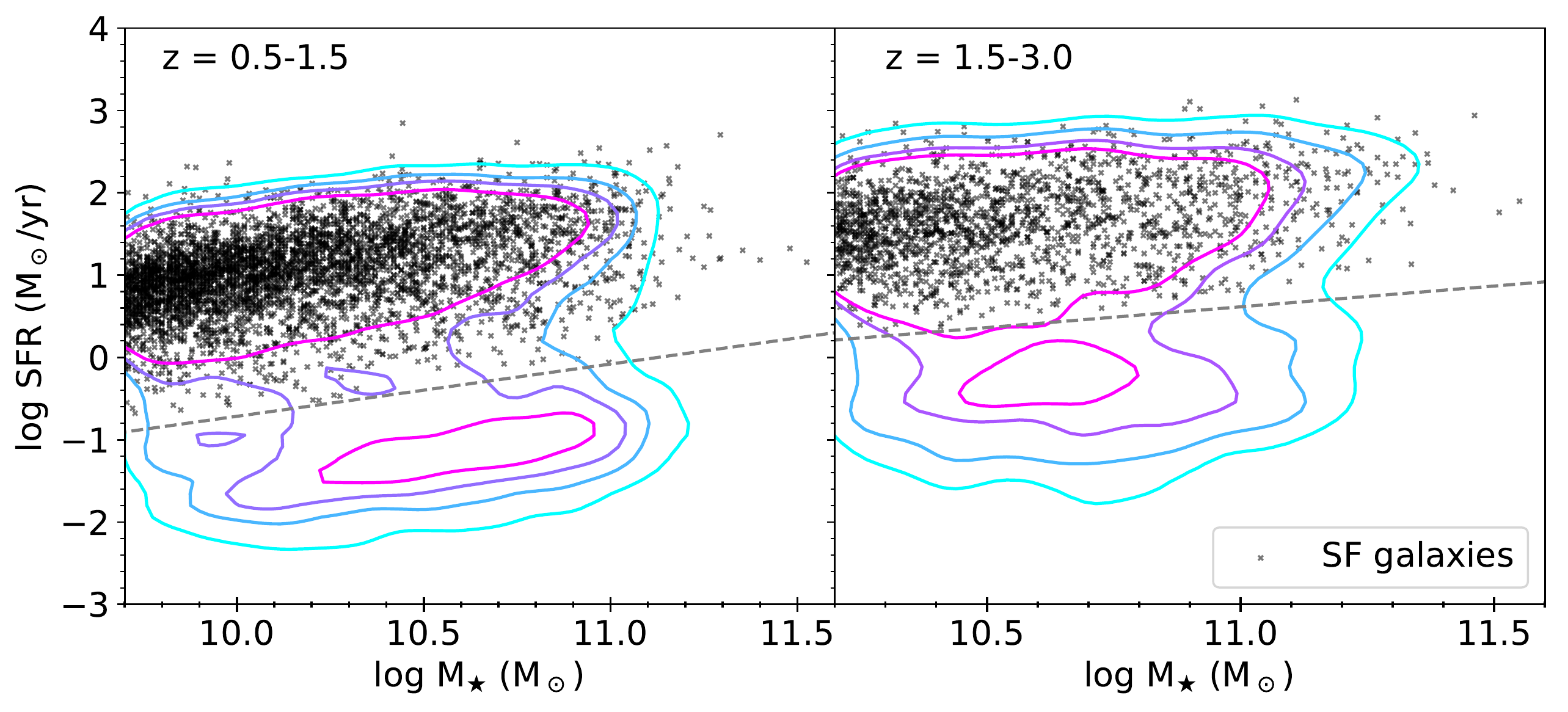}
\caption{SFR vs. stellar mass for galaxies in the low-redshift bin (left) and the high-redshift bin (right).
The contours encircle 68\%, 80\%, 90\%, and 95\% of galaxies in our sample. The black crosses mark the star-forming galaxies.
The gray dashed line in the left/right panel shows the division between SF galaxies and quiescent galaxies at $z = 0.5$/$z = 1.5$.
Our selection of star-forming galaxies roughly corresponds to galaxies lying above the minimum in the distributions of SFRs.}
\label{sfornot}
\end{center}
\end{figure*}

We use the line that is 1.3 dex below the star formation main sequence derived in \citet{Whitaker2012} at the appropriate redshift and stellar mass to divide star-forming (SF) galaxies from quiescent galaxies. We classify a galaxy as SF if its SFR value is above the line.
Our selection of star-forming galaxies roughly corresponds to galaxies lying above the local minimum in the distribution of SFRs between star-forming and quiescent galaxies (see Figure~\ref{sfornot}). We create a sample of 739 star-forming galaxies in the BD sample (hereafter ``SF BD'' in short), and a sample of 5662 star-forming galaxies in the Non-BD sample (hereafter ``SF Non-BD'' in short), where cold gas is surely available among galaxies.

\subsubsection{Measuring the host-galaxy compactness} \label{sssec-c}
\begin{figure*}
\begin{center}
\includegraphics[scale=0.6]{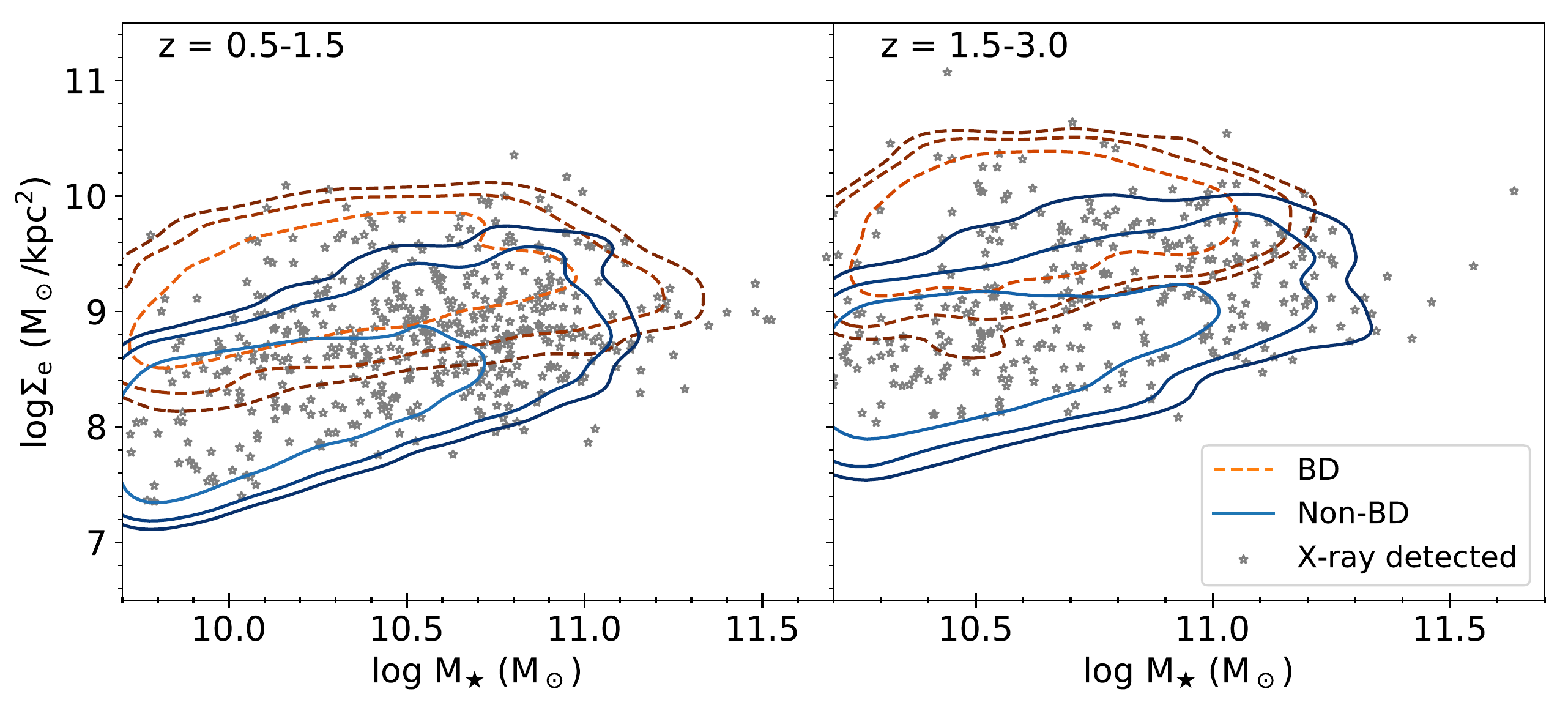}
\caption{Surface-mass density (\sigmae) vs. stellar mass for galaxies in the low-redshift bin (left) and the high-redshift bin (right).
The orange dashed contours encircle 68\%, 90\%, and 95\% of galaxies in the BD sample, and the blue solid contours encircle 68\%, 90\%, and 95\% of galaxies in the Non-BD sample. The gray stars mark the X-ray detected sources.
Galaxies in the BD sample generally have higher \sigmae\ than galaxies in the Non-BD sample.
}
\label{msigmae}
\end{center}
\end{figure*}

To measure the host-galaxy compactness, we first calculate the surface-mass density for galaxies in our sample as \sigmae~$ = 0.5 M_\star/\pi r_{\rm e}^2$. The effective radius $r_{\rm e}$ (measured along a galaxy's major axis) can be measured with a statistical uncertainty of 20$\%$ or better for galaxies with $H \lesssim 24.5$ \citep{vanderwel2012}.
Since \re\ is measured along a galaxy's major axis, note that the surface-mass density here is the surface-mass density when viewed face-on, where we assume approximately circular symmetry of galaxies.
The surface-mass density versus \mstar~is presented in Figure \ref{msigmae}.

We also calculate the projected central surface-mass density within 1 kpc ($\Sigma_1$) for galaxies in the sample.
Following \citet{Lee2018}, we numerically extrapolate $\Sigma_1$ from the best-fit S$\rm \acute{e}$rsic profile in \citet{vanderwel2012}:

\begin{equation} 
  I(r) = I_o \exp {\left \{ {-b_n \left[{\left(\frac{r}{r_e}\right)}^{1/n}-1\right]}\right \} } 
\end{equation}

\noindent
In the equation, $I(r)$ represents light intensity at a radius of $r$, and $I_o$ is the light intensity at \re. We take the asymptotic approximation for $b_n$ as a function of S$\rm \acute{e}$rsic index $n$ following \citet{CB1999}:
\begin{equation}
b_n \approx 2n - \frac{1}{3}+\frac{4}{405n}+\frac{46}{25515n^2}
\end{equation}

\noindent
Assuming a constant mass-to-light ratio throughout the galaxy, the projected central surface-mass density within 1 kpc ($\Sigma_1$) can be obtained with the following equation:
\begin{equation} 
\Sigma_{1} = \frac{\int_{0}^{1~\mathrm{kpc}}I(r)2\pi rdr}{\int_{0}^{\infty}I(r)2\pi rdr}\frac{L_{\mathrm{GALFIT}}}{L_{\mathrm{phot}}}\frac{M_{\star}}{\pi (1~\mathrm{kpc})^{2}}
\label{eq:s1}
\end{equation}

\noindent
Here, $L_{\mathrm{phot}}$ is the total luminosity adopted from the CANDELS catalogs \citep{Galametz2013,Guo2013,Stefanon2017,Barro2019} in the filter corresponding to the structural measurements, and $L_{\mathrm{GALFIT}}$ is the integrated luminosity from \texttt{GALFIT}. The $L_{\mathrm{GALFIT}}/L_{\mathrm{phot}}$ correction term is applied following Section~3.1 of \citet{vanDokkum2014}, with a median value of 1.11 and a scatter of 0.09 for objects in our sample. We note that the projected $\Sigma_1$ values are calculated assuming galaxies follow the measured S$\rm \acute{e}$rsic profiles in the central 1 kpc region.
This measurement of central mass density has its limitations, since the light profiles of some galaxies in the central 1 kpc deviate from the global S$\rm \acute{e}$rsic profile. Utilizing the galaxy cutouts, models, and fitting residuals provided along with \citet{vanderwel2012}, we found that only $\approx 2\%$ of galaxies in our sample have total fitting residuals in the central 1~kpc region greater than 20\% of the enclosed flux within 1~kpc. Thus, our use of the projected $\Sigma_1$ values should be acceptable generally given the fairly mild deviations from the global S$\rm \acute{e}$rsic profiles in the central 1 kpc regions of galaxies. 
We also verified that the analysis results in Section \ref{sec-ar} do not change qualitatively when excluding the $\approx 2\%$ of galaxies with $\geqslant 20\%$ deviations.

\begin{figure*}
\begin{center}
\includegraphics[scale=0.6]{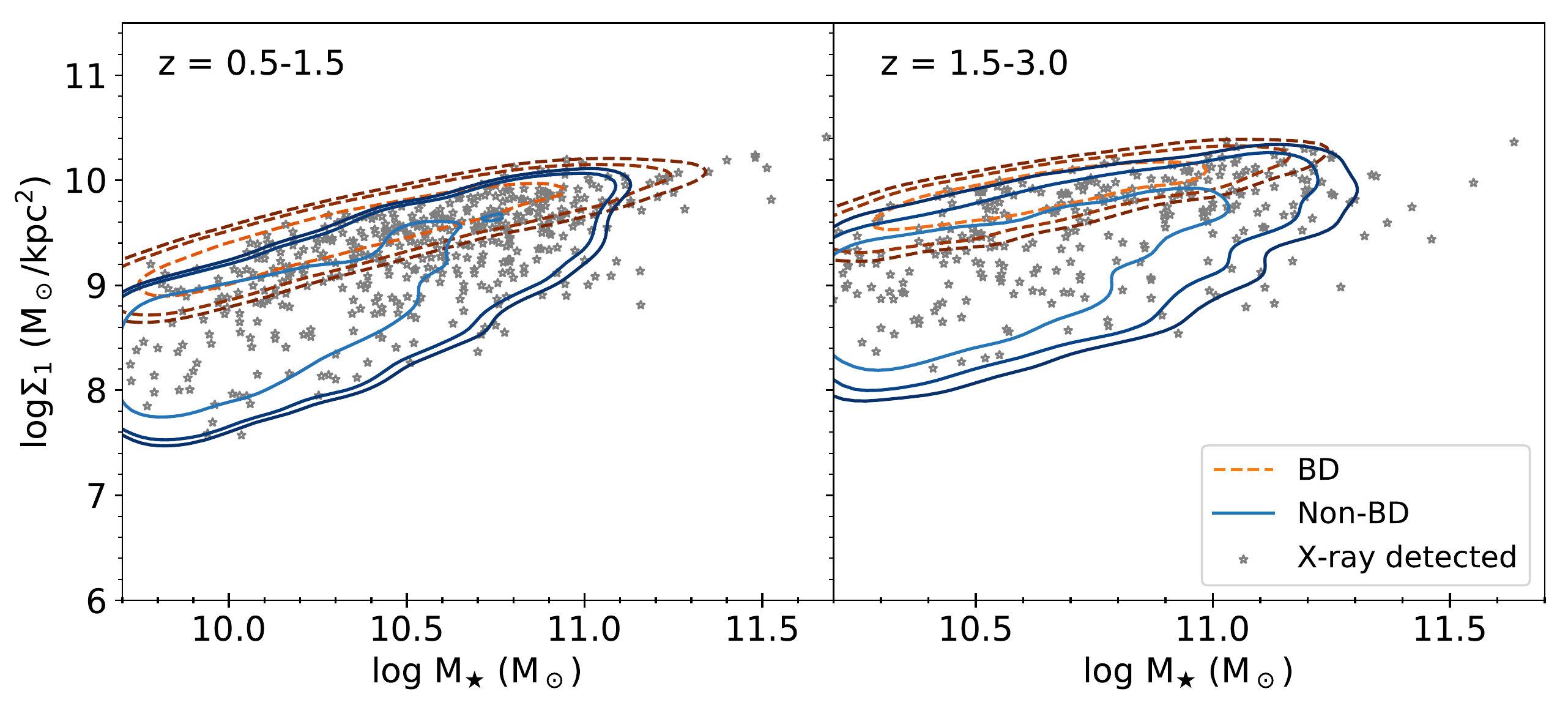}
\caption{Projected central surface-mass density within 1 kpc ($\Sigma_1$) vs. stellar mass for galaxies in the low-redshift bin (left) and the high-redshift bin (right).
The orange dashed contours encircle 68\%, 90\%, and 95\% of galaxies in the BD sample, and the blue solid contours encircle 68\%, 90\%, and 95\% of galaxies in the Non-BD sample.
The gray stars mark the X-ray detected sources.
Galaxies in the BD sample generally have higher \sigmaone\ than galaxies in the Non-BD sample.
}
\label{m-rho}
\end{center}
\end{figure*}

The projected central stellar density within 1 kpc versus mass for galaxies is presented in Figure \ref{m-rho}.
In general, our projected \sigmaone\ values are similar to those of \citet{Barro2017}, who measured $\Sigma_1$ values from stellar-mass profiles computed by fitting multiband SEDs derived from surface-brightness profiles in \textit{HST} bands, with a systematic offset of $\approx 0.1$ dex and a scatter of $\approx 0.3$ dex. This agreement further indicates that our assumption of a constant mass-to-light ratio roughly holds.\footnote{A caveat here is that SF galaxies may not follow this assumption as well as quiescent galaxies, as expected from their star-formation activity. For quiescent galaxies in our sample, the systematic offset of \sigmaone\ values compared with \citet{Barro2017} is $\approx 0.0$~dex, and the scatter is $\approx 0.1$~dex; for SF galaxies in our sample, the systematic offset is $\approx 0.2$~dex, and the scatter is $\approx 0.3$~dex. Also, as expected from the presence of pseudo-bulges in the Non-BD sample, SF Non-BD galaxies may not follow this assumption as well as SF BD galaxies: for galaxies in the SF BD sample, the systematic offset is $\approx 0.0$ dex, and the scatter is $\approx 0.2$ dex; for galaxies in the SF Non-BD sample, the systematic offset is $\approx 0.2$ dex, and the scatter is $\approx 0.3$ dex. Even still, these systematic offsets and scatters are acceptable since all analyses in this paper are performed based on sample-averaged values in relatively broad bins.} In addition, unlike the measured \sigmaone\ values, our projected $\Sigma_1$ values are relatively robust against possible AGN contamination since they are extrapolated from global \sersic\ profiles (although our \sigmaone\ values for X-ray AGNs are also similar to those of \citealt{Barro2017}).
We also define the central mass concentration parameter within 1 kpc ($C_1$)\footnote{Note that $C_1$ is different from the concentration parameter in the ``CAS'' definition \citep{Conselice2003}.}
which is independent of \mstar:
\begin{equation} 
C_{1} =  \frac{\int_{0}^{1~\mathrm{kpc}}I(r)2\pi rdr}{\int_{0}^{\infty}I(r)2\pi rdr}
\label{eq:c1}
\end{equation}

The uncertainties in \sigmaone\ and $C_1$ are propagated from the uncertainties of $n$ and \re. \citet{vanderwel2012} state that reliable measurements of basic size and shape parameters should be reached down to $H = 24.5$. 
For each object in the sample, we quantify the uncertainty of $C_1$ through computing 1000 $C_1$ values from \re\ values and $n$ values with random offsets.
The offsets of \re\ are randomly drawn from the Gaussian distribution that has the 1$\sigma$ measurement error of \re\ as the standard deviation; the offsets of $n$ are coupled with the random offsets generated for \re, as the errors of \re\ and $n$ are strongly correlated. The relation between \re\ errors and $n$ errors is adopted from Section 2.4 of \citet{Whitaker2017}.
We note that even for galaxies with $H \sim 24$--24.5, the median uncertainty of \hbox{log $C_1$} is $\approx 10\%$.
However, \citet{vanderwel2012} also suggest that $n$ could only reach the accuracy of $r_{\rm e}$ among galaxies with $H \sim 24.5$ when measured at $H \sim 23.5$. To assess this potential bias, we confirm that our results in Section \ref{sec-ar} do not change when limiting the analyses to $H < 23.5$ objects in the sample. 

\begin{figure*}
\begin{center}
\includegraphics[scale=0.35]{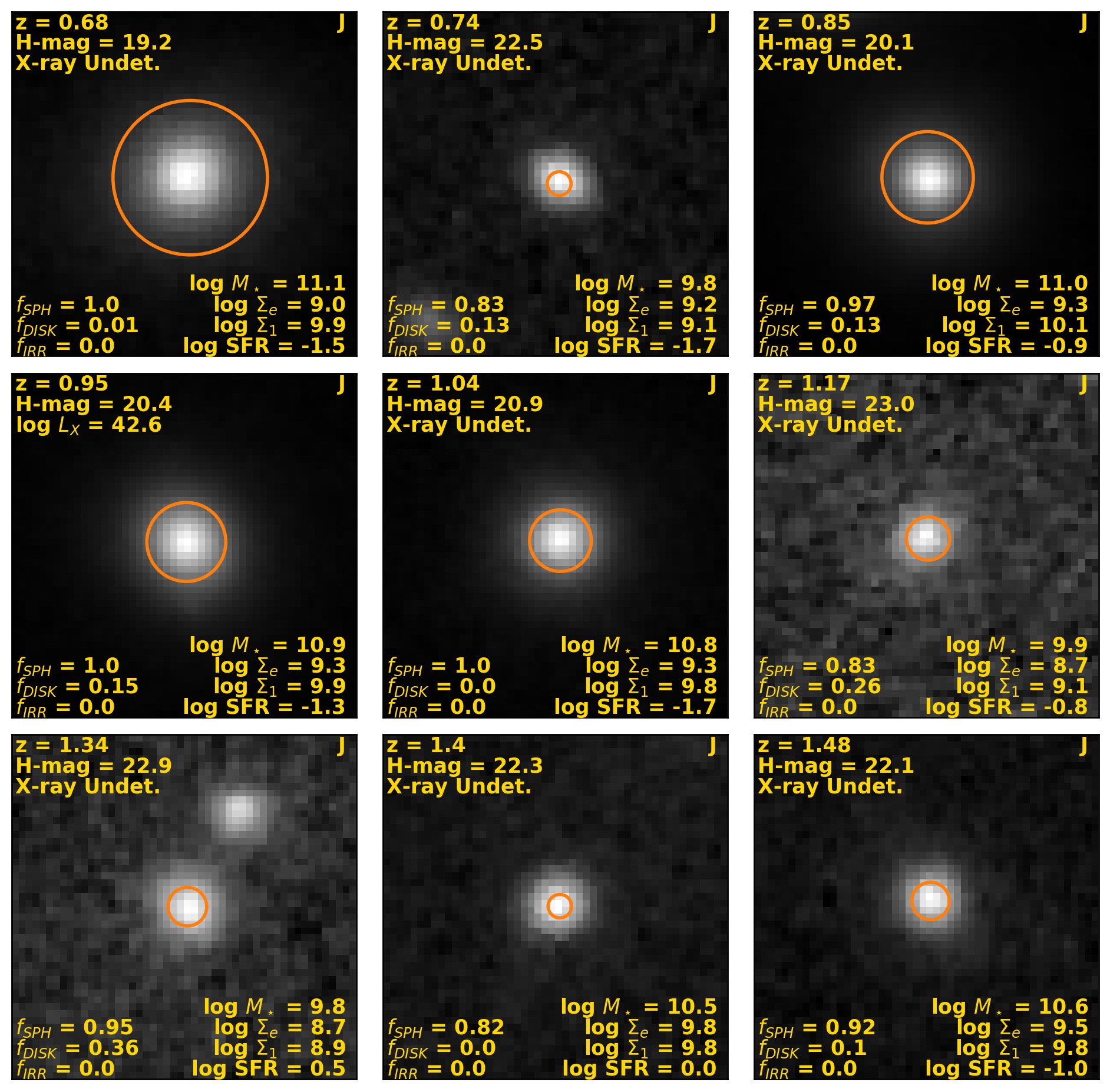}
~
\includegraphics[scale=0.35]{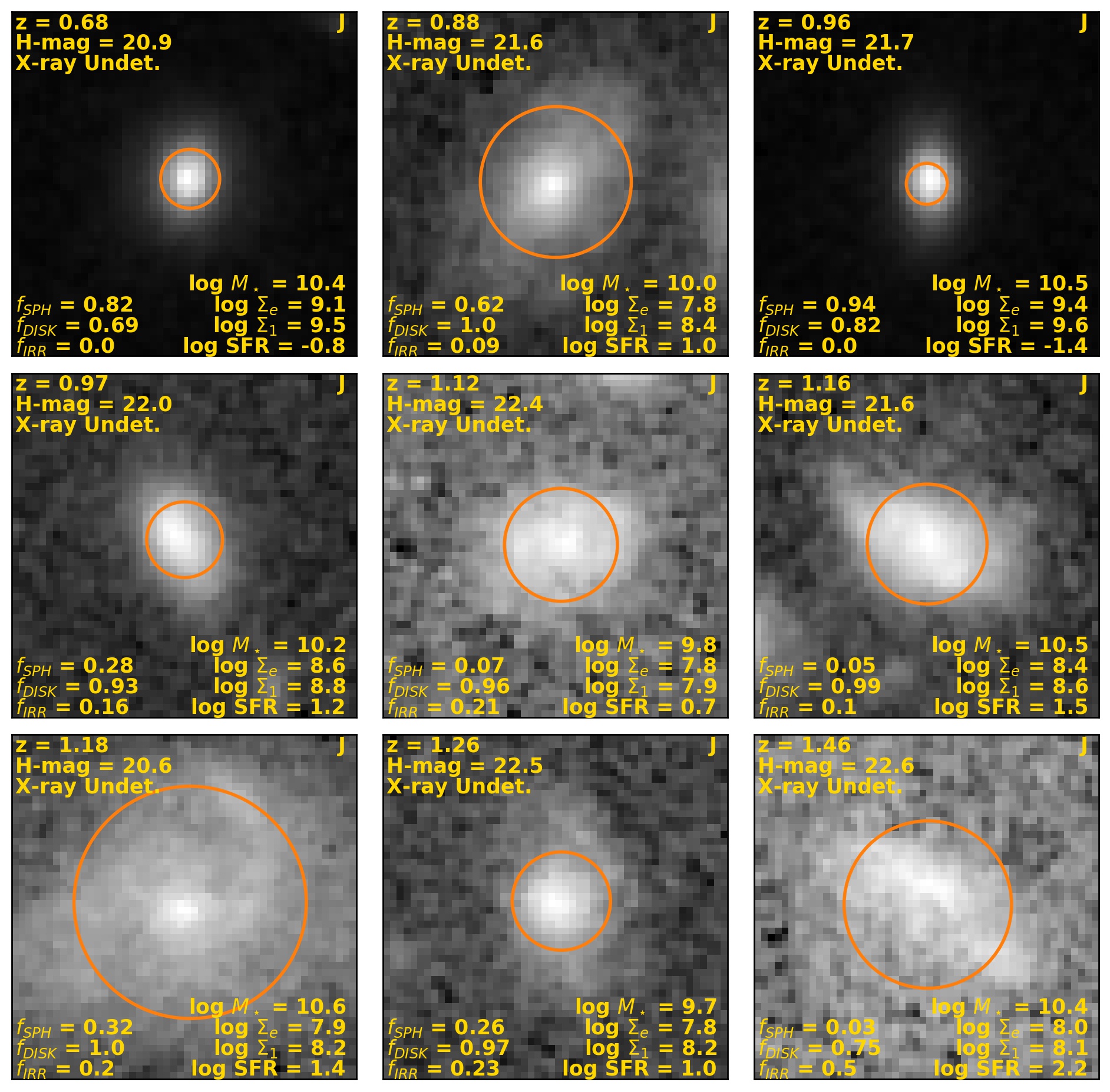}\\
\vspace{0.2cm}
\includegraphics[scale=0.35]{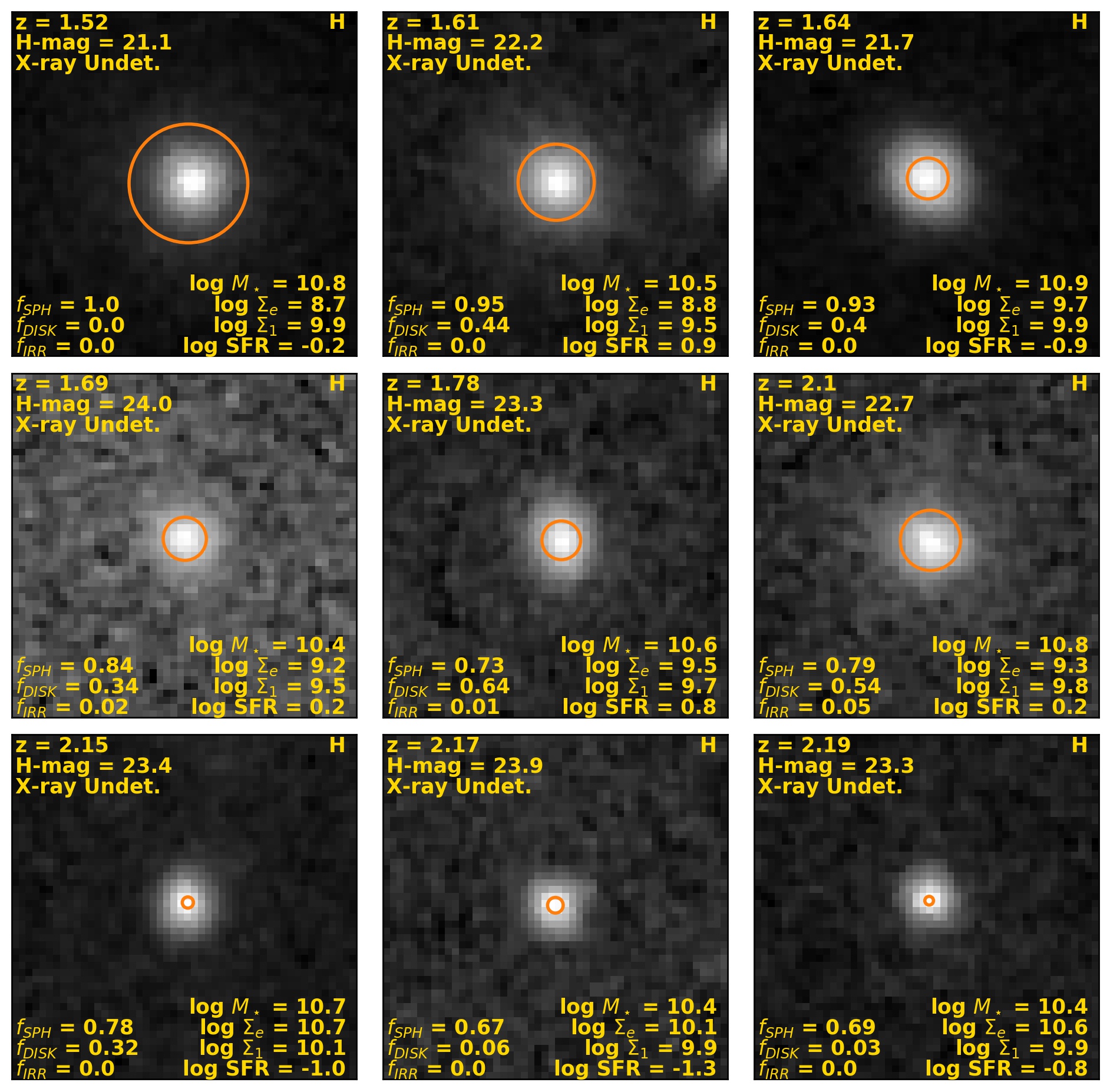}
~
\includegraphics[scale=0.35]{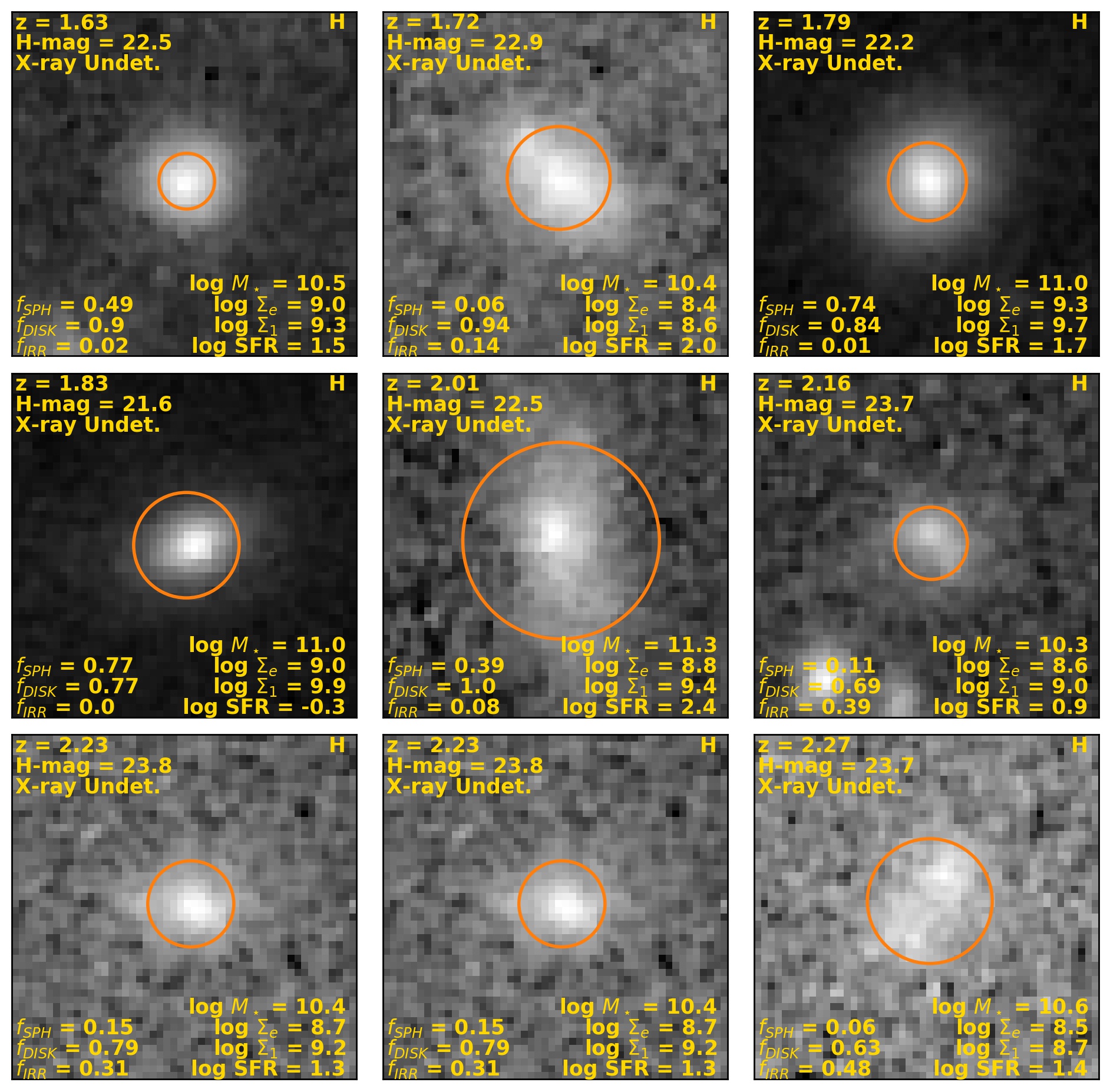}
\caption{Example $J$-band/$H$-band $3'' \times 3''$ cutouts with asinh normalization (for purposes of display) for galaxies at $z =$ 0.5--1.5/1.5--3. The galaxies are randomly selected, and they are placed at the center of each cutout. 
The orange circles show the effective radius ($r_{\rm e}$) from \citet{vanderwel2012} centered at the galaxy position; note \re\ is measured along a galaxy's major axis.
The left three columns are drawn from the BD sample, while the right three columns are drawn from the Non-BD sample.
The first three rows are drawn from the low-redshift bin ($z = 0.5$--1.5) and the last three rows are drawn from the high-redshift bin ($z = 1.5$--3).}
\label{morph}
\end{center}
\end{figure*}

In Figure~\ref{morph}, we show some random $J/H$-band cutouts for galaxies at $z = $ 0.5--1.5/1.5--3, with their properties (including redshift, morphology, \mstar, \sigmae, \sigmaone, and SFR) listed.

\section{Analyses and Results} \label{sec-ar}
In this section, we use the analysis methods described in Section~\ref{ssec-method} to study how BH growth relates to host-galaxy compactness, represented by \sigmae\ (in Section \ref{ssec-sigmae}) or \sigmaone\ (in Section \ref{ssec-sigma1}), when controlling for SFR or \mstar. While \sigmae\ and \sigmaone\ have both been used to represent the compactness of galaxies, \sigmae\ measures the mass-to-size ratio in the central 50\% of galaxies, and \sigmaone\ measures the mass-to-size ratio in the central 1~kpc of galaxies (see Section \ref{sssec-c}). Thus, testing how BH growth relates to both \sigmae\ and \sigmaone\ when controlling for SFR or \mstar\ can not only reveal if BH growth links with host-galaxy compactness fundamentally, but also if BH growth links more fundamentally with the compactness of the central 1~kpc regions of galaxies than the central 50\% regions of galaxies.

\subsection{Analysis methods} \label{ssec-method}
\citet{Yang2019} show that for bulge-dominated galaxies, $\rm \overline{BHAR}$ correlates with SFR when controlling for $M_\star$, while the converse does not hold true. They also show that for galaxies that are not dominated by bulges, $\rm \overline{BHAR}$ correlates with $M_\star$ when controlling for SFR, while the converse does not hold true. 
Thus, for the BD sample, we will study whether \bhar\ is mainly related to SFR or $\Sigma$ (see Sections \ref{sssec-bulge}/\ref{sssec-bulges1}). We will also confine the study to SF BD galaxies only in Sections \ref{sssec-bulge} and \ref{sssec-bulges1}. For galaxies in the Non-BD sample, we will study whether \bhar\ is mainly related to \mstar\ or $\Sigma$ (see Sections \ref{sssec-nonbulge}/\ref{sssec-nonbulge-s1}). Similarly, we will also confine the study to SF Non-BD galaxies only in Sections \ref{sssec-nonbulge} and \ref{sssec-nonbulge-s1}.
The motivation for confining analyses to SF galaxies only is that compactness may serve as an indicator of the amount of gas in the centers of galaxies when we know that there is cold gas available, and simulations predict that BH accretion is linked with the central gas density \citep[e.g.][]{Wellons2015,Habouzit2019}. Otherwise, if galaxies become quiescent, it is unlikely that compactness will indicate the central gas density.

For galaxies in the BD or SF BD (Non-BD or SF Non-BD) samples, we will first divide them into SFR (\mstar) bins with approximately the same number of sources per bin. We will also divide each SFR (\mstar) bin into two subsamples based on $\Sigma$. \bhar\ and its 1$\sigma$ confidence interval (obtained via bootstrapping) will be calculated for each bin and subsample, and presented in a plot of \bhar\ as a function of SFR (\mstar). 
We will also check if there is a significant difference in \bhar\ between subsamples ($\Delta \rm \overline{BHAR} = \rm \overline{BHAR}_{subsample~1}- \rm \overline{BHAR}_{subsample~2}$). The significance level of \dbhar\ is obtained by dividing it by its 1$\sigma$ uncertainty, which is derived from bootstrapping as (84th$-$16th percentile)/2 of the $\Delta \rm \overline{BHAR}$ distribution. For each bin, we will report the significance level of \dbhar\ between two subsamples on the plot if the level is $> 3\sigma$. 
We will then divide galaxies in the BD (Non-BD) sample into $\Sigma$ bins with approximately the same number of sources per bin. We will also divide each $\Sigma$ bin into two subsamples based on SFR (\mstar). Similarly, we will calculate \bhar\ and its 1$\sigma$ confidence interval for each bin and subsample, and present this in a plot of \bhar\ as a function of $\Sigma$. Significant \dbhar\ between two subsamples will be reported on the plot.
We note that when dividing a sample of galaxies into several bins with approximately the same number of sources per bin based on a certain galaxy property, we ensure that the bin size is large enough to provide reasonable statistical constraints on \bhar/\agnf.
The BD sample has 1539/673 galaxies at $z = 0.5$--1.5/1.5--3; we will utilize 3 bins in both redshift ranges, so each bin contains $\approx 500/200$ galaxies.
The SF BD sample has only 516/223 galaxies at $z = 0.5$--1.5/1.5--3; we will thus only use 1 bin in both redshift ranges, and will just report the result instead of showing the plot.
The Non-BD (SF Non-BD) sample has 4708/1922 (4045/1617) galaxies at $z = 0.5$--1.5/1.5--3; we will utilize 6/3 bins, so each bin contains $\approx 800/600$ ($\approx 700/500$) galaxies.
The relevant plots here for the BD, Non-BD, and SF Non-BD samples are Figures \ref{bulge_trend}, \ref{comp_trend}, and \ref{comp_sf_trend} when \sigmae\ is utilized to measure compactness;
Figures \ref{bulge_trend_s1}, \ref{comp_trend_s1}, and \ref{comp_trend_s1_sf} are relevant when \sigmaone\ is utilized to measure compactness.

We will repeat the analyses described above with AGN fraction ($f_{\rm AGN}$; the fraction of sources with log $L_{\rm X} > 42$)\footnote{We choose this ``log $L_{\rm X} > 42$'' criterion to select AGNs consistently with pervious works, including \citet{Kocevski2017}. We note that we cannot obtain a complete selection of objects with \hbox{log $L_{\rm X} > 42$} at \hbox{$z \sim 0.7$--3} considering the X-ray flux detection limits of COSMOS, UDS, and EGS \citep{Nandra2015,Civano2016,Kocevski2018}. Since we mainly utilize this criterion to probe the potential difference in $f_{\rm AGN}$ between different samples in our study, we do not necessarily require a complete \hbox{log $L_{\rm X} > 42$} selection: if a significant difference in the fraction of objects with \hbox{log $L_{\rm X} > 42$} is observed between two samples, given that AGNs with relatively low $L_{\rm X}$ can only be detected in relatively deep X-ray fields, the intrinsic difference in AGN fraction will be more significant (unless the differences in the fraction of low-$L_{\rm X}$ and high-$L_{\rm X}$ AGNs have different signs).}
instead of \bhar, which helps assess the prevalence of AGN activity instead of long-term average BH growth. \agnf\ and its 1$\sigma$ confidence interval (also obtained via bootstrapping) will be calculated for each bin and subsample, and presented in the relevant plots. The significance level of the difference in \agnf\ between two subsamples ($\Delta f_{\rm AGN} = f_{\rm AGN, subsample~1} -  f_{\rm AGN, subsample~2}$) is also calculated by dividing it by its 1$\sigma$ uncertainty that is obtained from bootstrapping as (84th$-$16th percentile)/2 of the $\Delta f_{\rm AGN}$ distribution. 
The relevant plots here for the BD, Non-BD, and SF Non-BD samples are Figures \ref{bulge_agnf}, \ref{nonbulge_agnf}, and \ref{nonbulge_sf_agnf} when \sigmae\ is utilized to measure compactness;
Figures \ref{bulge_agnf_s1}, \ref{nonbulge_agnf_s1}, and \ref{nonbulge_agnf_s1_sf} are relevant when \sigmaone\ is utilized to measure compactness.

We will perform PCOR analyses with \texttt{PCOR.R} in the \texttt{R} statistical package \citep{Kim2015} to assess if, for galaxies in the BD (Non-BD or SF Non-BD) sample, the \bhar-SFR relation (\hbox{\bhar-\mstar\ relation}) is still significant when controlling for $\Sigma$. We will also assess if the \bhar-$\Sigma$ relation is significant when controlling for SFR (\mstar).
We will bin sources based on both SFR (\mstar) and $\Sigma$, and calculate $\rm \overline{BHAR}$ for each bin. The bins for the $x$-axis/$y$-axis are chosen to include approximately the same numbers of sources.
Only bins with more than 50 objects will be utilized in the PCOR analyses to avoid large statistical uncertainties as well as potential systematic problems due to occasional
``outlier'' objects that could perturb a small sample.
Bins where \bhar\ does not have a lower limit $>$ 0 from bootstrapping will also be excluded from the PCOR analyses.
We will input the median log SFR (\mstar), median log~$\Sigma$, and log $\rm \overline{BHAR}$ of utilized bins to \texttt{PCOR.R}, to calculate the significance levels of the $\rm \overline{BHAR}$-SFR (\bhar-\mstar) relation when controlling for $\Sigma$ and the $\rm \overline{BHAR}$-$\Sigma$ relation when controlling for SFR (\mstar) with both the Pearson and Spearman statistics.
We will summarize the results of the PCOR analyses in tables (Table~\ref{pcortable} when \sigmae\ is utilized, and Table~\ref{pcortablecore} when \sigmaone\ is utilized).
We will use the parametric Pearson statistic to select significant results, and the nonparametric Spearman statistic will also be presented. Typically, the significance level obtained utilizing the Spearman statistic is qualitatively consistent with that obtained from the Pearson statistic.
For the PCOR analyses at $z = 0.5$--1.5/1.5--3, we will adopt a $3 \times 3$ grid for the BD sample, so that each bin contains $\approx 170/70$ sources on average; we will adopt a $5 \times 5$/$3 \times 3$ grid for the Non-BD (SF Non-BD) sample, so that each bin contains $\approx 190/210$ (160/180) sources on average.
As for the SF BD sample, we are not able to perform PCOR analyses due to its limited sample size.
For all the PCOR analyses in this work, $\gtrsim 98\%$ of sources in the sample are included with the utilized binning approach.
When a $5 \times 5$ grid is adopted, we will also perform tests with a $3 \times 3$ grid and a $4 \times 4$ grid. Typically, our results do not change qualitatively with the choice of grid; we will note in the text if a result is only significant with a $5 \times 5$ grid. 
We have also verified that our results do not change qualitatively with different binning approaches, e.g., binning based on equal intervals for the $x$-axis/$y$-axis, or binning on one axis first and then another axis to make each bin have approximately the same number of sources. 

\subsection{The relation between BH growth and \sigmae} \label{ssec-sigmae}
In this section, we study how BH growth relates to $\Sigma_{\rm e}$ (which measures host-galaxy compactness more globally compared with \sigmaone; see Section~\ref{sssec-c}) when controlling for SFR or \mstar\ among galaxies in the BD sample (see Section \ref{sssec-bulge}) and Non-BD sample (see Section \ref{sssec-nonbulge}), respectively.
Figures \ref{bulge_trend}--\ref{nonbulge_sf_agnf} are relevant for this subsection, and note we use a
consistent black-purple-orange color scheme for these figures.

\subsubsection{How does BH growth relate to \sigmae~for bulge-dominated galaxies?} \label{sssec-bulge}
\begin{figure}
\begin{center}
\includegraphics[scale=0.27]{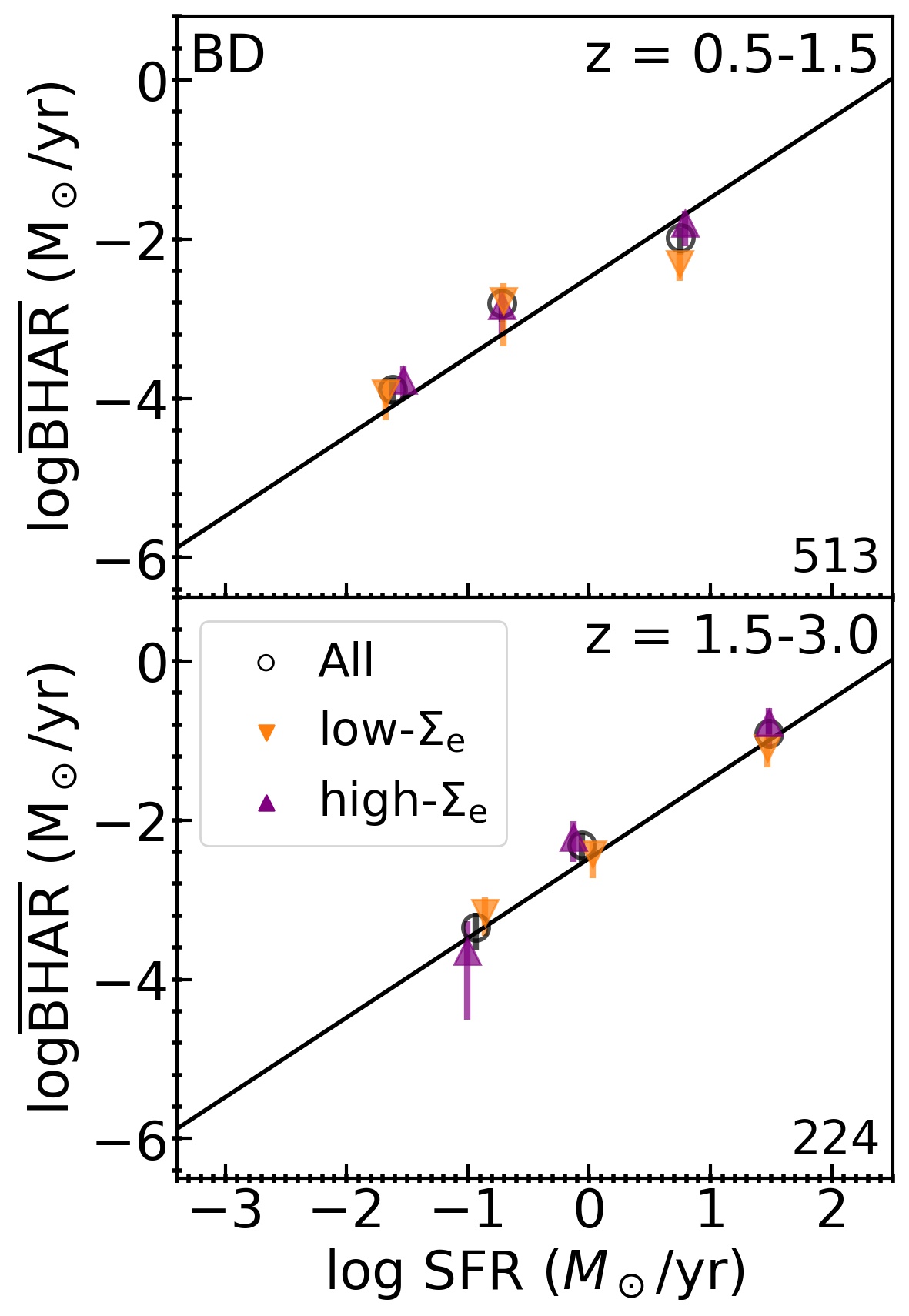}
\includegraphics[scale=0.27]{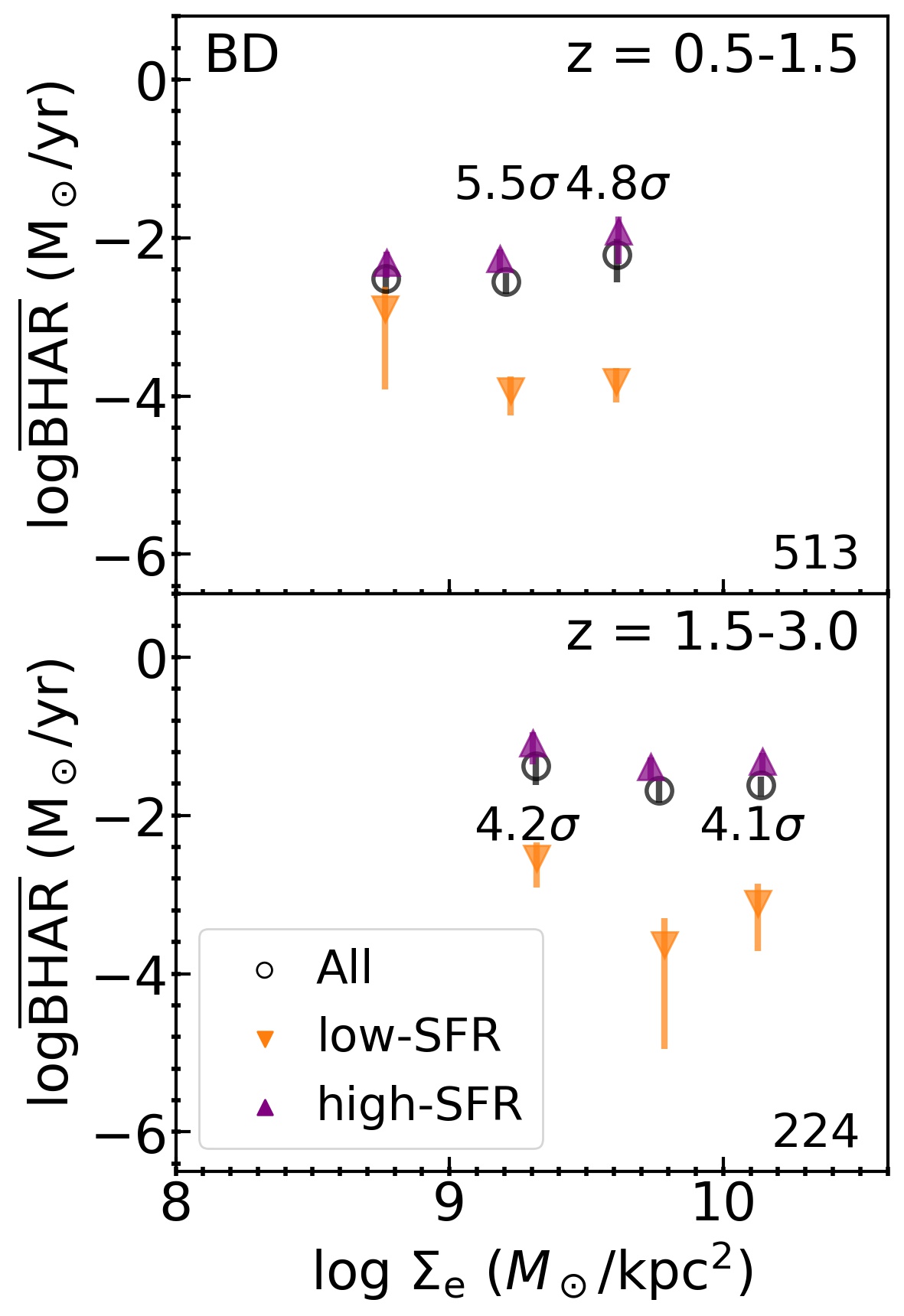}
\caption{$\rm \overline{BHAR}$ vs. SFR (left) and $\Sigma_{\rm e}$ (right) for galaxies in the BD sample. The horizontal position of each data point indicates the median SFR/$\Sigma_{\rm e}$ (left/right) of the sources in the bin. Each SFR/$\Sigma_{\rm e}$ sample (black circles) is further divided into two subsamples with $\Sigma_{\rm e}$/SFR above (purple upward-pointing triangles) and below (orange downward-pointing triangles) the median $\Sigma_{\rm e}$/SFR of the sample, respectively. The error bars represent the 1$\sigma$ confidence interval of \bhar~from bootstrapping. The significance levels of the differences between \bhar~in the subsamples are labeled at the position of the bin if the level is $> 3\sigma$.
The number in the bottom-right corner represents the number of objects in each SFR/$\Sigma_{\rm e}$ bin.
The black solid lines in the left panel represent the best-fit \bhar-SFR relation in Yang et al. (2019) with slope fixed to unity. We can see that while the \bhar-SFR relation is close to that obtained in \citet{Yang2019}, \bhar\ does not vary with \sigmae.
}
\label{bulge_trend}
\end{center}
\end{figure}

We plot $\rm \overline{BHAR}$ as a function of SFR and $\Sigma_{\rm e}$ in Figure \ref{bulge_trend} for galaxies in the BD sample.
Each SFR/\sigmae\ bin is further divided into two subsamples with \sigmae/SFR above or below the median \sigmae/SFR, and the \bhar\ values of these subsamples are shown on the plot as well.
We can see that for galaxies in the BD sample, there is no obvious \bhar-\sigmae~relation (in the right panel of Figure \ref{bulge_trend}), and for a given SFR, the differences in $\Sigma_{\rm e}$ do not cause \textit{any} significant differences in $\rm \overline{BHAR}$ (in the left panel of Figure \ref{bulge_trend}). This qualitatively indicates that $\rm \overline{BHAR}$ does not depend on \sigmae.
Given that we define high/low-\sigmae\ subsamples based on median \sigmae\ values, it is possible that the difference in \bhar\ associated with \sigmae\ might only be revealed by subsamples of extreme \sigmae. Considering this, we confirm that even when defining $\Delta \rm \overline{BHAR}$ as the difference between \bhar\ of a subsample of galaxies with \sigmae\ greater than the 75th percentile of the \sigmae\ distribution and a subsample of galaxies with \sigmae\ less than the 25th percentile of the \sigmae\ distribution, we do not observe significant $\Delta \rm \overline{BHAR}$ associated with \sigmae.
To test the point that \bhar\ does not depend on \sigmae\ in the BD sample further, we bin sources based on both SFR and \sigmae\ (with the binning approach described in Section \ref{ssec-method}), and use the median log SFR, median log~$\Sigma_{\rm e}$, and log $\rm \overline{BHAR}$ of bins to perform PCOR analyses. The results are summarized in Table~\ref{pcortable}. 
While the \bhar-SFR relation is significant as expected when controlling for \sigmae, $\rm \overline{BHAR}$ does not correlate with $\Sigma_{\rm e}$ significantly when controlling for SFR in the BD sample.

\begin{figure}
\begin{center}
\includegraphics[scale=0.28]{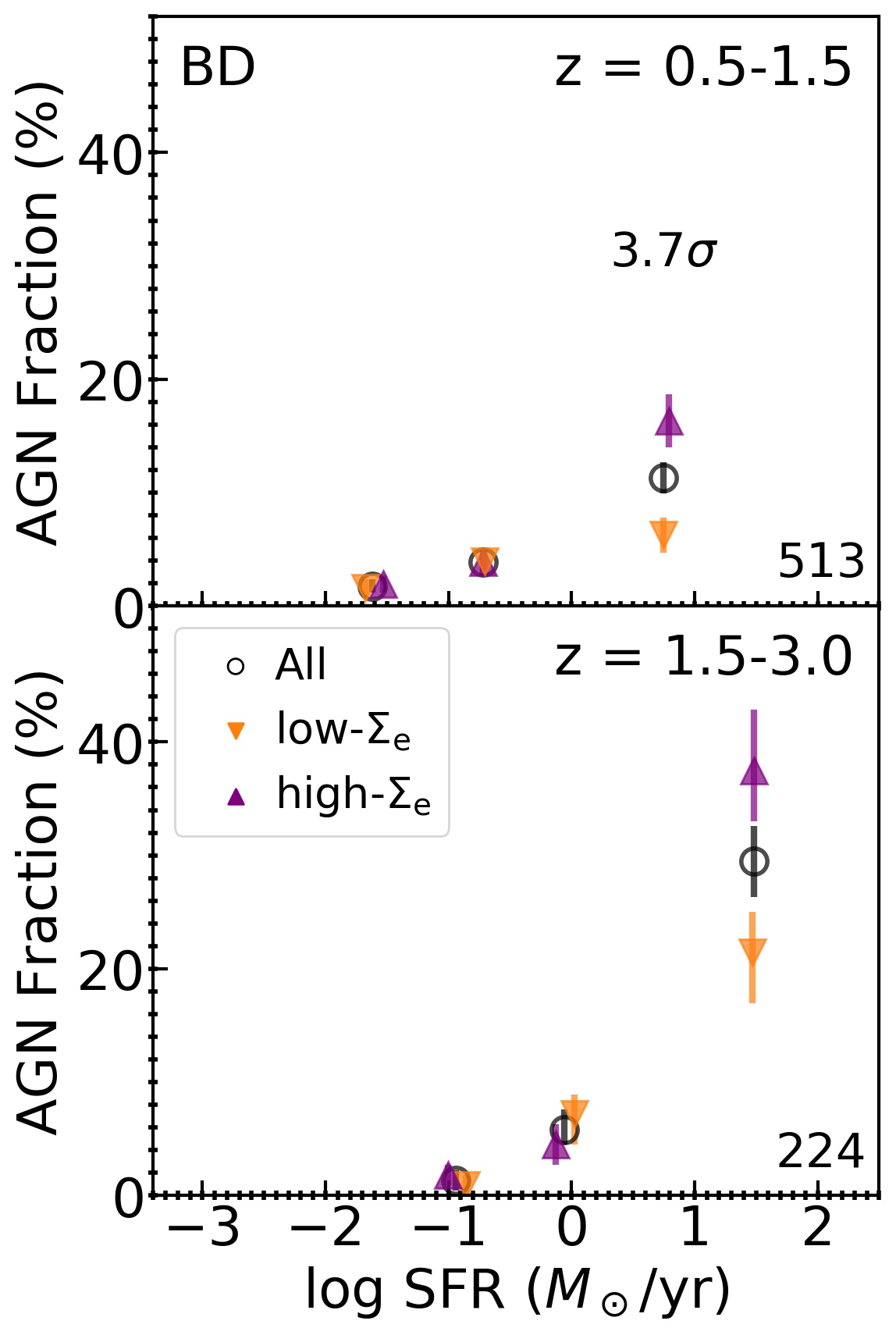}
\includegraphics[scale=0.28]{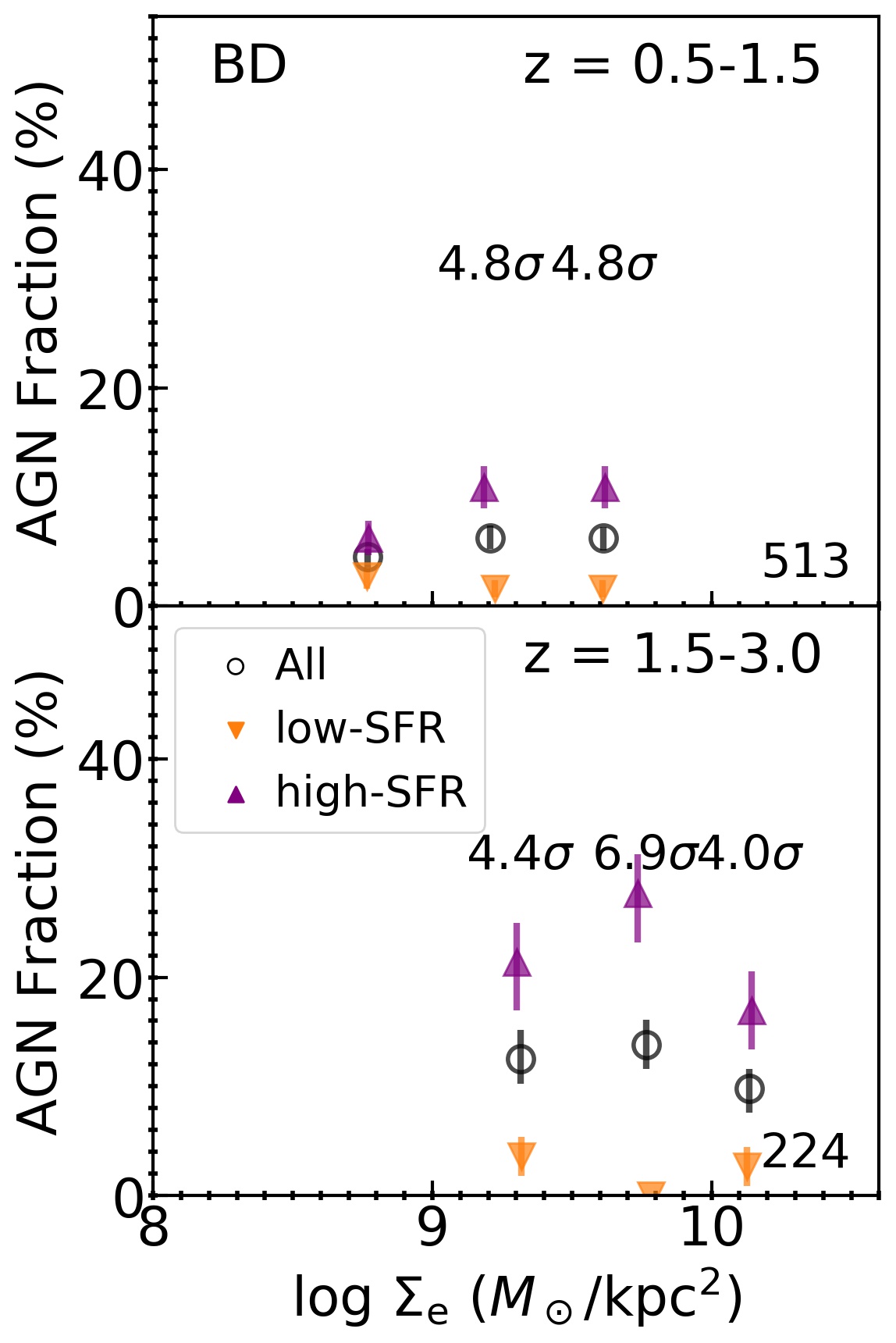}
\caption{AGN fraction vs. SFR (left) and $\Sigma_{\rm e}$ (right) for galaxies in the BD sample. The horizontal position of each data point indicates the median SFR/$\Sigma_{\rm e}$(left/right) of the sources in the bin. Each SFR/$\Sigma_{\rm e}$ sample (black circles) is further divided into two subsamples with $\Sigma_{\rm e}$/SFR above (purple upward-pointing triangles) and below (orange downward-pointing triangles) the median $\Sigma_{\rm e}$/SFR of the sample, respectively. The error bars represent the 1$\sigma$ confidence interval of AGN fraction from bootstrapping. The significance levels of the differences between AGN fraction in the subsamples are labeled at the position of the bin if the level is $> 3\sigma$.
The number in the bottom-right corner represents the number of objects in each SFR/$\Sigma_{\rm e}$ bin.
We can see that while AGN fraction varies with SFR, it does not vary significantly with \sigmae.}
\label{bulge_agnf}
\end{center}
\end{figure}

We also investigate how AGN fraction relates to \sigmae~when controlling for SFR for the BD sample. 
In Figure \ref{bulge_agnf}, we plot AGN fraction as a function of SFR and $\Sigma_{\rm e}$ for galaxies in the BD sample. The bins and subsamples in Figure \ref{bulge_agnf} are the same as those of Figure~\ref{bulge_trend}. We can see that while AGN fraction does not vary significantly with \sigmae~(in the right panel of Figure \ref{bulge_agnf}), it rises at the high-SFR end (in the left panel of Figure \ref{bulge_agnf}). Also, the $f_{\rm AGN}$ differences associated with \sigmae~when controlling for SFR are not significant except for one bin with the highest SFR at $z = 0.5$--1.5, as can be seen in the left panel of Figure~\ref{bulge_agnf}. If we consider the Bonferroni correction to counteract the problem of multiple comparisons (see Section~\ref{sec-intro}; since we are testing 6 hypotheses together here, we require the difference to be significant at $> 3.5 \sigma$), this 3.7$\sigma$ difference is still significant.

Could this suggest a dependence of AGN fraction on \sigmae\ among SF galaxies in the BD sample? Due to the limited number of SF BD galaxies, we calculate the significance level of $\Delta f_{\rm AGN}$ for all SF BD galaxies at $z = 0.5$--1.5/1.5--3 when splitting into high/low-\sigmae\ subsamples, which is 3.7$\sigma$/2.6$\sigma$. In terms of $\Delta \rm \overline{BHAR}$, the significance levels at both $z = 0.5$--1.5 and $z = 1.5$--3 are below $3\sigma$.
We also note that when splitting all SF BD galaxies into high/low-\mstar\ subsamples, the significance level of $\Delta f_{\rm AGN}$ is 6.3$\sigma$/2.5$\sigma$ at \hbox{$z = 0.5$--1.5/1.5--3}, and the significance level of $\Delta \rm \overline{BHAR}$ is 6.4$\sigma$/3.7$\sigma$. 
Interestingly, when splitting all SF BD galaxies into high/low-SFR subsamples, the \dbhar/\dagnf\ between two subsamples in both redshift ranges are not significant.
As mentioned in Section \ref{ssec-method}, the sample size of SF BD galaxies is too small to perform PCOR analyses to disentangle the relative roles of \mstar\ and \sigmae\ effects. However, we note that the influence of \mstar\ is more significant than the influence of \sigmae\ in both \bhar\ and $f_{\rm AGN}$.

\begin{table}
\begin{center}
\caption{$p$-values (significances) of 
	partial correlation analyses for the \bhar-\sigmae\ relation}
\label{pcortable}
\begin{tabular}{ccccc}\hline\hline
Relation &  Pearson & Spearman  \\ \hline\hline
\multicolumn{3}{c}{BD: $0.5 \leqslant z < 1.5$ ($3\times 3$ bins)} \\ \hline
\bhar-SFR                   &  $\boldsymbol{3\times 10^{-4}~(3.6 \sigma)}$  & $\boldsymbol{1\times 10^{-5}~(4.4\sigma)}$ \\
\bhar-$\Sigma_{\rm_e}$ & $0.29~(1.1\sigma)$  & $0.68~(0.4\sigma)$ \\ \hline
\multicolumn{3}{c}{BD: $1.5 \leqslant  z < 3$ ($3\times 3$ bins)} \\ \hline
\bhar-SFR                   &  $\boldsymbol{2\times 10^{-3}~(3.2\sigma)}$  & $0.01~(2.5\sigma)$ \\
\bhar-$\Sigma_{\rm_e}$ & $0.44~(0.8\sigma)$  & $0.28~(1.1\sigma)$ \\ \hline\hline
\multicolumn{3}{c}{Non-BD: $0.5 \leqslant  z < 1.5$ ($5\times 5$ bins)} \\ \hline
\bhar-$M_\star$              &  $\boldsymbol{3\times 10^{-3}~(3.0\sigma)}$  & $\boldsymbol{8\times 10^{-4}~(3.4\sigma)}$\\
\bhar-$\Sigma_{\rm_e}$ & $0.42~(0.8\sigma)$  & $0.86~(0.2\sigma)$  \\ \hline
\multicolumn{3}{c}{Non-BD: $1.5 \leqslant z < 3$ ($3\times 3$ bins)} \\ \hline
\bhar-$M_\star$             &  $\boldsymbol{3 \times 10^{-3}~(3.0 \sigma)}$  & $\boldsymbol{3 \times 10^{-3}~(3.0\sigma)}$ \\
\bhar-$\Sigma_{\rm_e}$ &  $0.40~(0.8\sigma)$  & $0.46~(0.7\sigma)$ \\ \hline\hline
\multicolumn{3}{c}{SF Non-BD: $0.5 \leqslant  z < 1.5$ ($5\times 5$ bins)} \\ \hline
\bhar-$M_\star$              &  $\boldsymbol{1\times 10^{-3}~(3.2\sigma)}$  & $6 \times 10^{-3}~(2.7\sigma)$\\
\bhar-$\Sigma_{\rm_e}$ & $0.13~(1.5\sigma)$  & $0.37~(0.9\sigma)$  \\ \hline
\bhar-$M_\star$              &  $\boldsymbol{1\times 10^{-8}~(5.7\sigma)}$  & $\boldsymbol{2\times 10^{-7}~(5.2\sigma)}$\\
\bhar-$r_{\rm_e}$ & $0.08~(1.7\sigma)$  & $0.11~(1.6\sigma)$  \\ \hline
\multicolumn{3}{c}{SF Non-BD: $1.5 \leqslant z < 3$ ($3\times 3$ bins)} \\ \hline
\bhar-$M_\star$             &  $0.02~(2.4 \sigma)$  & $0.02~(2.4\sigma)$ \\
\bhar-$\Sigma_{\rm_e}$ &  $0.35~(0.9\sigma)$  & $0.93~(0.1\sigma)$ \\ \hline
\bhar-$M_\star$             &  $\boldsymbol{7 \times 10^{-4}~(3.4 \sigma)}$  & $\boldsymbol{1 \times 10^{-3}~(3.2\sigma)}$ \\
\bhar-$r_{\rm_e}$ &  $0.48~(0.7\sigma)$  & $0.36~(0.9\sigma)$ \\ \hline\hline
\end{tabular}                                         
\end{center}
\end{table}

Thus, for galaxies in the BD sample, \sigmae\ has no apparent relation to either the long-term average BH growth or the prevalence of AGN activity.

\subsubsection{How does BH growth relate to \sigmae~for galaxies that are not bulge-dominated?} \label{sssec-nonbulge}

\begin{figure}
\begin{center}
\includegraphics[scale=0.27]{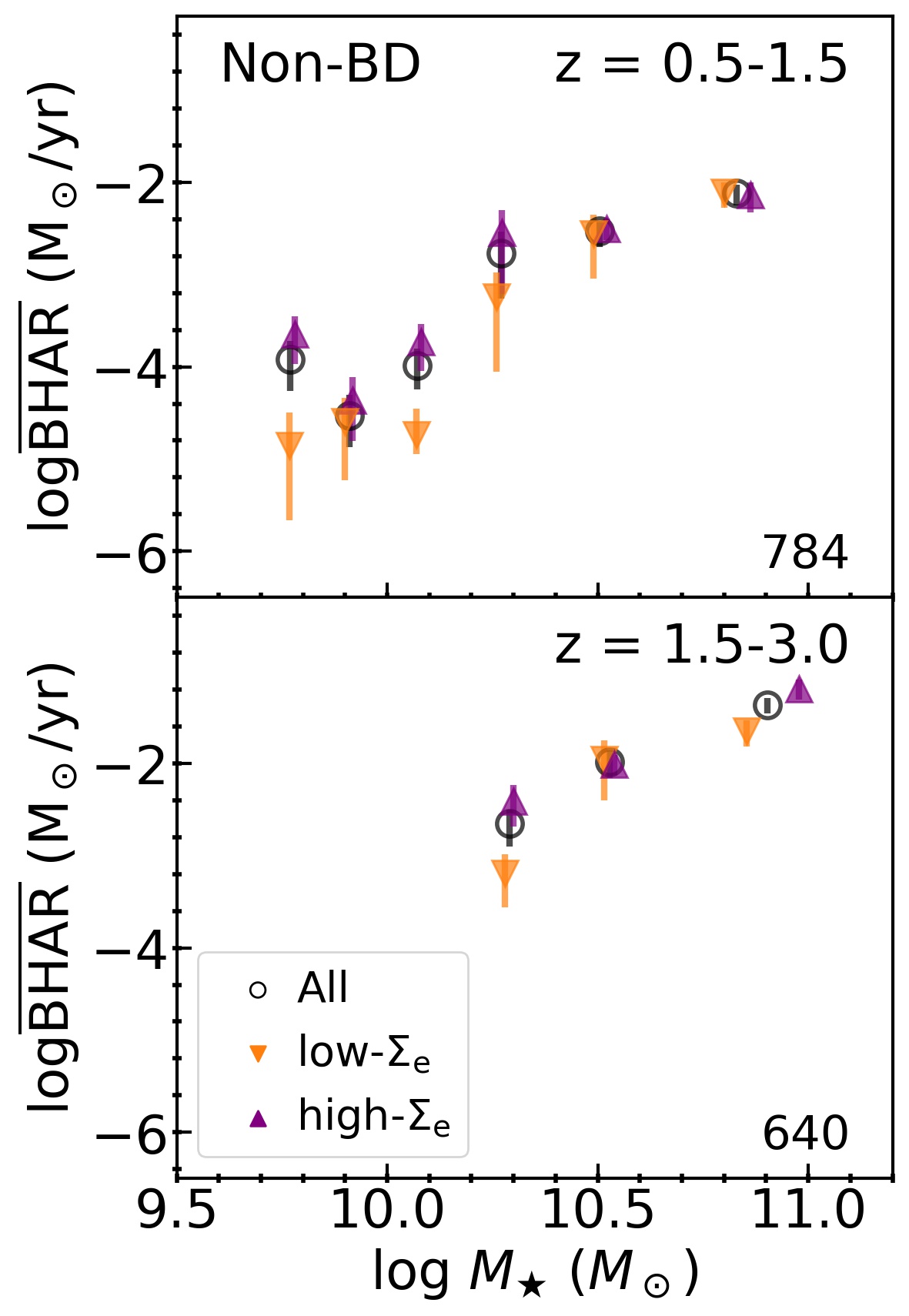}
\includegraphics[scale=0.27]{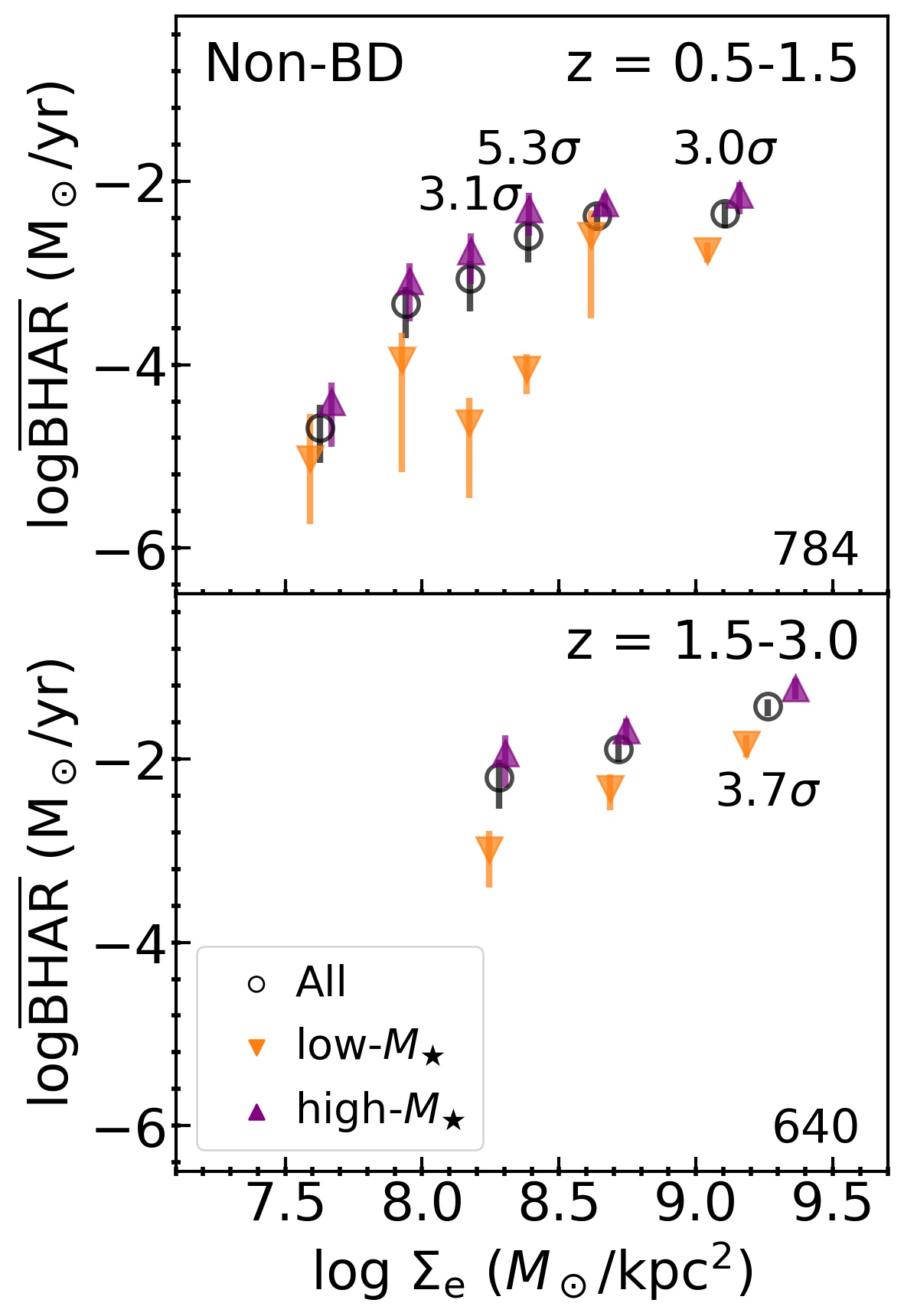}
\caption{$\rm \overline{BHAR}$ vs. $M_\star$ (left) and $\Sigma_{\rm e}$ (right) for galaxies in the Non-BD sample. The horizontal position of each data point indicates the median $M_\star$/$\Sigma_{\rm e}$(left/right) of the sources in the bin. Each $M_\star$/$\Sigma_{\rm e}$ sample (black circles) is further divided into two subsamples with $\Sigma_{\rm e}$/$M_\star$ above (purple upward-pointing triangles) and below (orange downward-pointing triangles) the median $\Sigma_{\rm e}$/$M_\star$ of the sample, respectively. The error bars represent the 1$\sigma$ confidence interval of \bhar~from bootstrapping. The significance levels of the differences between \bhar~in the subsamples are labeled at the position of the bin if the level is $> 3\sigma$. The number in the bottom-right corner represents the number of objects in each \mstar/$\Sigma_{\rm e}$ bin. While we can see both the \bhar-\mstar\ and \bhar-\sigmae\ relations, $\Delta \rm \overline{BHAR}$ values associated with \mstar\ are generally noticeable (in the right panel) and all $\Delta \rm \overline{BHAR}$ values associated with \sigmae\ are not significant (in the left panel).}
\label{comp_trend}
\end{center}
\end{figure}

In Figure \ref{comp_trend}, we plot $\rm \overline{BHAR}$ as a function of $M_\star$ and $\Sigma_{\rm e}$ for the Non-BD sample. 
Each \mstar/\sigmae\ bin is further divided into two subsamples with \sigmae/\mstar\ above or below the median \sigmae/\mstar, and the \bhar\ values of these subsamples are shown on the plot as well.
We can see that while both the \bhar-\mstar~relation and \bhar-\sigmae~relation exist with non-zero slope (which is expected given the degeneracy between \mstar\ and \sigmae\ in Figure~\ref{msigmae}), in most cases the differences in $M_\star$ for a given $\Sigma_{\rm e}$ (in the right panel) are linked with noticeable differences in $\rm \overline{BHAR}$, and the differences in $\Sigma_{\rm e}$ for a given \mstar~(in the left panel) do not lead to significant differences in $\rm \overline{BHAR}$. 
We confirm that even when defining $\Delta \rm \overline{BHAR}$ as the difference between \bhar\ of a subsample of galaxies with \sigmae\ greater than the 75th percentile of the \sigmae\ distribution and a subsample of galaxies with \sigmae\ less than the 25th percentile of the \sigmae\ distribution, we do not observe significant $\Delta \rm \overline{BHAR}$ linked with \sigmae.

We then perform PCOR analyses to test quantitatively if the \bhar-\sigmae\ relation is a secondary manifestation of the \bhar-\mstar\ relation.
We bin sources based on both $M_\star$ and \sigmae~and calculate $\rm \overline{BHAR}$ for each bin. The median log $M_\star$, median log \sigmae, and log~$\rm \overline{BHAR}$ of these bins are used for PCOR analyses to calculate the significance levels of the $\rm \overline{BHAR}$-$M_\star$ relation when controlling for \sigmae~and the $\rm \overline{BHAR}$-\sigmae~relation when controlling for \mstar. The results are summarized in Table \ref{pcortable}.
We can see that while \bhar~significantly depends on \mstar~as expected when controlling for \sigmae, $\rm \overline{BHAR}$ does not correlate significantly with \sigmae~when controlling for \mstar. Thus, the \bhar-\sigmae~relation among galaxies in the Non-BD sample is not fundamental.

\begin{figure}
\begin{center}
\includegraphics[scale=0.28]{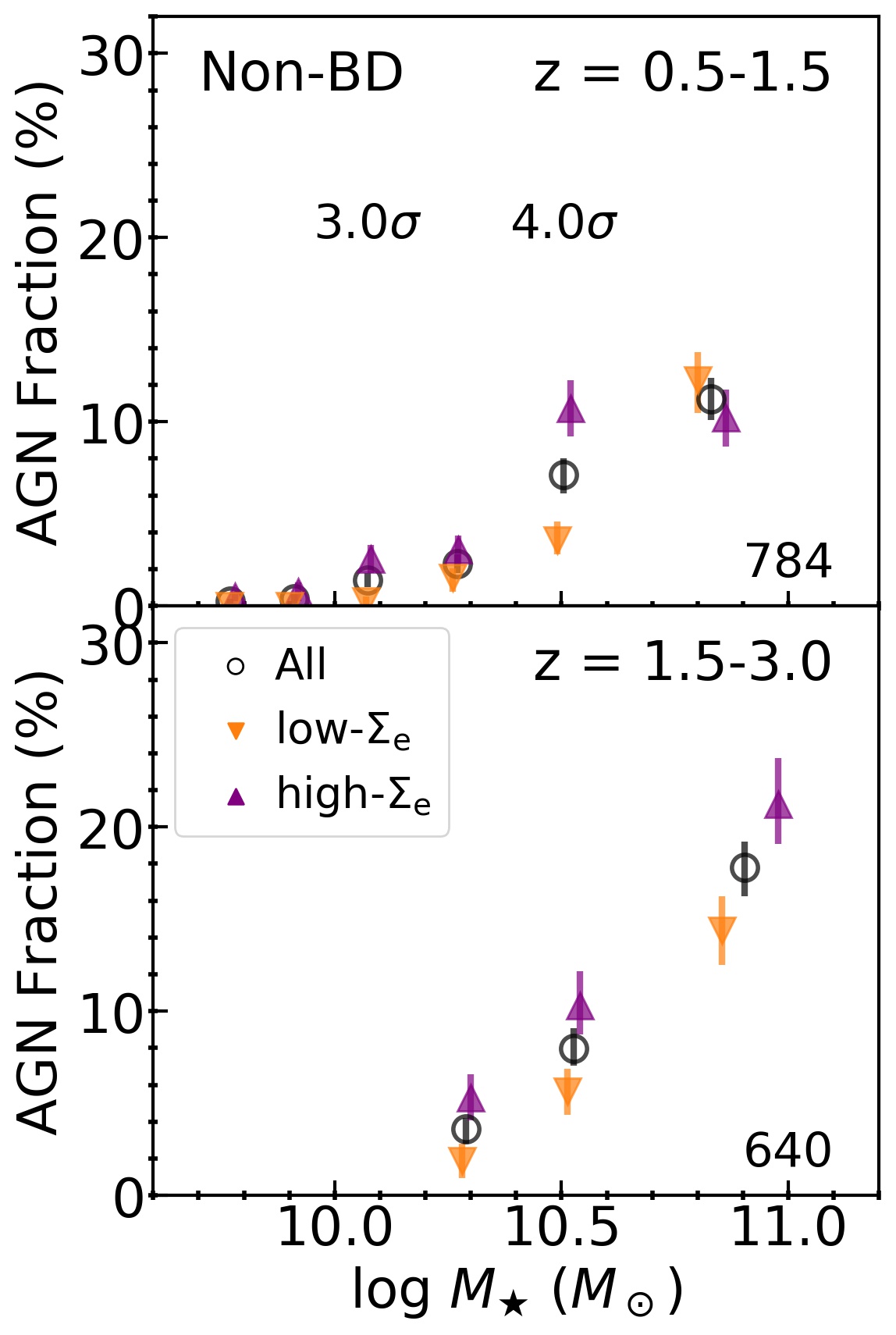}
\includegraphics[scale=0.28]{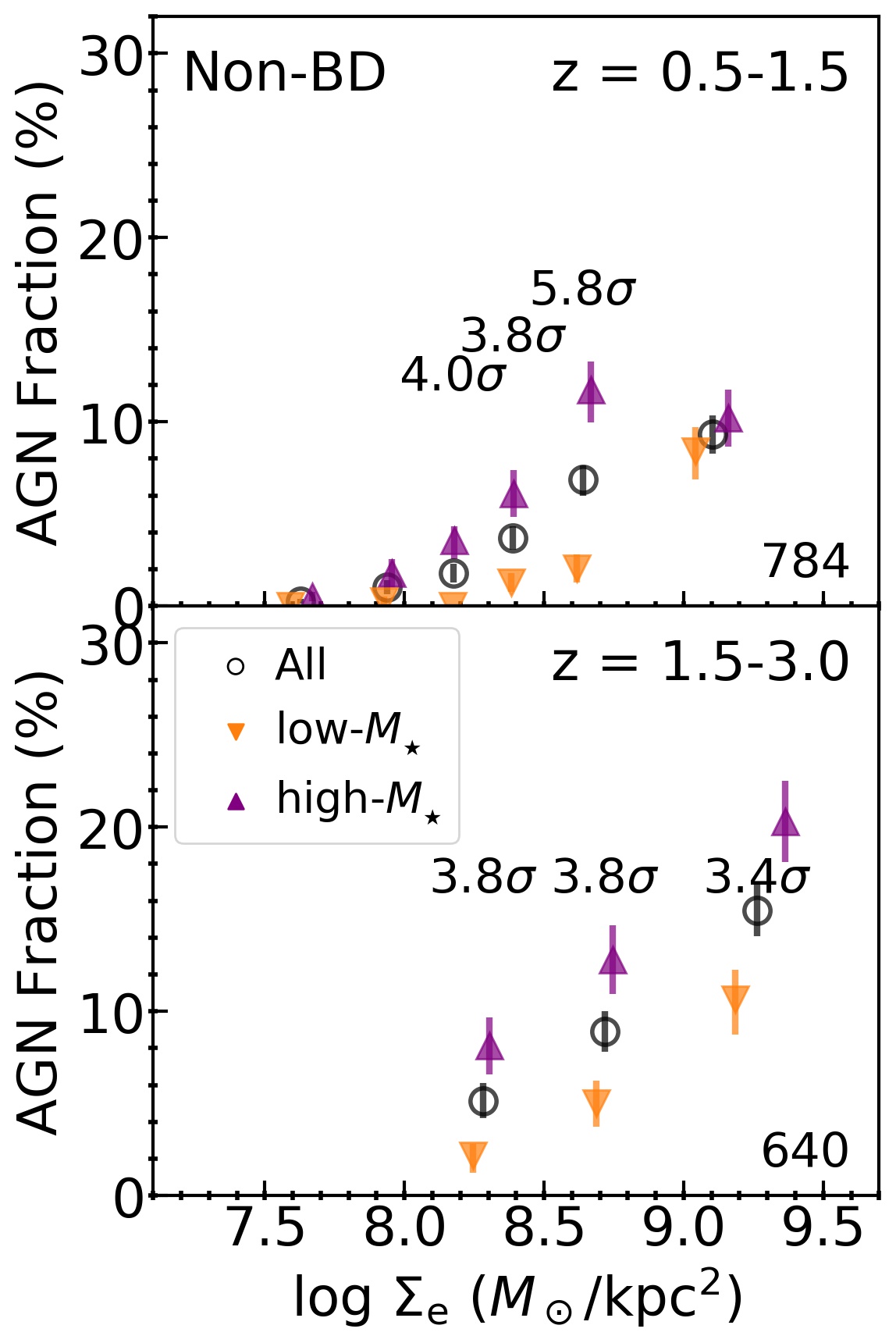}
\caption{AGN fraction vs. \mstar~(left) and $\Sigma_{\rm e}$ (right) for galaxies in the Non-BD sample. The horizontal position of each data point indicates the median \mstar/$\Sigma_{\rm e}$(left/right) of the sources in the bin. Each \mstar/$\Sigma_{\rm e}$ sample (black circles) is further divided into two subsamples with $\Sigma_{\rm e}$/\mstar~above (purple upward-pointing triangles) and below (orange downward-pointing triangles) the median $\Sigma_{\rm e}$/\mstar~of the sample, respectively. The error bars represent the 1$\sigma$ confidence interval of AGN fraction from bootstrapping. The significance levels of the differences between AGN fraction in the subsamples are labeled at the position of the bin if the level is $> 3\sigma$.
The number in the bottom-right corner represents the number of objects in each \mstar/$\Sigma_{\rm e}$ bin.
We can see that $\Delta f_{\rm AGN}$ values associated with \mstar\ are generally noticeable (in the right panel), and almost all $\Delta f_{\rm AGN}$ values associated with \sigmae\ are not significant (in the left panel) considering the Bonferroni correction except for one bin with log\mstar\ $\approx 10.5$ at $z = 0.5-1.5$.}
\label{nonbulge_agnf}
\end{center}
\end{figure}

We also investigate how the prevalence of AGN relates to \sigmae~when controlling for \mstar~for the Non-BD sample. 
In Figure \ref{nonbulge_agnf}, we plot AGN fraction as a function of \mstar~and $\Sigma_{\rm e}$ for galaxies in the Non-BD sample. 
The bins and subsamples in Figure \ref{nonbulge_agnf} are the same as those of Figure \ref{comp_trend}. We can see that, similar to the case for \bhar, the differences in $M_\star$ for a given $\Sigma_{\rm e}$ (in the right panel) are generally linked with noticeable differences in AGN fraction, and the differences in $\Sigma_{\rm e}$ for a given \mstar~(in the left panel) are not. 
Interestingly, for one bin with median log\mstar~$\approx 10.5$ at $z = 0.5$--1.5, $\Delta f_{\rm AGN}$ has a significance level of 4.0$\sigma$.
Even when the Bonferroni correction is considered (since we are testing 9 hypotheses together here, we require the difference to be significant at $> 3.6 \sigma$), this difference is still significant. However, as can be seen in Figure~\ref{comp_trend}, the $\Delta$\bhar~for this bin is not significant (0.2$\sigma$). We find that the difference in AGN fraction here is mainly a result of a higher fraction of low-$L_{\rm X}$ AGN ($L_{\rm X} = 10^{42-43}$ erg s$^{-1}$) among high-\sigmae~galaxies than low-\sigmae~galaxies in this mass range. At the same time, the fraction of high-$L_{\rm X}$ AGN ($L_{\rm X} > 10^{43}$ erg s$^{-1}$) does not significantly vary with \sigmae~in this mass range, leading to the lack of difference in \bhar.
We note that this difference in AGN fraction linked with \sigmae~when log\mstar~$\approx 10.5$ at $z = 0.5$--1.5 is not caused by any potential dependence of AGN fraction on SFR: for this \mstar~bin, the difference in AGN fraction linked with SFR is not significant ($0~\sigma$).
We will discuss the possible reason for this significant \dagnf\ associated with \sigmae\ that only occurs within certain mass ranges in Section \ref{wce}.

\begin{figure}
\begin{center}
\includegraphics[scale=0.27]{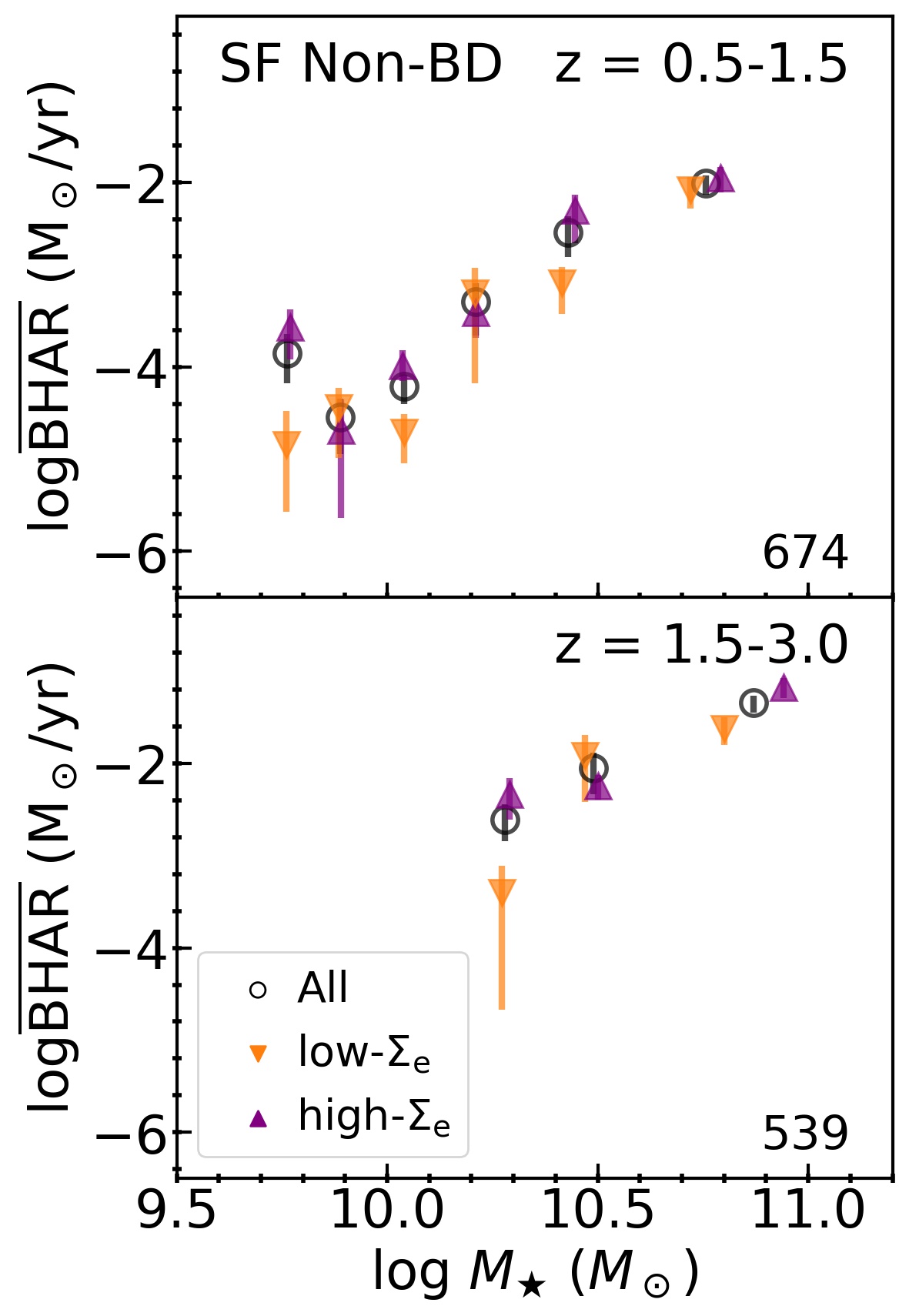}
\includegraphics[scale=0.27]{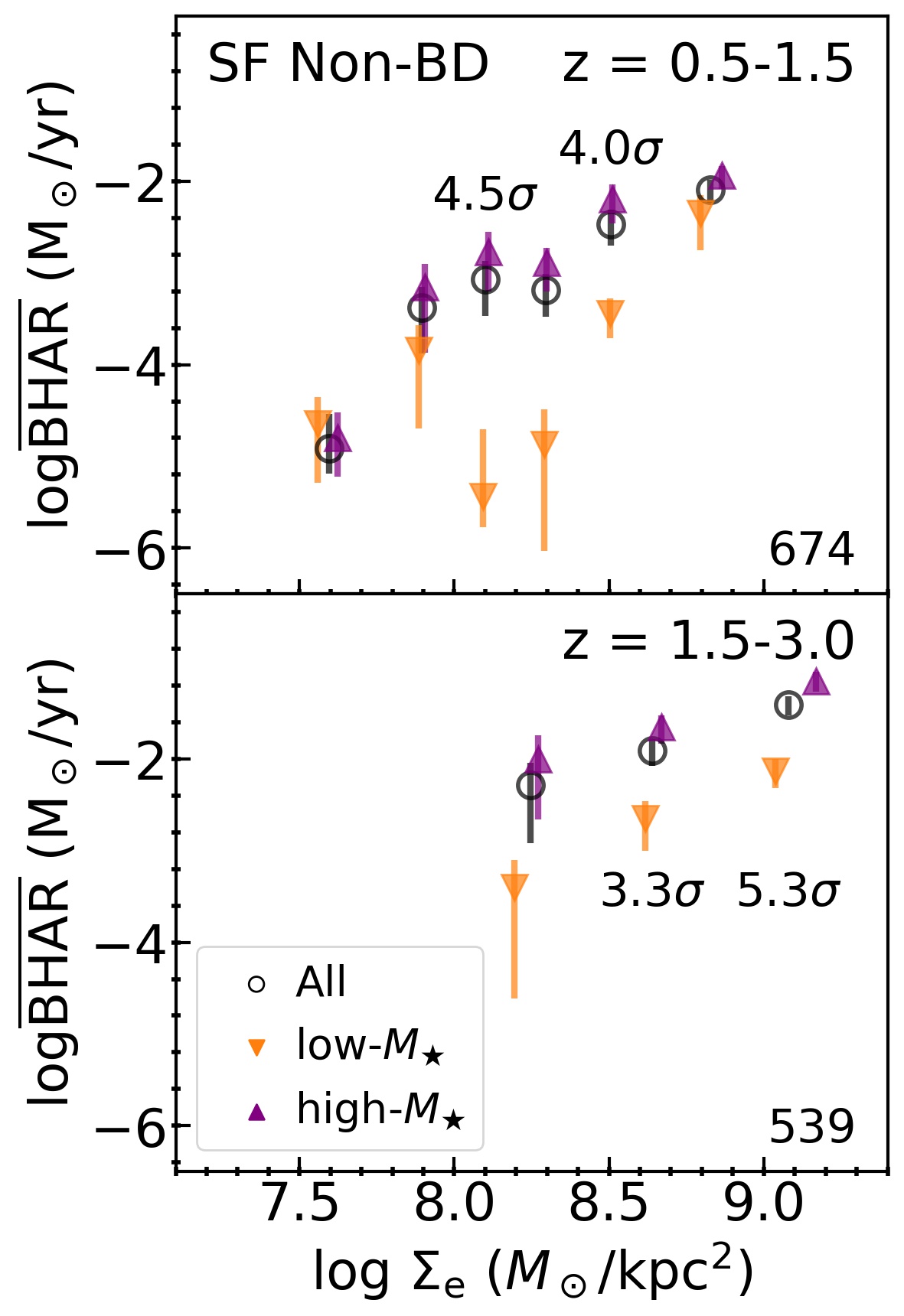}
\caption{Similar to Figure \ref{comp_trend}, but for galaxies in the SF Non-BD sample. $\Delta \rm \overline{BHAR}$ values associated with \mstar\ are generally noticeable (in the right panel) and all $\Delta \rm \overline{BHAR}$ values associated with \sigmae\ are not significant (in the left panel).}
\label{comp_sf_trend}
\end{center}
\end{figure}

\begin{figure}
\begin{center}
\includegraphics[scale=0.28]{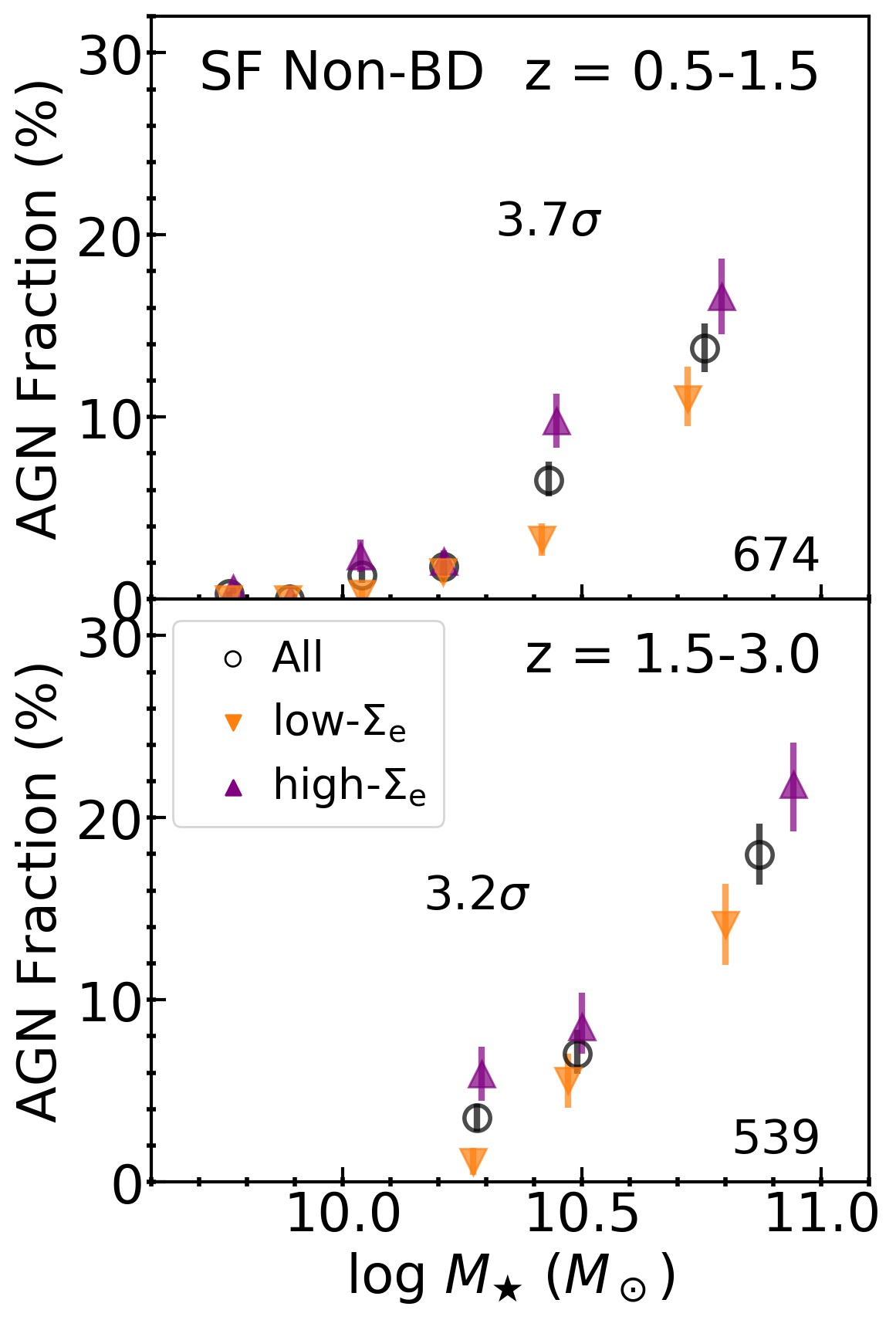}
\includegraphics[scale=0.28]{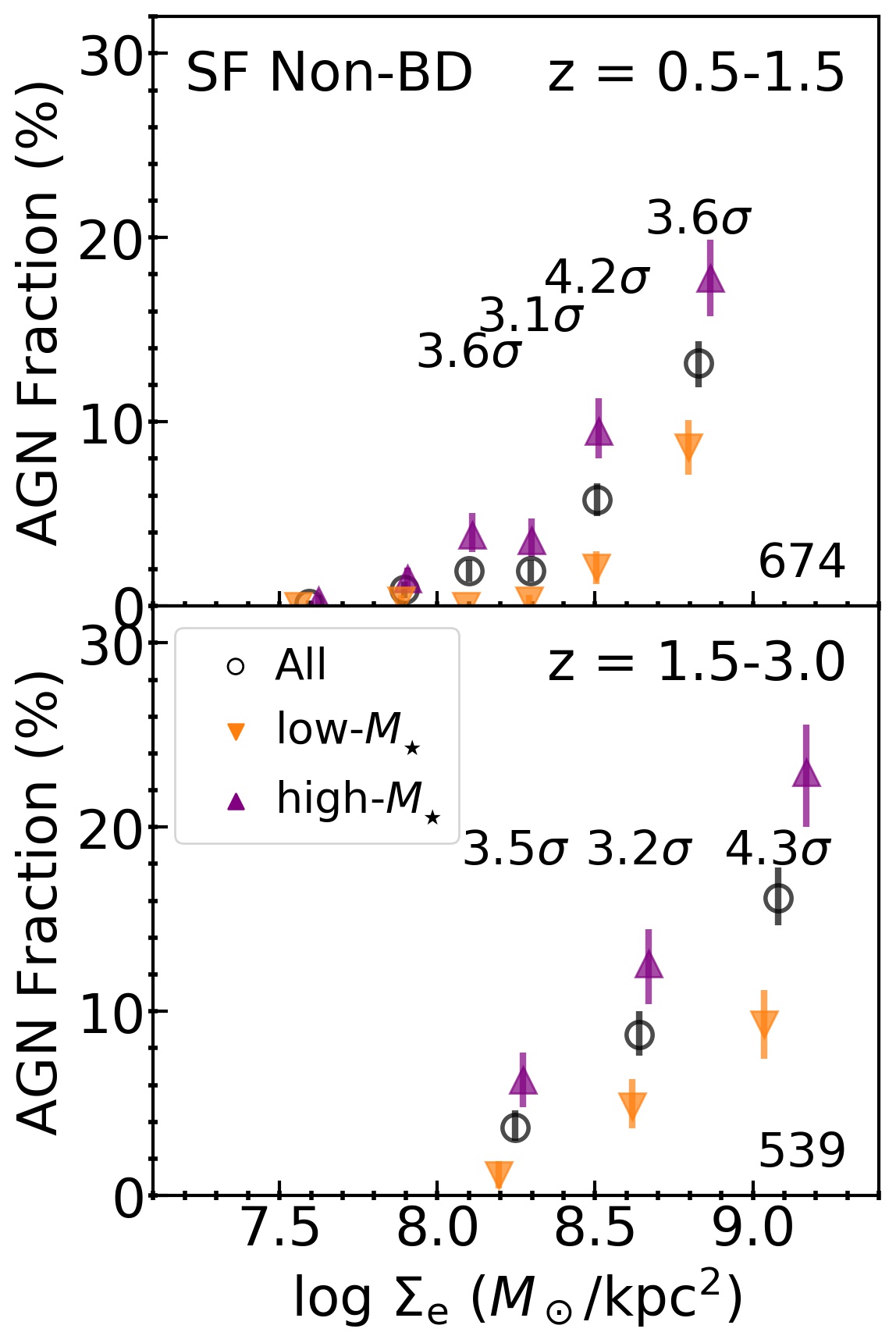}
\caption{Similar to Figure \ref{nonbulge_agnf}, but for galaxies in the SF Non-BD sample. We can see that $\Delta f_{\rm AGN}$ values associated with \mstar\ are generally noticeable (in the right panel), and almost all $\Delta f_{\rm AGN}$ values associated with \sigmae\ are not significant (in the left panel) considering the Bonferroni correction (significant if $> 3.6 \sigma$ when 9 hypotheses are tested together) except for one bin with log\mstar\ $\approx 10.4$ at $z = 0.5$--1.5. The bin with log\mstar\ $\approx 10.3$ at $z =1.5$--3 also has nearly significant \dagnf\ associated with \sigmae.}
\label{nonbulge_sf_agnf}
\end{center}
\end{figure}

We also confined the objects under investigation to be only SF galaxies in the Non-BD sample to study the relation between BH growth and \sigmae, where \sigmae\ may serve as an indicator of the gas density within \re. 
The \bhar/$f_{\rm AGN}$ as a function of \mstar/\sigmae\ among SF Non-BD galaxies is presented in Figures \ref{comp_sf_trend} and \ref{nonbulge_sf_agnf}. Similar to the results for galaxies in the Non-BD sample, a $\Delta \rm \overline{BHAR}$ link with \sigmae\ is not significant in any \mstar\ bin. A $\Delta f_{\rm AGN}$ link with \sigmae\ is only significant (at $3.7\sigma$) for one bin with median log\mstar\ $\approx 10.4$ at \hbox{$z = 0.5$--1.5}. This mass range is similar to that of the \mstar\ bin where a $4.0\sigma$ \dagnf\ associated with \sigmae\ is observed for the Non-BD sample at $z = 0.5$--1.5.
The significance levels of the $\rm \overline{BHAR}$-$M_\star$ relation and the $\rm \overline{BHAR}$-\sigmae~relation obtained from PCOR analyses for galaxies in the SF Non-BD sample are summarized in Table \ref{pcortable}: the \bhar-\sigmae\ relation is not significant when controlling for \mstar. However, the \bhar-\mstar\ relation is also not always significant (though it is still more significant than the \bhar-\sigmae\ relation), probably due to the degeneracy between \mstar\ and \sigmae\ among SF galaxies (e.g. see Figure~2 of \citealt{Barro2017}). Thus, for galaxies in the SF Non-BD sample, we further test if the \bhar-\re\ relation is significant when controlling for \mstar, which can reveal if \bhar\ truly depends on \sigmae, as log \sigmae\ $=$ log\mstar\ $-$ 2 $\times$ log \re\ $+~Constant$ from the definition $\Sigma_{\rm e} = 0.5M_\star/\pi r_{\rm e}^2$. The results are also summarized in Table \ref{pcortable}. We find that the \bhar-\re\ relation is not significant when controlling for \mstar, suggesting that the \bhar-\sigmae\ relation is also not fundamental among SF Non-BD galaxies. We note that previous studies found significantly elevated BH growth among high-\sigmae\ galaxies compared with low-\sigmae\ galaxies, and we will explain how this result compares with our findings in Section \ref{highsigmae}.

%%%%%%%%%%%%%%%%%%%%%%%%
%%%%%%%%%%%%%%%%%%%%%%%%

\subsection{The relation between BH growth and \sigmaone} \label{ssec-sigma1}

In this section, we perform the same analyses as those in Section~\ref{ssec-sigmae}, but now utilizing the projected central surface-mass density, \sigmaone, to represent the host-galaxy compactness. As noted in Section \ref{sec-intro}, \sigmaone\ has the potential of being a more effective indictor of BH growth compared with \sigmae. Thus, we will test if BH growth indeed has a fundamental dependence on host-galaxy compactness that can only be effectively revealed by \sigmaone, given the failure to find a fundamental \bhar-\sigmae\ relation in Section \ref{ssec-sigmae}.
Figures \ref{bulge_trend_s1}--\ref{comp_pcor_sf_s1} are relevant for this subsection, and note we use a consistent \hbox{black-blue-red} color scheme for these figures.

\subsubsection{How does BH growth relate to \sigmaone\ for the bulge-dominated galaxies?} \label{sssec-bulges1}

\begin{figure}
\begin{center}
\includegraphics[scale=0.27]{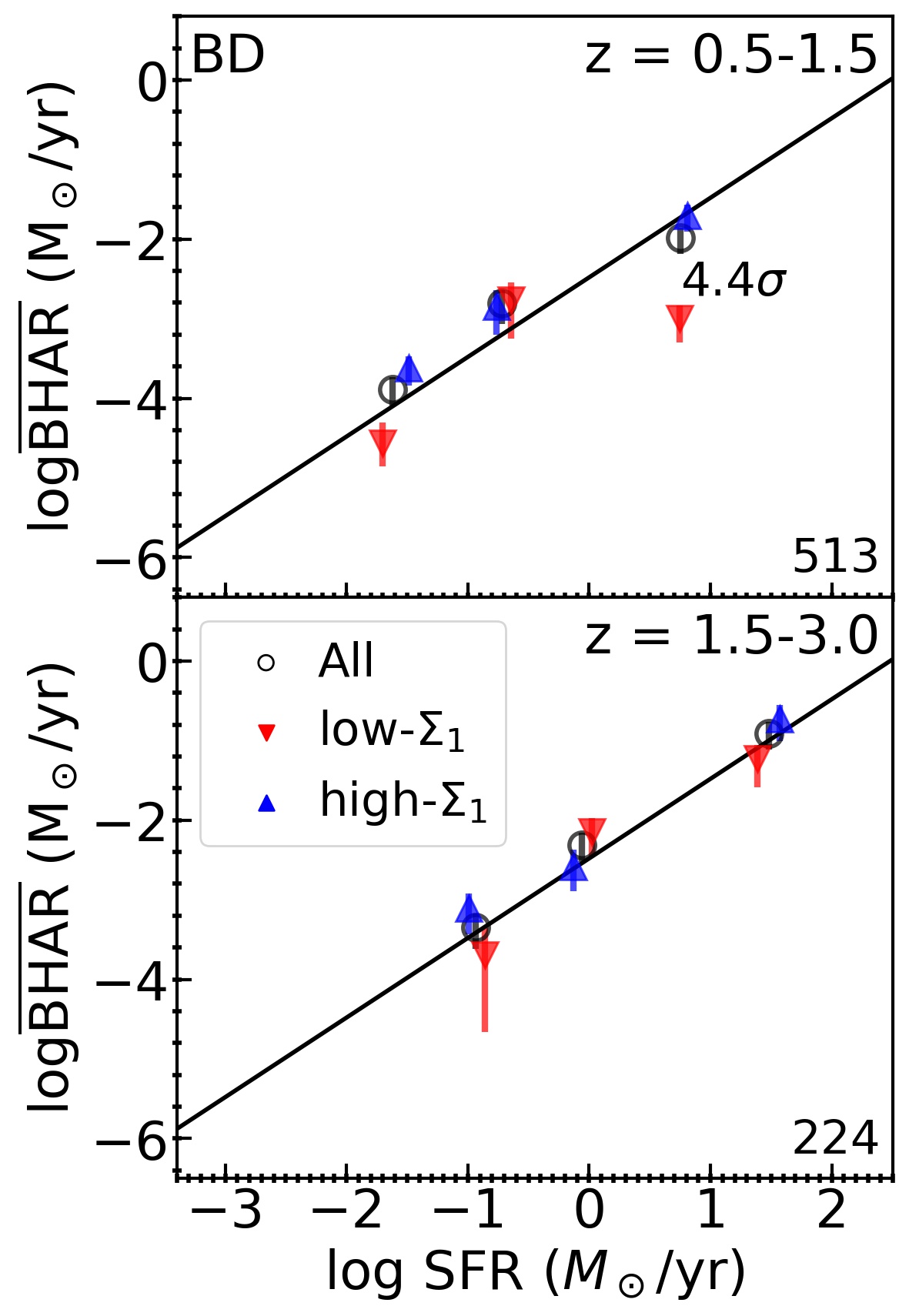}
\includegraphics[scale=0.27]{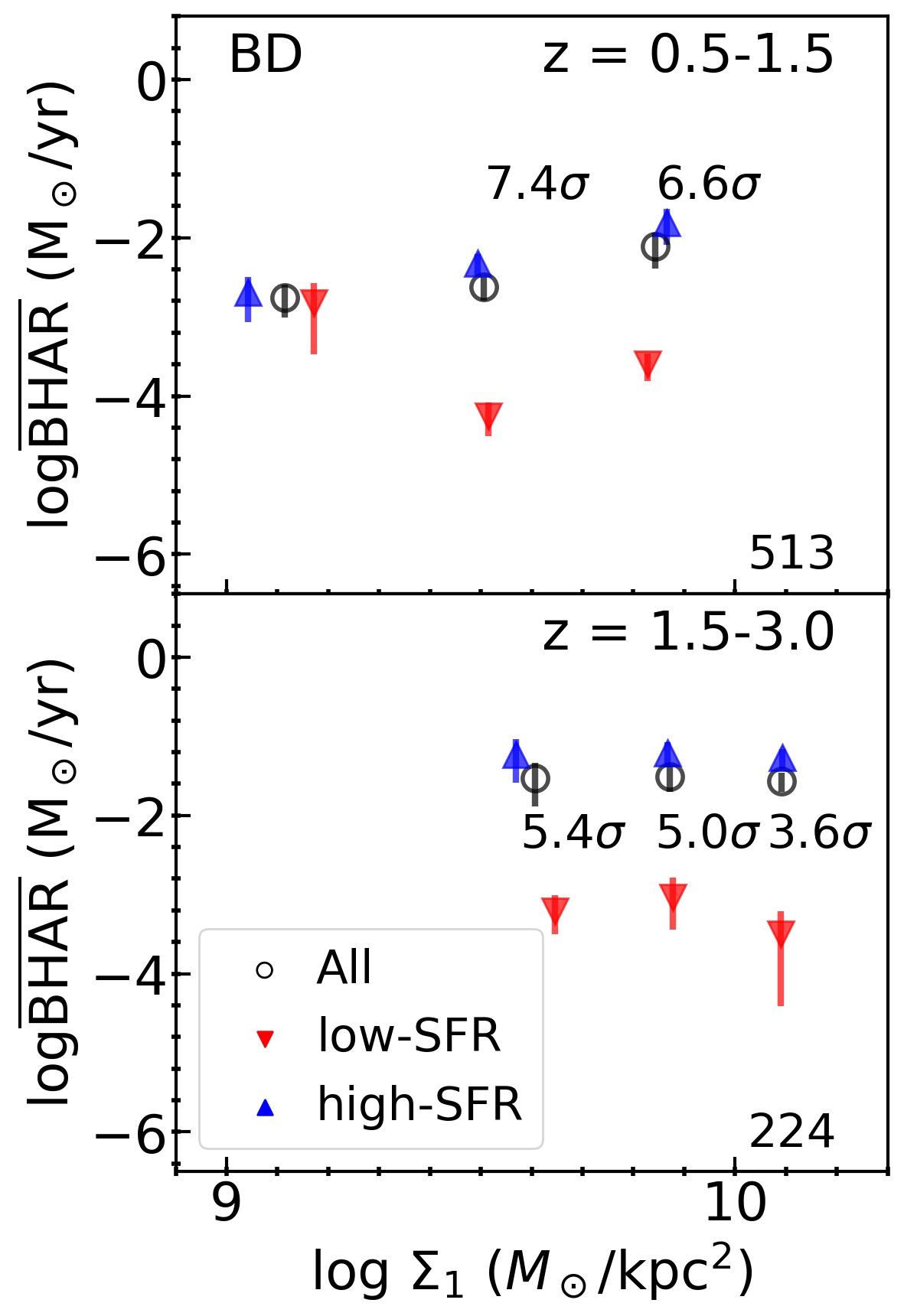}
\caption{$\rm \overline{BHAR}$ vs. SFR (left) and \sigmaone\ (right) for galaxies in the BD sample. The horizontal position of each data point indicates the median SFR/\sigmaone\ (left/right) of the sources in the bin. Each SFR/\sigmaone\ sample (black circles) is further divided into two subsamples with \sigmaone/SFR above (blue upward-pointing triangles) and below (red downward-pointing triangles) the median \sigmaone/SFR of the sample, respectively. The error bars represent the 1$\sigma$ confidence interval of \bhar~from bootstrapping. The significance levels of the differences between \bhar~in the subsamples are labeled at the position of the bin if the level is $> 3\sigma$.
The number in the bottom-right corner represents the number of objects in each SFR/\sigmaone\ bin.
The black solid lines in the left panel represent the best-fit \bhar-SFR relation in \citet{Yang2019} with slope fixed to unity. We can see that \bhar\ does not vary substantially with \sigmaone.
}
\label{bulge_trend_s1}
\end{center}
\end{figure}

\begin{figure}
\begin{center}
\includegraphics[scale=0.28]{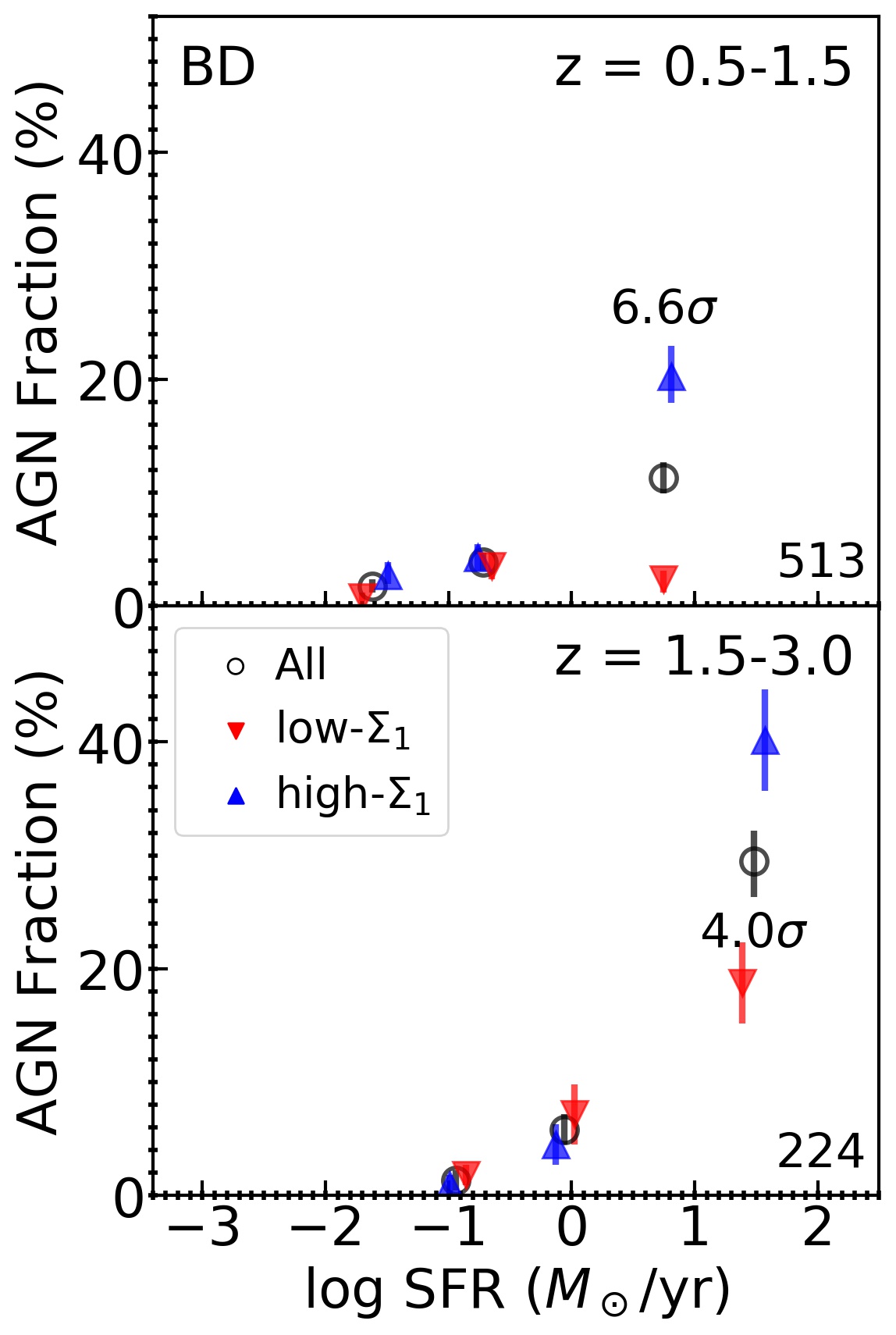}
\includegraphics[scale=0.28]{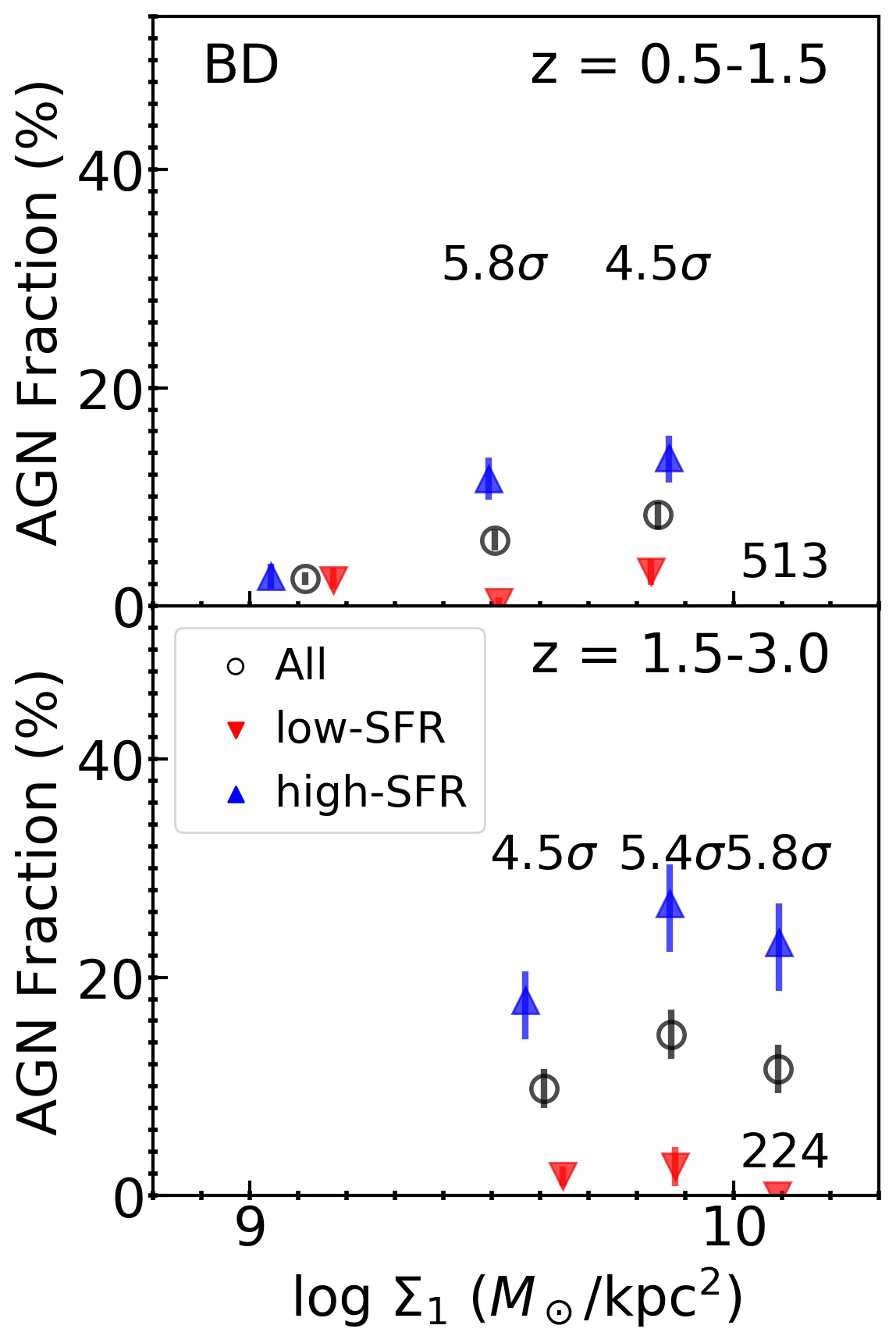}
\caption{AGN fraction vs. SFR (left) and  \sigmaone\ (right) for galaxies in the BD sample. The horizontal position of each data point indicates the median SFR/ \sigmaone\ (left/right) of the sources in the bin. Each SFR/\sigmaone\ sample (black circles) is further divided into two subsamples with \sigmaone/SFR above (blue upward-pointing triangles) and below (red downward-pointing triangles) the median  \sigmaone/SFR of the sample, respectively. The error bars represent the 1$\sigma$ confidence interval of AGN fraction from bootstrapping. The significance levels of the differences between AGN fraction in the subsamples are labeled at the position of the bin if the level is $> 3\sigma$.
The number in the bottom-right corner represents the number of objects in each SFR/\sigmaone\ bin.
We can see that the differences in \sigmaone\ do not cause significant differences in \agnf\ except for the highest SFR bin at both redshift ranges.}
\label{bulge_agnf_s1}
\end{center}
\end{figure}

We plot $\rm \overline{BHAR}$ as a function of SFR and $\Sigma_{1}$ in Figure \ref{bulge_trend_s1} for galaxies in the BD sample. 
Each SFR/\sigmaone\ bin is further divided into two subsamples with \sigmaone/SFR above or below the median \sigmaone/SFR, and the \bhar\ values of these subsamples are shown on the plot as well.
Similarly, we plot $f_{\rm AGN}$ as a function of SFR and $\Sigma_{1}$ in Figure~\ref{bulge_agnf_s1}. The bins and subsamples of Figure~\ref{bulge_agnf_s1} are the same as those of Figure~\ref{bulge_trend_s1}.
We can see that for all galaxies in the BD sample, there is no obvious \bhar-\sigmaone\ relation (in the right panel of Figure~\ref{bulge_trend_s1}). For a given SFR, the differences in \sigmaone\ do not cause significant differences in $\rm \overline{BHAR}$ except for the highest SFR bin at $z = 0.5$--1.5 (in the left panel of Figure~\ref{bulge_trend_s1}), and do not cause significant differences in \agnf\ except for the highest SFR bin at both $z = 0.5$--1.5 and $z = 1.5$--3 (in the left panel of Figure~\ref{bulge_agnf_s1}). 

We thus confine our attention to SF BD galaxies, and calculate the significance level of $\Delta \rm \overline{BHAR}$ ($\Delta f_{\rm AGN}$) for all SF BD galaxies in the low/high-redshift bin when splitting into two subsamples by \sigmaone\ value, which is 4.3$\sigma$/1.7$\sigma$ (5.7$\sigma$/4.0$\sigma$). We note that $\Delta \rm \overline{BHAR}$/$\Delta f_{\rm AGN}$ associated with \sigmaone\ in the SF BD sample is more significant than that associated with \sigmae\ (see Section \ref{sssec-bulge}). However, we still cannot conclude whether \sigmaone\ or \mstar\ plays a more fundamental role here, as high/low-\mstar\ subsamples also have significant $\Delta \rm \overline{BHAR}$/$\Delta f_{\rm AGN}$ (see Section \ref{sssec-bulge}), and the sample size of SF BD galaxies is too small to disentangle the relative roles of \mstar\ and \sigmaone\ effects.

We also performed PCOR analyses to test the significance level of the \bhar-SFR relation when controlling for \sigmaone, and the significance level of the \bhar-\sigmaone\ relation when controlling for SFR in the BD sample. The results are summarized in Table \ref{pcortablecore}. The \bhar-\sigmaone\ relation is not significant when controlling for SFR for bulge-dominated galaxies. 

\begin{table}
\begin{center}
\caption{$p$-values (significances) of partial correlation analyses for the \bhar-\sigmaone\ relation}
\label{pcortablecore}
\begin{tabular}{ccccc}\hline\hline
Relation &  Pearson & Spearman \\\hline\hline
\multicolumn{3}{c}{BD: $0.5 \leqslant z < 1.5$ ($3\times 3$ bins)} \\ \hline 
\bhar-SFR                   &  $\boldsymbol {2\times 10^{-3}~(3.1\sigma)}$  & $\boldsymbol {3\times 10^{-4}~(3.6\sigma)}$  \\
\bhar-$\Sigma_1$ & $0.19~(1.3\sigma)$  & $0.69~(0.4\sigma)$ \\ \hline
\bhar-SFR                    &  $\boldsymbol {1\times 10^{-3}~(3.3\sigma)}$  & $\boldsymbol {2\times 10^{-4}~(3.7\sigma)}$  \\
\bhar-$C_1$ & $0.56~(0.6\sigma)$  & $0.44~(0.8\sigma)$ \\ \hline
\multicolumn{3}{c}{BD: $1.5 \leqslant  z < 3$ ($3\times 3$ bins)} \\ \hline
\bhar-SFR                   &  $\boldsymbol {5\times 10^{-4}~(3.5\sigma)}$  & $\boldsymbol {1\times 10^{-3}~(3.2\sigma)}$ \\
\bhar-$\Sigma_1$ & $0.23~(1.2\sigma)$  & $0.86~(0.2\sigma)$ \\ \hline
\bhar-SFR                   &  $\boldsymbol {2\times 10^{-4}~(3.8\sigma)}$  & $\boldsymbol {2\times 10^{-3}~(3.1\sigma)}$  \\
\bhar-$C_1$ & $0.70~(0.4\sigma)$  & $0.17~(1.4\sigma)$ \\ \hline\hline
\multicolumn{3}{c}{Non-BD: $0.5 \leqslant  z < 1.5$ ($5\times 5$ bins)} \\ \hline
\bhar-$M_\star$              &  $0.03~(2.2\sigma)$  & $0.16~(1.4\sigma)$  \\
\bhar-$\Sigma_1$           & $4 \times 10^{-3}~(2.9\sigma)$  & $0.02~(2.4\sigma)$  \\ \hline
\bhar-$M_\star$              &  $ \boldsymbol {8\times 10^{-8}~(5.4\sigma)}$  & $\boldsymbol {2\times 10^{-7}~(5.2 \sigma)}$  \\
\bhar-$C_1$ & $0.01~(2.6\sigma)$ & $0.02~(2.3\sigma)$  \\ \hline
\multicolumn{3}{c}{Non-BD: $1.5 \leqslant  z < 3$ ($3\times 3$ bins)} \\ \hline
\bhar-$M_\star$             &  $0.06~(1.9\sigma)$  & $0.36~(0.9\sigma)$  \\
\bhar-$\Sigma_1$ &  $0.11~(1.6\sigma)$  & $0.27~(1.1\sigma)$  \\ \hline
\bhar-$M_\star$             &  $ \boldsymbol {2 \times 10^{-3}~(3.1 \sigma)}$  & $\boldsymbol {4 \times 10^{-4}~(3.5\sigma)}$  \\
\bhar-$C_1$ &  $0.08~(1.8\sigma)$  & $0.03~(2.1\sigma)$  \\ \hline\hline
\multicolumn{3}{c}{SF Non-BD: $0.5 \leqslant  z < 1.5$ ($5\times 5$ bins)} \\ \hline
\bhar-$M_\star$              & 0.02 $(2.3\sigma)$  & 0.01 $(2.4\sigma)$  \\
\bhar-$\Sigma_1$ & $0.01~(2.5\sigma)$ & $0.11~(1.6\sigma)$  \\ \hline
\bhar-$M_\star$              &  $ \boldsymbol {4\times 10^{-6}~(4.6\sigma)}$  & $\boldsymbol {2\times 10^{-6}~(4.8\sigma)}$  \\
\bhar-$C_1$ & $\boldsymbol {3\times 10^{-3}~(3.0\sigma)}$ & $\boldsymbol {3 \times 10^{-3}~(3.0\sigma)}$  \\ \hline
\multicolumn{3}{c}{SF Non-BD: $1.5 \leqslant  z < 3$ ($3\times 3$ bins)} \\ \hline
\bhar-$M_\star$             &  $0.03~(2.2\sigma)$  & $0.05~(2.0\sigma)$  \\
\bhar-$\Sigma_1$ &  $0.21~(1.3\sigma)$  & $0.92~(0.1\sigma)$  \\ \hline
\bhar-$M_\star$             &  $4 \times 10^{-3}~(2.9 \sigma)$  & $\boldsymbol {2 \times 10^{-3}~(3.2\sigma)}$  \\
\bhar-$C_1$ &  $0.29~(1.1\sigma)$  & $0.56~(0.6\sigma)$  \\ \hline\hline
\end{tabular}                                         
\end{center}
\end{table}

\subsubsection{How does BH growth relate to \sigmaone\ for galaxies that are not bulge-dominated?} \label{sssec-nonbulge-s1}

\begin{figure}
\begin{center}
\includegraphics[scale=0.27]{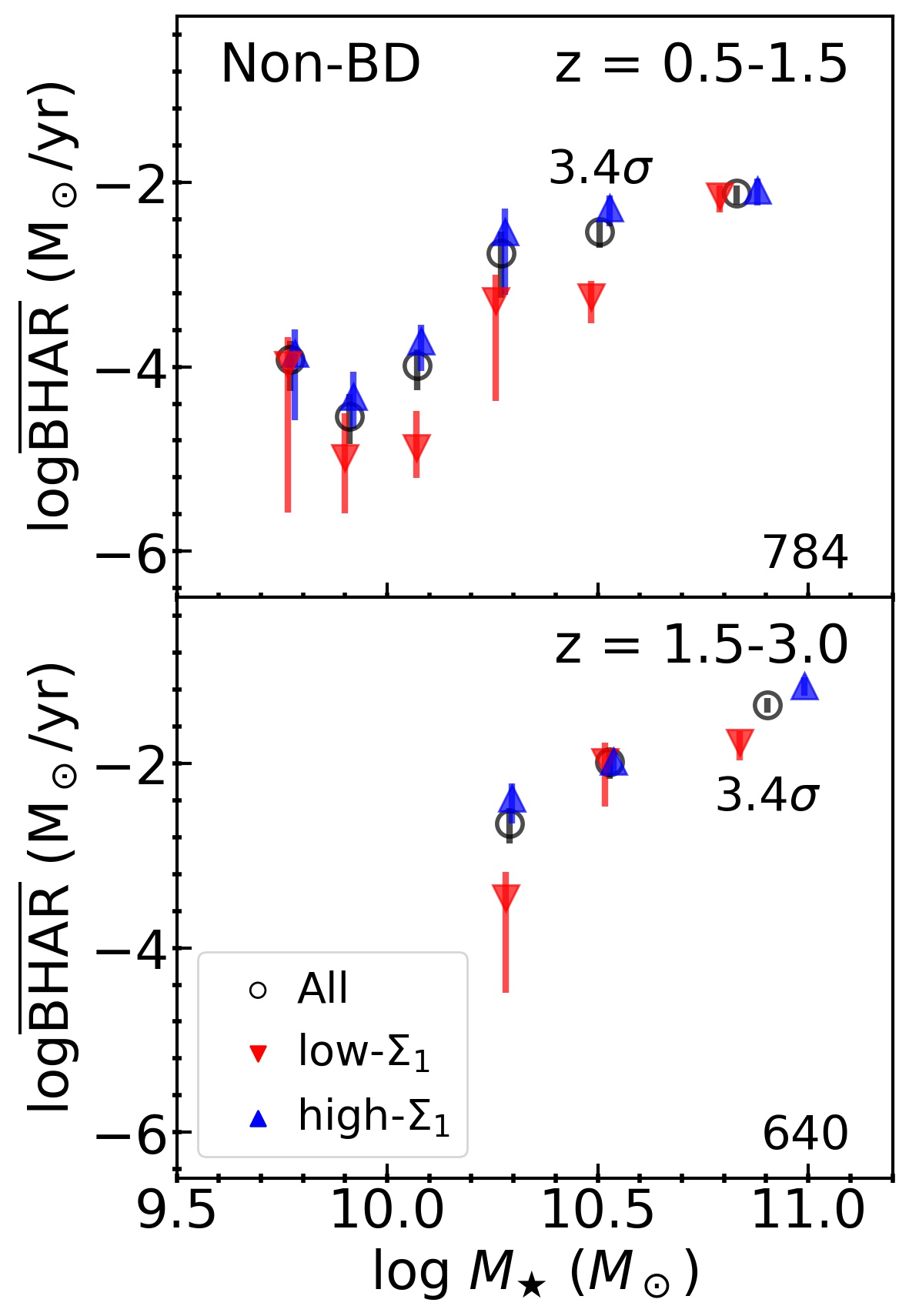}
\includegraphics[scale=0.27]{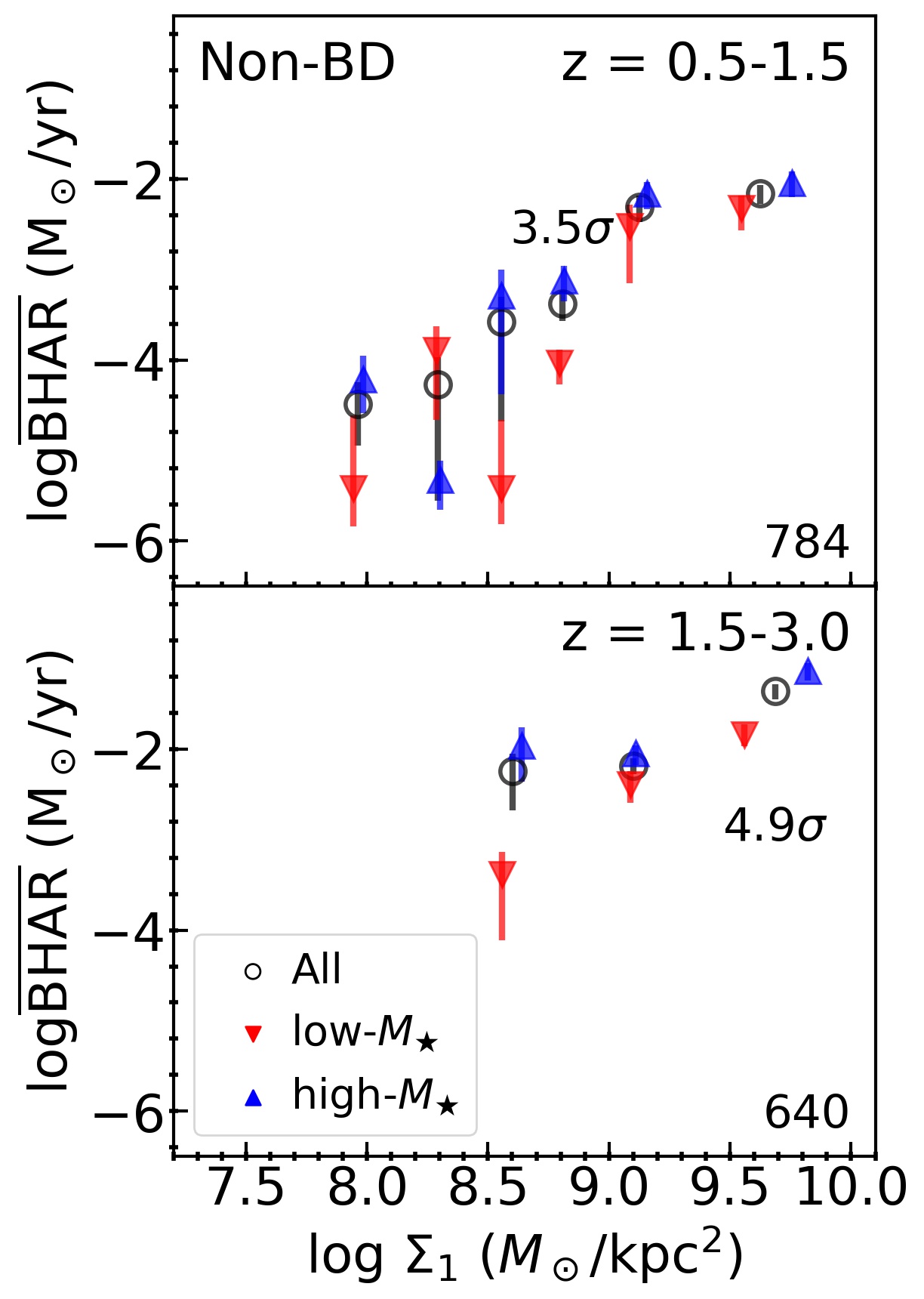}
\caption{$\rm \overline{BHAR}$ vs. $M_\star$ (left) and \sigmaone\ (right) for galaxies in the Non-BD sample. The horizontal position of each data point indicates the median $M_\star$/\sigmaone\ (left/right) of the sources in the bin. Each $M_\star$/\sigmaone\ sample (black circles) is further divided into two subsamples with \sigmaone/$M_\star$ above (blue upward-pointing triangles) and below (red downward-pointing triangles) the median \sigmaone/$M_\star$ of the sample, respectively. The error bars represent the 1$\sigma$ confidence interval of \bhar~from bootstrapping. The significance levels of the differences between \bhar~in the subsamples are labeled at the position of the bin if the level is $> 3\sigma$.
The number in the bottom-right corner represents the number of objects in each \mstar/\sigmaone\ bin.
Noticeable $\Delta \rm \overline{BHAR}$ values are associated with both \mstar\ and \sigmaone.}
\label{comp_trend_s1}
\end{center}
\end{figure}

\begin{figure}
\begin{center}
\includegraphics[scale=0.28]{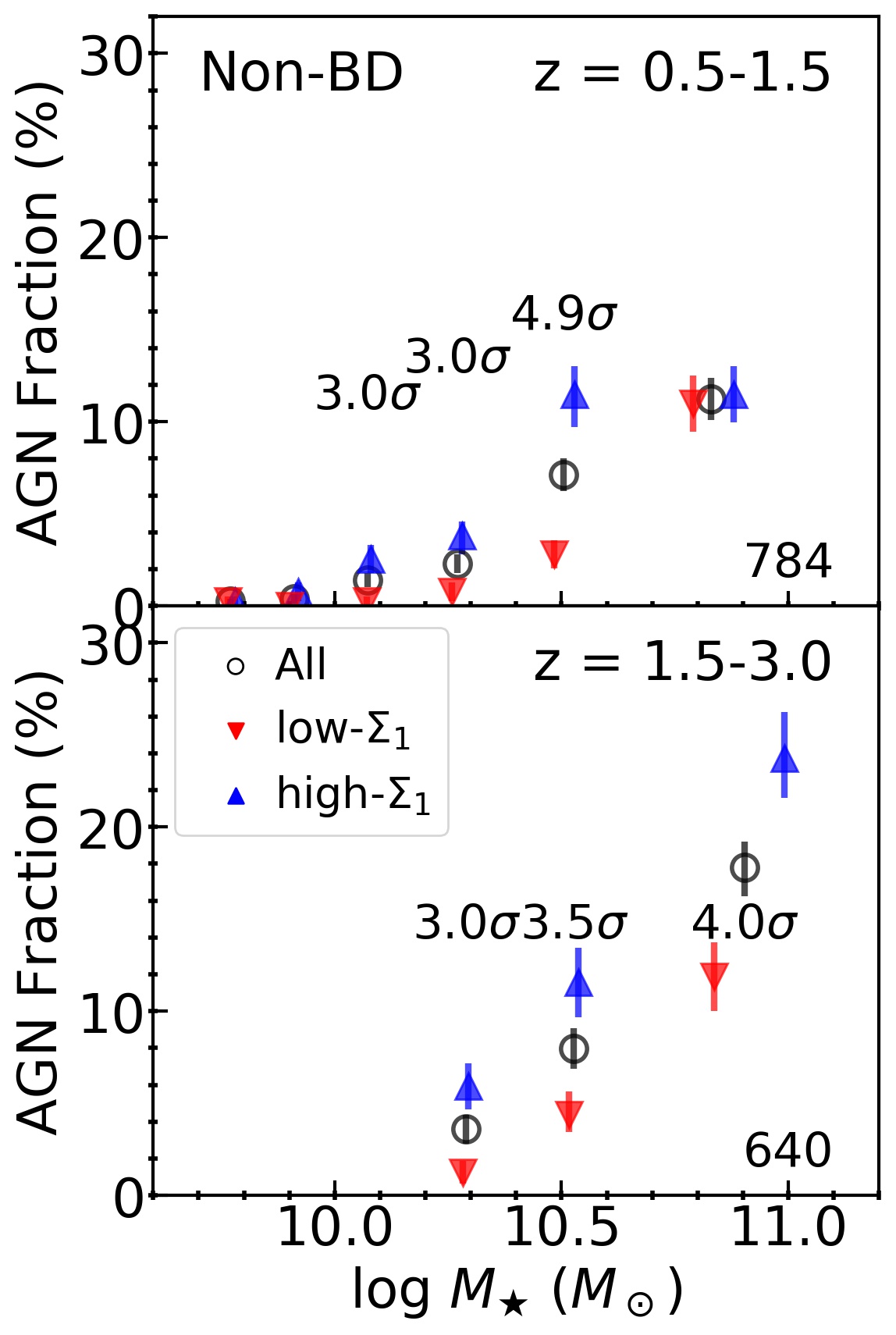}
\includegraphics[scale=0.28]{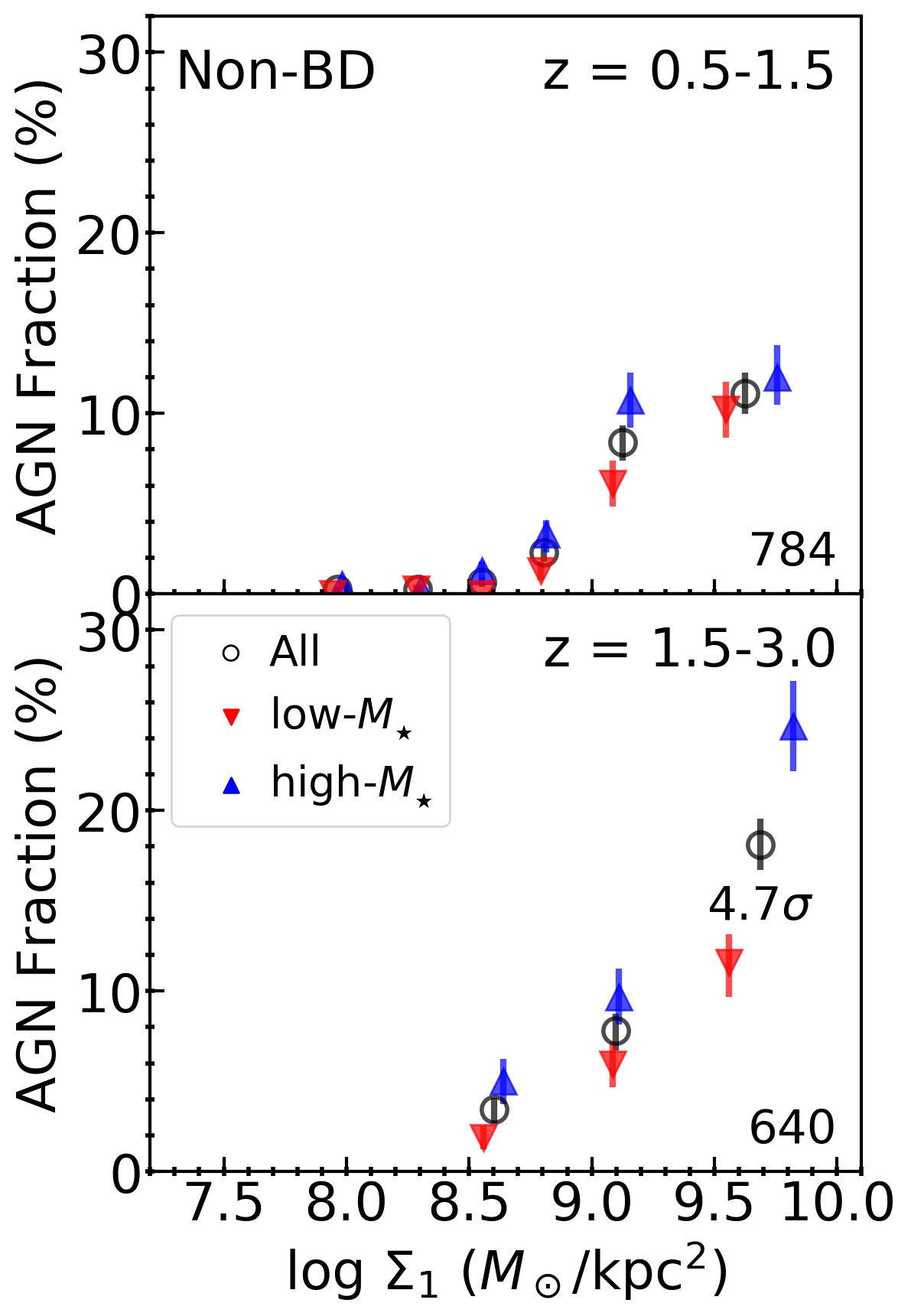}
\caption{AGN fraction vs. \mstar~(left) and \sigmaone\ (right) for galaxies in the Non-BD sample. The horizontal position of each data point indicates the median \mstar/\sigmaone\ (left/right) of the sources in the bin. Each \mstar/\sigmaone\ sample (black circles) is further divided into two subsamples with \sigmaone/\mstar~above (blue upward-pointing triangles) and below (red downward-pointing triangles) the median \sigmaone/\mstar~of the sample, respectively. The error bars represent the 1$\sigma$ confidence interval of AGN fraction from bootstrapping. The significance levels of the differences between AGN fraction in the subsamples are labeled at the position of the bin if the level is $> 3\sigma$.
The number in the bottom-right corner represents the number of objects in each \mstar/\sigmaone\ bin.
Noticeable \dagnf\ values are associated with \sigmaone\ mostly.}
\label{nonbulge_agnf_s1}
\end{center}
\end{figure}

\begin{figure}
\begin{center}
\includegraphics[scale=0.27]{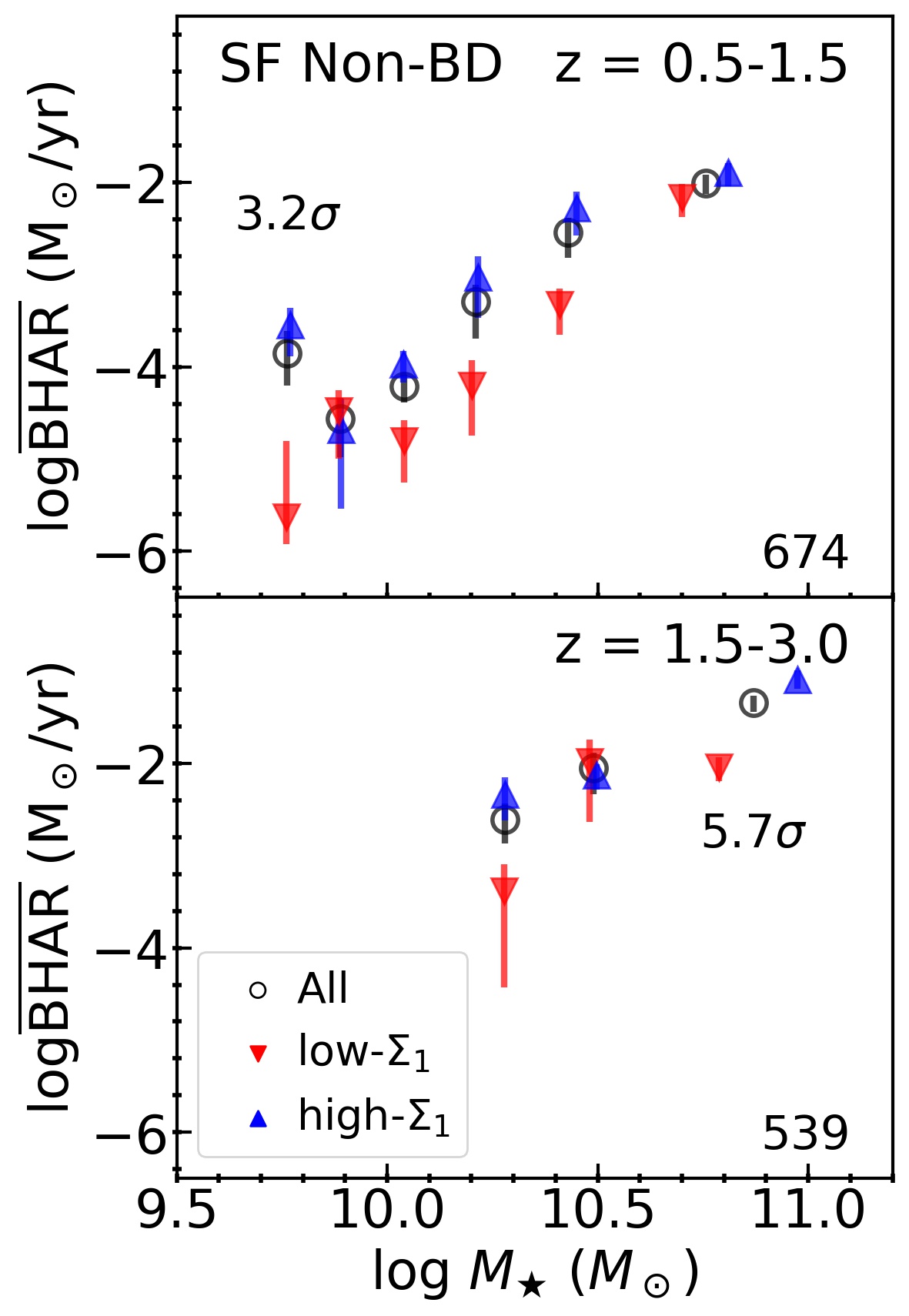}
\includegraphics[scale=0.27]{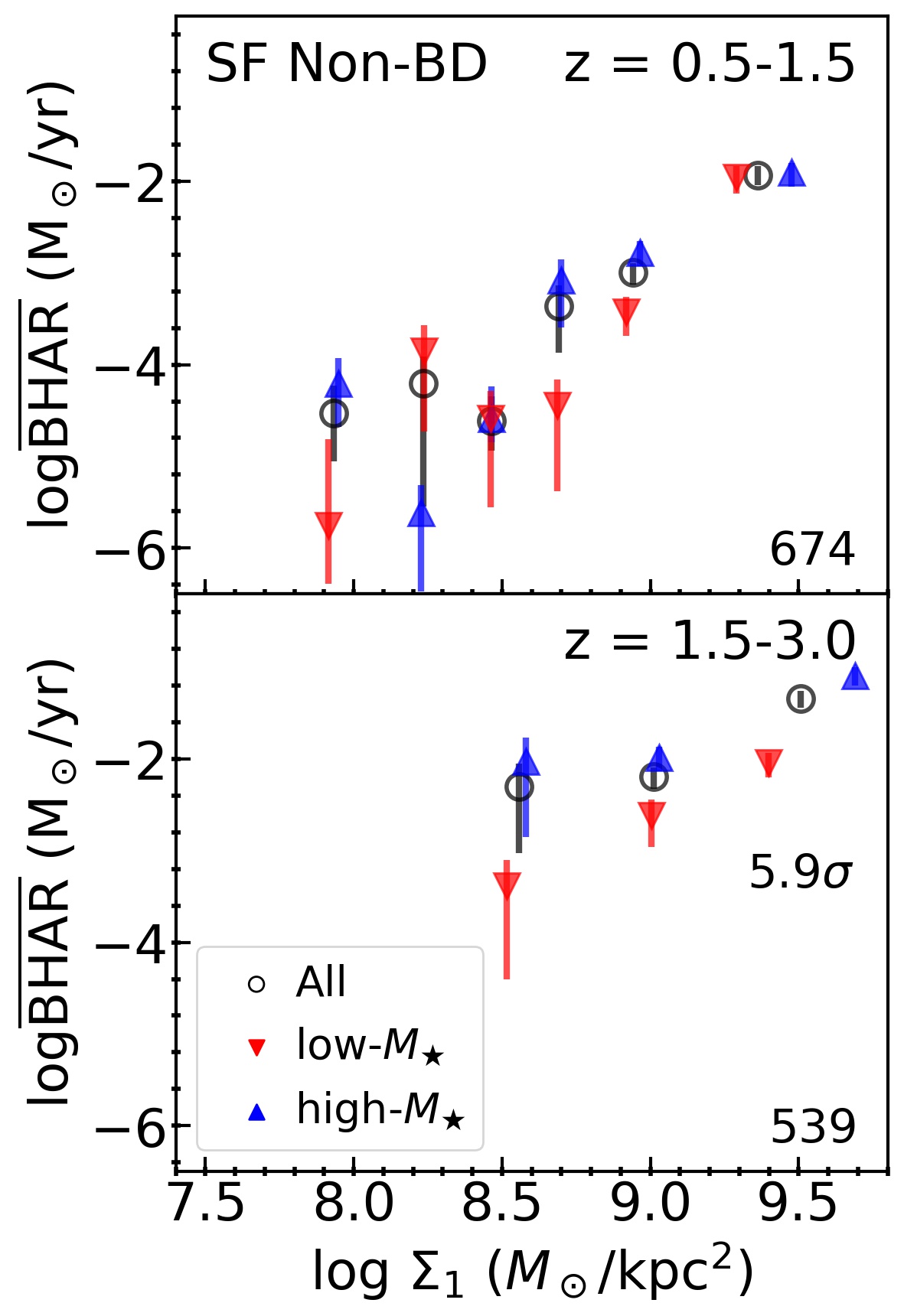}
\caption{Similar to Figure \ref{comp_trend_s1}, but for galaxies in the SF Non-BD sample. Noticeable $\Delta \rm \overline{BHAR}$ values are associated with both \mstar\ and \sigmaone.}
\label{comp_trend_s1_sf}
\end{center}
\end{figure}

\begin{figure}
\begin{center}
\includegraphics[scale=0.28]{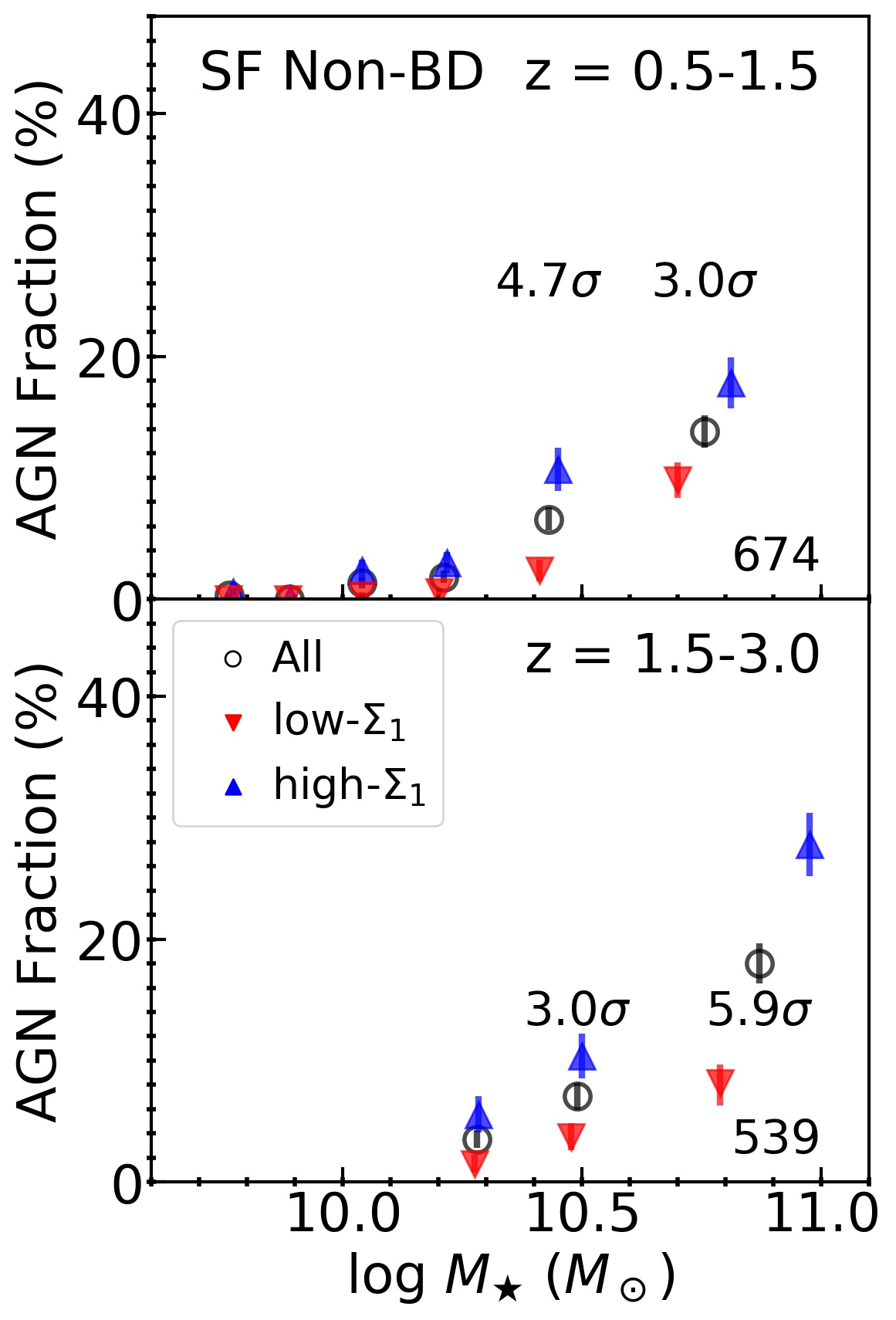}
\includegraphics[scale=0.28]{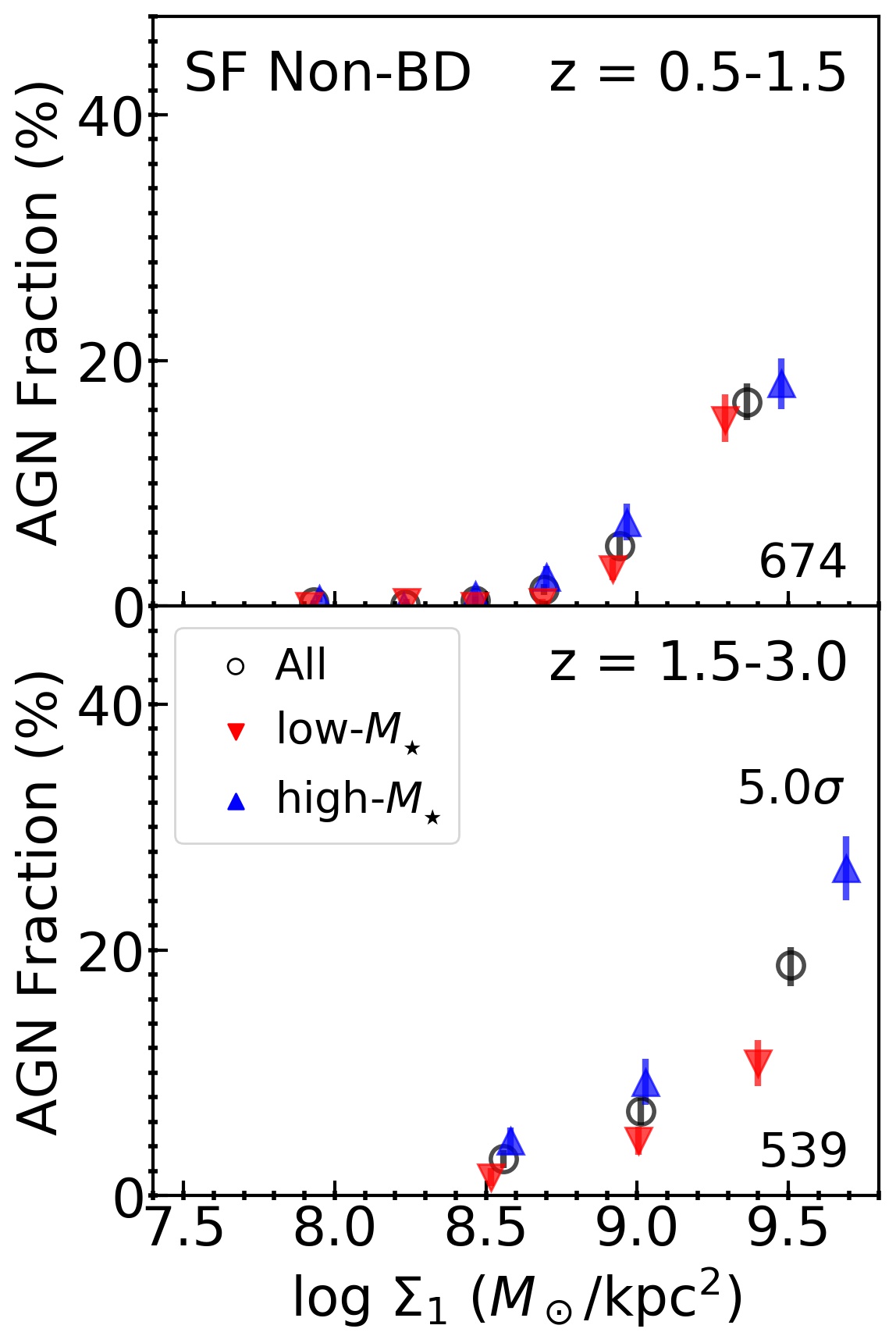}
\caption{Similar to Figure \ref{nonbulge_agnf_s1}, but for galaxies in the SF Non-BD sample. Noticeable \dagnf\ values are associated with \sigmaone\ mostly.}
\label{nonbulge_agnf_s1_sf}
\end{center}
\end{figure}

In Figures \ref{comp_trend_s1}/\ref{comp_trend_s1_sf}, we plot $\rm \overline{BHAR}$ as a function of $M_\star$ and \sigmaone\ for galaxies in the Non-BD/SF Non-BD sample.
Each \mstar/\sigmaone\ bin is further divided into two subsamples with \sigmaone/\mstar\ above or below the median \sigmaone/\mstar, and the \bhar\ values of these subsamples are shown on the plot as well.
We can see that for both the Non-BD and SF Non-BD samples, differences in $M_\star$ for a given \sigmaone\ (in the right panel) and differences in \sigmaone\ for a given \mstar\ (in the left panel) can both cause noticeable differences in \bhar. 
We also plot AGN fraction as a function of \mstar\ and \sigmaone\ for galaxies in the Non-BD/SF Non-BD sample in Figures \ref{nonbulge_agnf_s1}/\ref{nonbulge_agnf_s1_sf}.
The bins and subsamples in Figures \ref{nonbulge_agnf_s1}/\ref{nonbulge_agnf_s1_sf} are the same as those of Figures \ref{comp_trend_s1}/\ref{comp_trend_s1_sf}. 
We can see that, for massive galaxies with log \mstar\ $\gtrsim 10$ in the left panel of Figure~\ref{nonbulge_agnf_s1}, almost all the mass bins have \dagnf\ associated with \sigmaone\ at a $\gtrsim 3.0 \sigma$ significance level (except for the highest mass bin at $z= 0.5$--1.5), though only two bins satisfy the $3.6\sigma$ criterion after considering the Bonferroni correction.
When we confine the analysis to SF galaxies in the Non-BD sample, the highest mass bin at $z= 0.5$--1.5 also shows a hint of \dagnf\ (at 3.0$\sigma$) associated with \sigmaone\ (see the left panel of Figure \ref{nonbulge_agnf_s1_sf}). In contrast, significant \dagnf\ associated with \mstar\ can only be seen in one \sigmaone\ bin (in the right panels of Figures \ref{nonbulge_agnf_s1}/\ref{nonbulge_agnf_s1_sf}). These results naturally raise the question: is the \bhar-\sigmaone\ relation more fundamental than the \bhar-\mstar\ relation for both the Non-BD and SF Non-BD samples?

\begin{figure*}
\begin{center}
\includegraphics[scale=0.48]{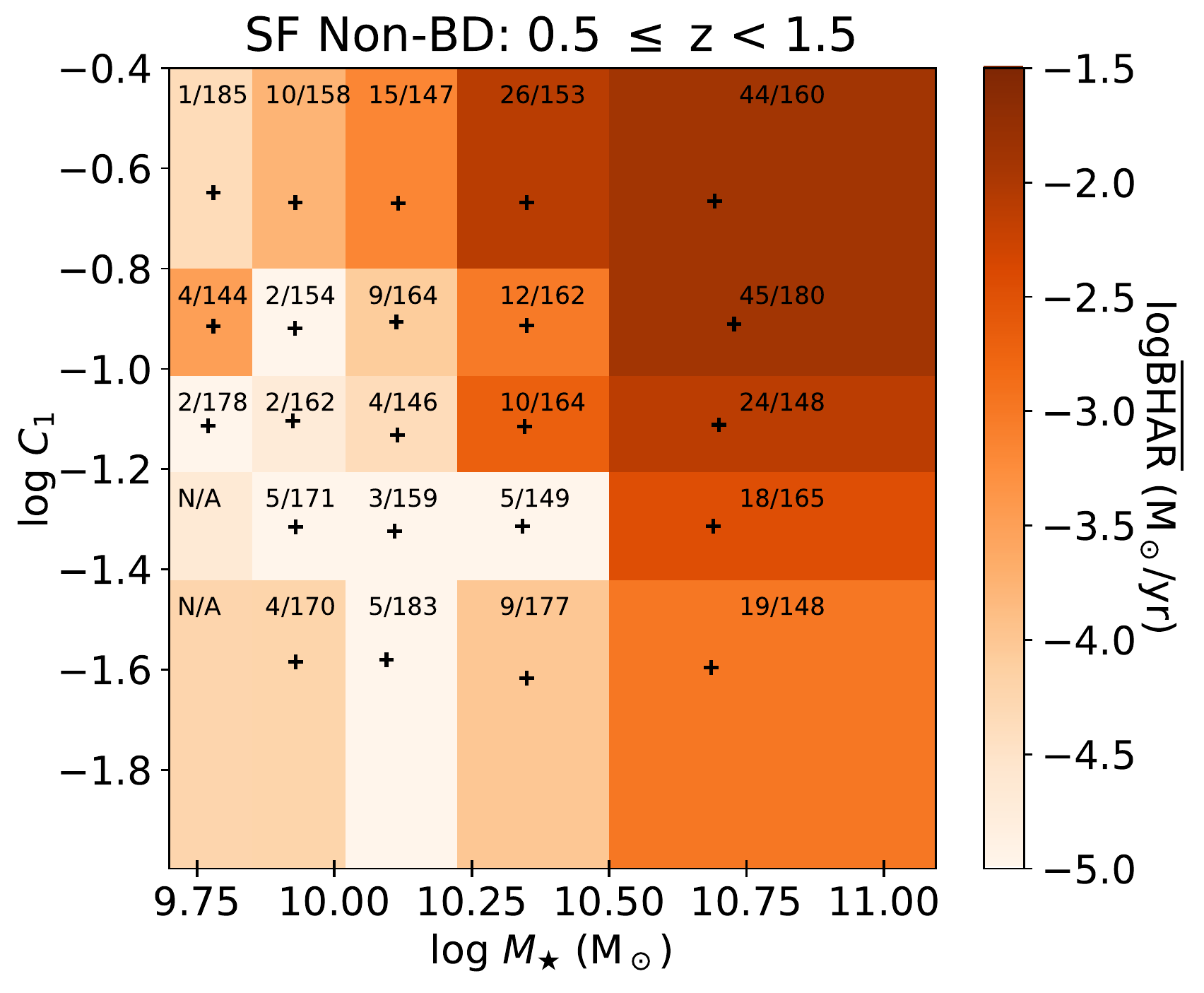}
\includegraphics[scale=0.48]{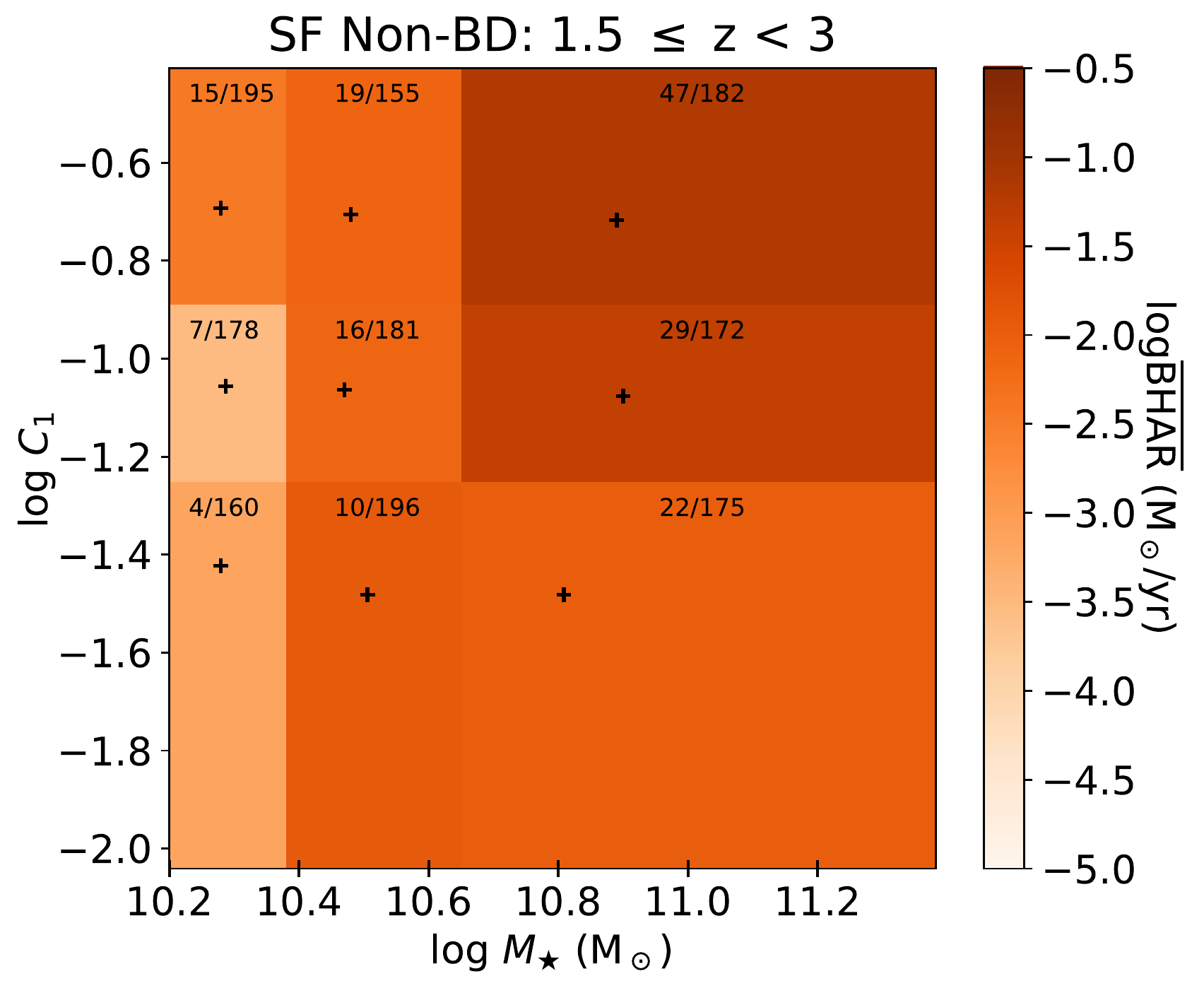}
\caption{Color-coded $\rm \overline{BHAR}$ in different bins of $M_\star$ and \cone\ for galaxies in the SF Non-BD sample. The black plus sign indicates the median $M_\star$ and \cone\ of the sources in each bin. The median log $M_\star$, median log \cone, and log $\rm \overline{BHAR}$ are the inputs to our PCOR analyses. 
For each bin, the number of X-ray detected galaxies and the total number of galaxies are listed.
For bins where \bhar\ does not have a lower limit $>0$ from bootstrapping, `N/A' is shown instead.
For a given \cone, the \bhar-\mstar~relation is overall noticeable, while the \bhar-\cone\ relation for a given \mstar\ is more noticeable at $z = 0.5$--1.5 than at $z = 1.5$--3.}
\label{comp_pcor_sf_s1}
\end{center}
\end{figure*}

We then perform PCOR analyses to assess if the \bhar-\mstar\ relation is simply a secondary manifestation of the \bhar-\sigmaone\ relation for both the Non-BD and SF Non-BD samples. We bin sources based on \mstar\ and \sigmaone, and use the median log \mstar, median log \sigmaone, and log \bhar\ of each bin as the input to the PCOR analyses.
The results are summarized in Table \ref{pcortablecore}. We note that neither the \bhar-\mstar\ nor \bhar-\sigmaone\ relations are significant for both the Non-BD and SF Non-BD samples, probably due to the high level of degeneracy between \mstar\ and $\Sigma_{1}$ (see Figure \ref{m-rho}). Thus, we are not able to conclude which of the \bhar-\mstar\ and \bhar-\sigmaone\ relations is the primary one for the Non-BD/SF Non-BD samples.
We further test if the \bhar-$C_1$ relation is significant when controlling for \mstar, to determine if \bhar\ truly depends on \sigmaone\ (\cone\ is the percentage of mass concentrated in the central 1~kpc and is independent of \mstar; log \cone\ $\approx$ log \sigmaone\ $-$ log \mstar\ $+$ \textit{Constant}, see Equations \ref{eq:s1} and \ref{eq:c1}). However, we note that when performing the PCOR analysis between \bhar, \mstar, and \cone, we will not be able to test if the \bhar-\mstar\ relation is a manifestation of the \bhar-\sigmaone\ relation.
As can be seen in Table \ref{pcortablecore}, the \bhar-\mstar\ relation becomes significant when the influence of \mstar\ in \sigmaone\ is removed for both the Non-BD and SF Non-BD samples. 
For the Non-BD sample, the \bhar-\cone\ relation is not significant when controlling for \mstar, suggesting that the \bhar-\sigmaone\ relation not fundamental in this sample.
For the SF Non-BD sample at $z = 0.5$--1.5, the \bhar-\cone\ relation is just significant at 3.0$\sigma$ when controlling for \mstar. 
At the same time, for the SF Non-BD sample at $z = 1.5$--3, the \bhar-\cone\ relation is not significant when controlling for \mstar.
We present the bins divided by \mstar\ and \cone\ of galaxies in the SF Non-BD sample utilized in the PCOR analyses in Figure \ref{comp_pcor_sf_s1}, with color-coded \bhar.
In the left panel of Figure \ref{comp_pcor_sf_s1}, we can directly observe apparent \bhar-\cone\ relations for a given \mstar\ at $z = 0.5$--1.5, especially at log \mstar\ $>$ 10. 

The above results indicate that, at least for the SF Non-BD sample at $z = 0.5$--1.5, the \bhar-\sigmaone\ relation is not likely to be only a secondary manifestation of the \bhar-\mstar\ relation. 
A larger sample will be needed to test if this statement holds indisputably for all redshift ranges, and if the \bhar-\sigmaone\ relation is indeed more fundamental than the \bhar-\mstar\ relation for the SF Non-BD sample. We will further discuss the observed link between BH growth and \sigmaone\ in Section \ref{ssec-dis1},
and we will also discuss the possibility that \bhar\ only truly depends on \sigmaone\ among massive galaxies (as indicated by Figure \ref{comp_pcor_sf_s1}) in Section \ref{sssec-103}.

\section{Discussion} \label{sec-discuss}

\subsection{The limited power of \sigmae} \label{ssec-dise}

In Section \ref{ssec-sigmae}, we found that BH growth does not fundamentally depend on \sigmae\ in general.
In Section~\ref{sssec-bulge}, we did not find a fundamental \bhar-\sigmae\ relation when controlling for SFR among galaxies in the BD sample; in Section \ref{sssec-nonbulge}, we did not find a fundamental \bhar-\sigmae\ relation when controlling for \mstar\ among galaxies in the Non-BD sample, even when considering only SF galaxies.
In Section~\ref{highsigmae}, we will discuss how these results compare with other results in the literature that have claimed elevated BH growth associated with \sigmae.
We will then discuss in Section~\ref{wce} the observed potential association between AGN fraction and \sigmae\ in a characteristic mass range at $z = 0.5$--1.5 among Non-BD galaxies and the possible reason for it.

\subsubsection{Comparison with other results in the literature} \label{highsigmae}
A correlation between $L_{\rm X}$ and compactness (defined as $M_\star/r_{\rm e} ^{1.5}$) has been found in \citet{Rangel2014}, utilizing a sample of 268 galaxies with \mstar\ $>$ 10$^{10.5}$ $M_\odot$ at $1.4 < z < 3$.
However, the lack of a fundamental link between \bhar\ and \sigmae\ (or \re) demonstrated in our work indicates that this correlation is not fundamental.
We found in Section \ref{sssec-nonbulge} that among Non-BD (or SF Non-BD) galaxies, \bhar\ does not significantly depend on \sigmae\ when controlling for \mstar;
in Appendix \ref{sssec-allsigmae}, we found that even when we do not distinguish between BD galaxies and Non-BD galaxies, no fundamental \bhar-\sigmae\ relation is obtained. 
The above results also hold true when limiting our analyses to galaxies with \mstar\ $>$ 10$^{10.5}$~$M_\odot$ at \hbox{$z = 1.5$--3}.
The \citet{Rangel2014} results likely arise due to the dependence of their compactness parameter on \mstar, since \mstar\ has a strong apparent link with BH growth \citep[e.g.][]{Yang2017,Yang2018a}.

We also note that in \citet{Kocevski2017}, the AGN fraction in massive ``high-\sigmae'' SF galaxies was found to be significantly higher than that in a mass-matched sample of ``low-\sigmae'' SF galaxies at $1.4 < z < 3$.\footnote{\label{cef}In \citet{Kocevski2017}, ``high-\sigmae'' SF galaxies are SF galaxies that satisfy the relation log \sigmae\ $> (-0.52~\times$ log\mstar\ $-~10.5) + 9.91 - 0.3$; ``low-\sigmae'' SF galaxies are SF galaxies that do not satisfy this relation.}
Given that we find the \dagnf\ association with \sigmae\ when controlling for \mstar\ is not significant among SF Non-BD galaxies at \hbox{$z = 1.5$--3} (see the lower left panel of Figure \ref{nonbulge_sf_agnf}), why is elevated BH growth among ``high-\sigmae'' SF galaxies compared with mass-matched ``low-\sigmae'' SF galaxies observed in \citet{Kocevski2017}?

We first notice that \citet{Kocevski2017} do not distinguish between bulge-dominated galaxies and galaxies that are not dominated by bulges.
We find that we also observe elevated BH growth associated with \sigmae\ in our sample if we do not distinguish between BD galaxies and Non-BD galaxies.
In our $z = 1.5$--3 sample, 216 SF galaxies satisfy the criterion of being ``high-\sigmae'' following \citet{Kocevski2017} (see our Footnote~\ref{cef} for the \citealt{Kocevski2017} definition of ``high-\sigmae'' galaxies), with median log\mstar\ $\approx 10.9$ and median log \sigmae\ $\approx 9.7$.
For each of these 216 galaxies, we select one ``low-\sigmae'' SF galaxy in our $z = 1.5$--3 sample that has the closest \mstar\ value to it (not allowing duplications) to constitute a mass-matched ``low-\sigmae'' sample with median log \sigmae\ $\approx 8.9$.
We find that the AGN fraction among these ``high-\sigmae'' SF galaxies is $33.3^{+3.3}_{-3.3}\%$, and the AGN fraction in the mass-matched sample of ``low-\sigmae'' SF galaxies is $18.1^{+2.9}_{-2.9}\%$. The difference in AGN fraction is significant at 3.5$\sigma$, consistent with the \citet{Kocevski2017} results.

However, if we only consider the 105 of these 216 SF galaxies that are not dominated by bulges (with median log\mstar\ $\approx 11.0$ and median log \sigmae\ $\approx 9.5$), we find that the AGN fraction among these ``high-\sigmae'' SF Non-BD galaxies is $28.0^{+4.0}_{-4.0}\%$, and the AGN fraction in the mass-matched sample of ``low-\sigmae'' SF galaxies with median log \sigmae\ $\approx 8.9$ is $26.0^{+4.0}_{-4.0}\%$. The significance of the difference in AGN fraction is only 0.4$\sigma$, consistent with the limited power of \sigmae\ presented in Section~\ref{sssec-nonbulge}.

Thus, the high AGN fraction found by \citet{Kocevski2017} among ``high-\sigmae'' SF galaxies may not be due to high \sigmae\ values per se, but rather due to the presence of many SF bulges ($\approx 50\%$) which generally have high \sigmae\ values and high levels of BH growth \citep[e.g.][]{Silverman2008,Yang2019}. 
\citet{Yang2019} argue that the high level of BH growth among SF bulges can be explained by the \bhar-SFR relation among bulge-dominated galaxies. As can be seen in Table~\ref{pcortable}, for galaxies in the BD sample, the \bhar-SFR relation is significant while the \bhar-\sigmae\ relation is not. Even when only SF bulges are considered, we do not observe a significant difference in \bhar\ associated with \sigmae\ (see Section~\ref{sssec-bulge}). These findings further support the idea that, among bulge-dominated galaxies, \bhar\ is fundamentally related to SFR rather than \sigmae.

%\vspace{0.5cm}
We also note that the correlation between $L_{\rm X}$ and compactness found in \citet{Rangel2014} and the elevated BH growth among \hbox{``high-\sigmae''} SF galaxies found in \citet{Kocevski2017} may ultimately reflect a \bhar-\sigmaone\ relation existing among all SF galaxies.\footnote{In the Appendix of \citet{Kocevski2017}, they also found elevated AGN fraction associated with \sigmaone. However, they did not try to distinguish the relative roles of \sigmae\ and \sigmaone\ in predicting BH growth.}
We will discuss this \bhar-\sigmaone\ relation for the overall SF galaxy population in Section \ref{allsf-sigma1}.

\subsubsection{Potential association between AGN fraction and \sigmae\ in a characteristic mass range: the effects of wet compaction events?} \label{wce}

The only place where a significant difference in BH growth associated with \sigmae\ can been seen among Non-BD/SF Non-BD galaxies is for the log\mstar~$\approx$ 10.5/10.4 bin at \hbox{$z = 0.5$--1.5} in terms of \dagnf\ (see the left panels of Figures \ref{nonbulge_agnf}/\ref{nonbulge_sf_agnf}), at 4.0$\sigma$/$3.7\sigma$.  
When using the Bonferroni correction to adjust the required significance level for these \dagnf\ values in Section~\ref{sssec-nonbulge}, we consider the number of tests to be the number of \mstar\ bins in the Non-BD/SF Non-BD sample. However, if we are more conservative and treat the number of tests as the total number of \mstar\ bins in Figures~\ref{bulge_trend}, \ref{comp_trend}, and \ref{comp_sf_trend} (24), we can only call a difference significant if the level is $> 3.9 \sigma$. In this case, it is less certain that the \dagnf\ associated with \sigmae\ in a characteristic mass range is not due to statistical fluctuations.

If \agnf\ is indeed associated with \sigmae\ in this characteristic mass range, this could possibly be explained by a scenario where BH growth is triggered by the high gas density during a wet compaction event \citep[e.g.][]{Wellons2015,Habouzit2019} which changes the \re\ of galaxies at the critical halo mass $M_{\rm halo} \sim 10^{12} M_\odot$.
It has been suggested that, below the critical halo mass $M_{\rm halo} \sim 10^{12} M_\odot$, supernova feedback is efficient at evacuating the core and BH growth is thus suppressed \citep[e.g.][]{DS1986,Dekel2017,Kocevski2017,Dekel2019}. Once the halo reaches the critical mass,  the compressed gas during wet compaction events triggered among disks \citep{Dekel2014} can overcome supernova feedback and activate BH growth.  After that, the BH continues to grow and regulates the accretion itself. Thus, BH growth will not be linked with \sigmae\ significantly when $M_{\rm halo} \gtrsim 10^{12}~M_\odot$. For $M_{\rm halo} \sim 10^{12}~M_\odot$, the corresponding \mstar~is $\sim 10^{10.4-10.5} M_\odot$ at $z \approx 0.5$ \hbox{\citep[e.g.][]{Legrand2018}}, which is consistent with the characteristic mass we observed. The corresponding \mstar~is $\sim 10^{10} M_\odot$ at $z \approx 2$, which can also explain why we do not observe significant differences in AGN fraction linked with \sigmae~at $z = 1.5$--3: our \mstar-complete sample does not include galaxies with log\mstar~$< 10.2$ in this redshift range (and we do observe a $3.2 \sigma$ significance for \dagnf\ at log\mstar~$\approx 10.3$ for the SF Non-BD sample in Figure~\ref{nonbulge_sf_agnf}). It is not clear from this scenario why the triggered AGNs have low $L_{\rm X}$ (as found in Section \ref{sssec-nonbulge}, the relevant AGNs mainly have $L_{\rm X} = 10^{42-43}$ erg s$^{-1}$). This may be due to the limited gas content at $z = 0.5$--1.5.

\subsection{The relevance of \sigmaone\ to BH growth} \label{ssec-dis1}
In Section \ref{sssec-nonbulge-s1}, we found significant \dagnf\ associated with \sigmaone\ in the Non-BD and SF Non-BD samples at $z = 0.5$--3 (see Figures \ref{nonbulge_agnf_s1} and \ref{nonbulge_agnf_s1_sf}), in contrast to the overall non-significant \dagnf\ associated with \sigmae\ (see Figures \ref{nonbulge_agnf} and \ref{nonbulge_sf_agnf}). 
The \bhar-$C_1$ relation has a 3.0$\sigma$ significance when controlling for \mstar\ for the SF Non-BD sample at $z = 0.5$--1.5 (see Section~\ref{sssec-nonbulge-s1} and Table~\ref{pcortablecore}), suggesting that the \bhar-\sigmaone\ relation is not likely just a secondary manifestation of the primary \bhar-\mstar\ relation at least in this regime.
In Section~\ref{allsigma1-other}, we will discuss the physical implications of this \bhar-\sigmaone\ relation and its possible existence in a broader regime.
In Section~\ref{allsf-sigma1}, we will study the \bhar-\sigmaone\ relation for the overall SF galaxy population when controlling for \mstar. This is motivated by the discussion in Section~\ref{allsigma1-other} proposing that if the \bhar-SFR relation of SF BD galaxies is reflecting the same underlying link as the \bhar-\sigmaone\ relation, there is no need to distinguish between SF BD and SF Non-BD galaxies.
In Section~\ref{sssec-103}, we will study the \bhar-\sigmaone\ relation among SF galaxies when $M_{\rm halo} \gtrsim 10^{12} M_\odot$, as theoretical ideas argue that BH growth will be suppressed by supernova feedback when $M_{\rm halo} \lesssim 10^{12} M_\odot$.

\subsubsection{The \bhar-\sigmaone\ relation as a link between BH growth and the central gas density within 1~kpc?} \label{allsigma1-other}

As can be seen in Table~\ref{pcortablecore}, the \bhar-\cone\ relation has a 3.0$\sigma$ significance for the SF Non-BD sample at $z = 0.5$--1.5.
For the Non-BD sample in general at $z = 0.5$--1.5, the \bhar-$C_1$ relation is not significant when controlling for \mstar.
This suggestive confirmation of the \bhar-\sigmaone\ relation only among SF galaxies in the Non-BD sample at $z = 0.5$--1.5 indicates that if the \bhar-\sigmaone\ relation truly exists among SF Non-BD galaxies, it may not be reflecting a link between BH growth and the central stellar-mass density within 1~kpc.
Instead, it may reflect a link between BH growth and the central gas density within 1~kpc, with the rough assumption that the \mstar-to-gas ratios of galaxies are the same. As mentioned in Section \ref{ssec-method}, \sigmaone\ can only serve as an indicator of the central gas density for galaxies that are actively forming stars since when galaxies become quiescent, it is unclear that \sigmaone\ can trace gas conditions.

It is reasonable to speculate that the \bhar-\sigmaone\ relation also exists among SF BD galaxies, as indicated by the significant difference in BH growth associated with \sigmaone\ for such systems (see Section \ref{sssec-bulges1}). 
However, as can be seen in Section~\ref{sssec-bulge}, a significant difference in BH growth is also associated with \mstar, and the current sample size of SF BD galaxies is too small to perform PCOR analyses to disentangle the relative roles of \mstar\ and \sigmaone\ effects.
If a significant \bhar-\sigmaone\ relation can be confirmed when controlling for both SFR and \mstar\ among SF BD galaxies, a straightforward explanation might be found for local BH ``monsters'' (see Section \ref{sec-intro}) by attributing their unexpectedly large $M_{\rm BH}$ values to elevated BH growth linked with the compactness of host galaxies in the central region.
As we discussed before, the \bhar-\sigmaone\ relation could be considered as a manifestation of the link between BH growth and the amount of gas in the vicinity of the central BH.
This underlying link may also be the one reflected by the \bhar-SFR relation among bulges.
Specifically, we know that galaxies in the BD sample are generally compact, with a median \re\ of $1.5/1.1~\rm kpc$ in the low/high-redshift bin.
Thus, it is possible that the SFR of bulges is substantially correlated with the total amount of cold gas available in the central $\sim$1 kpc region, and the \bhar-SFR relation of bulges is actually a secondary manifestation of an underlying relation between BH growth and the amount of gas in the vicinity of the central BH. 
When considering the possibility that the \bhar-\sigmaone\ and \bhar-SFR relations may reflect the same underlying link among SF bulges, there is no need to distinguish between BD and Non-BD galaxies when testing the significance of the \bhar-\sigmaone\ relation among all SF galaxies, and we only need to control for \mstar. We will perform such PCOR analyses for the overall SF galaxy population in Section \ref{allsf-sigma1}.

\subsubsection{The \bhar-\sigmaone\ relation for the overall SF galaxy population} \label{allsf-sigma1}

\begin{figure*}
\begin{center}
\includegraphics[scale=0.48]{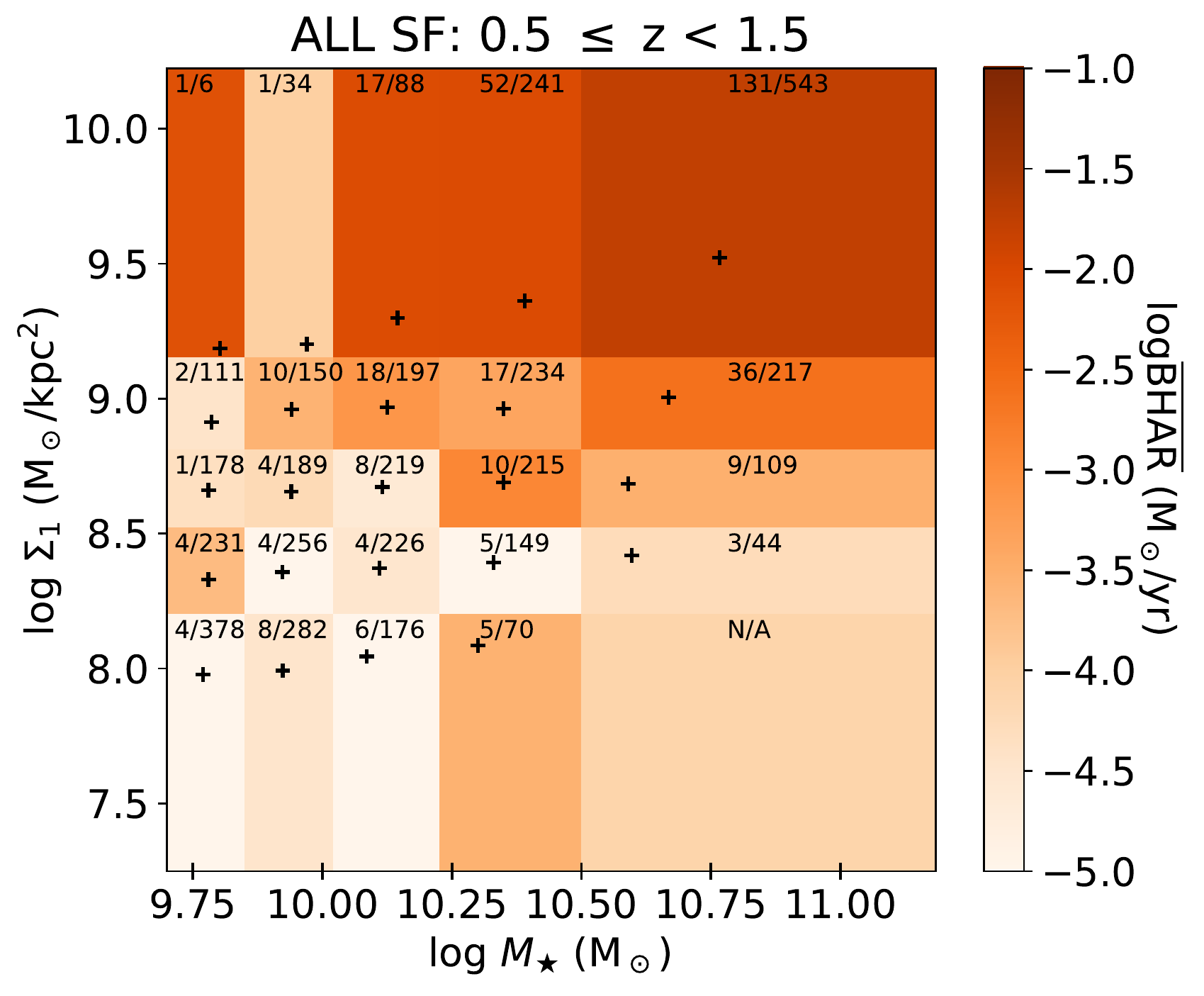}
\includegraphics[scale=0.48]{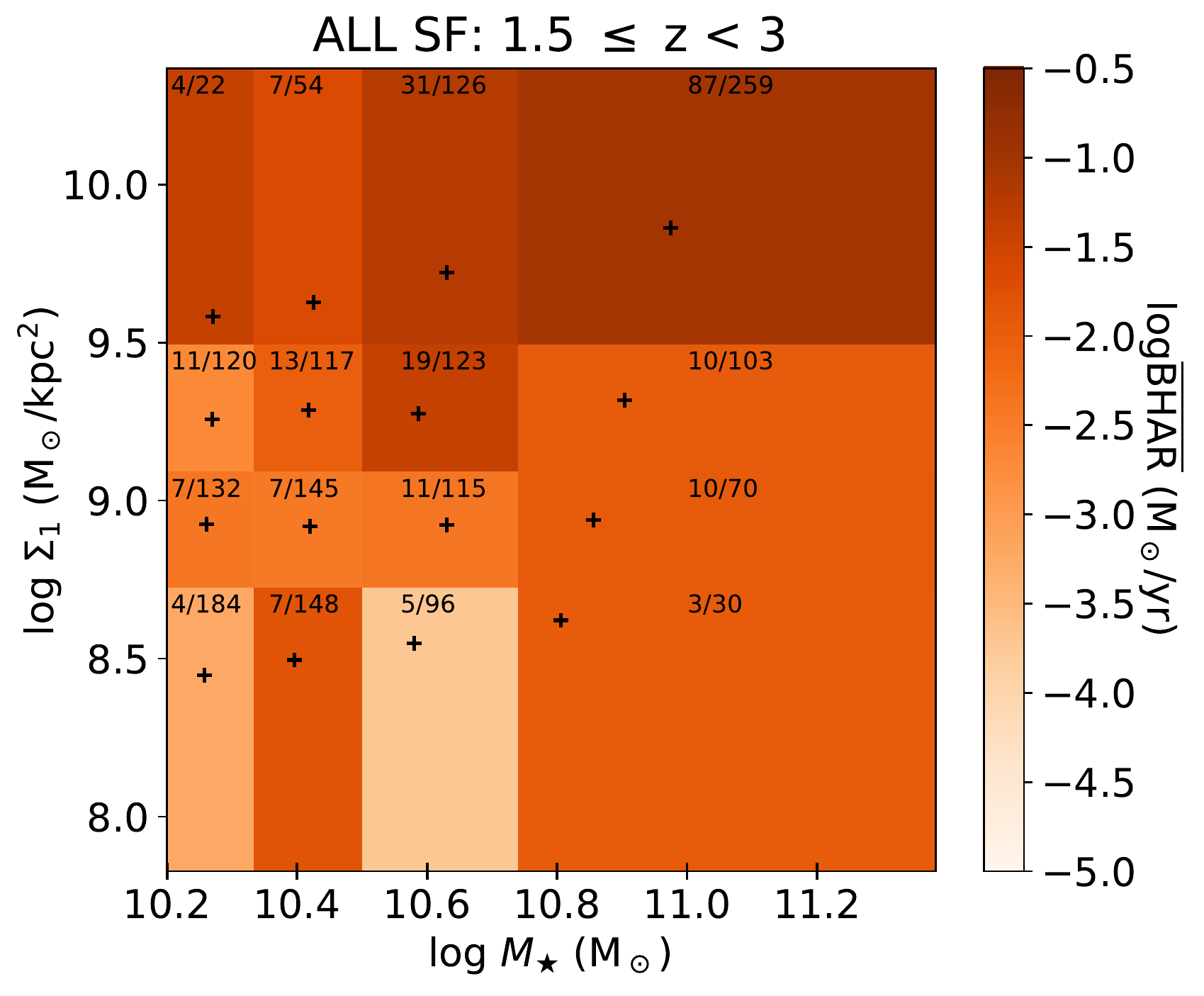}
\caption{Color-coded $\rm \overline{BHAR}$ in different bins of $M_\star$ and \sigmaone\ for all the SF galaxies in the sample. The black plus sign indicates the median $M_\star$ and \sigmaone\ of the sources in each bin. The median log $M_\star$, median log \sigmaone, and log $\rm \overline{BHAR}$ are the inputs to our PCOR analyses. For each bin, the number of X-ray detected galaxies and the total number of galaxies are listed.
For bins where \bhar\ does not have a lower limit $>0$ from bootstrapping, `N/A' is shown instead.
The \bhar-\sigmaone~relation is overall more noticeable than the \bhar-\mstar\ relation.
}
\label{all_pcor_sf_s1}
\end{center}
\end{figure*}

\begin{figure*}
\begin{center}
\includegraphics[scale=0.48]{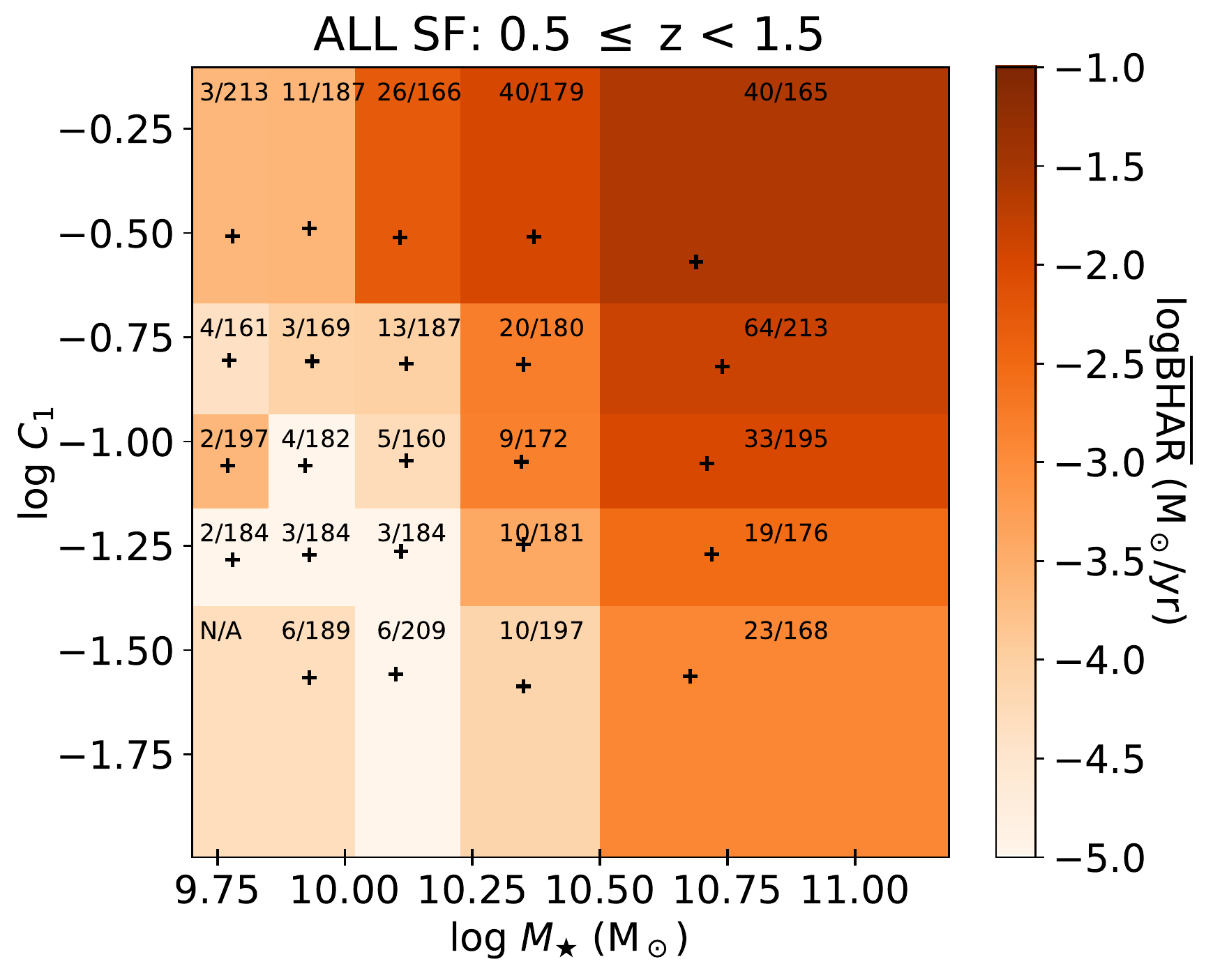}
\includegraphics[scale=0.48]{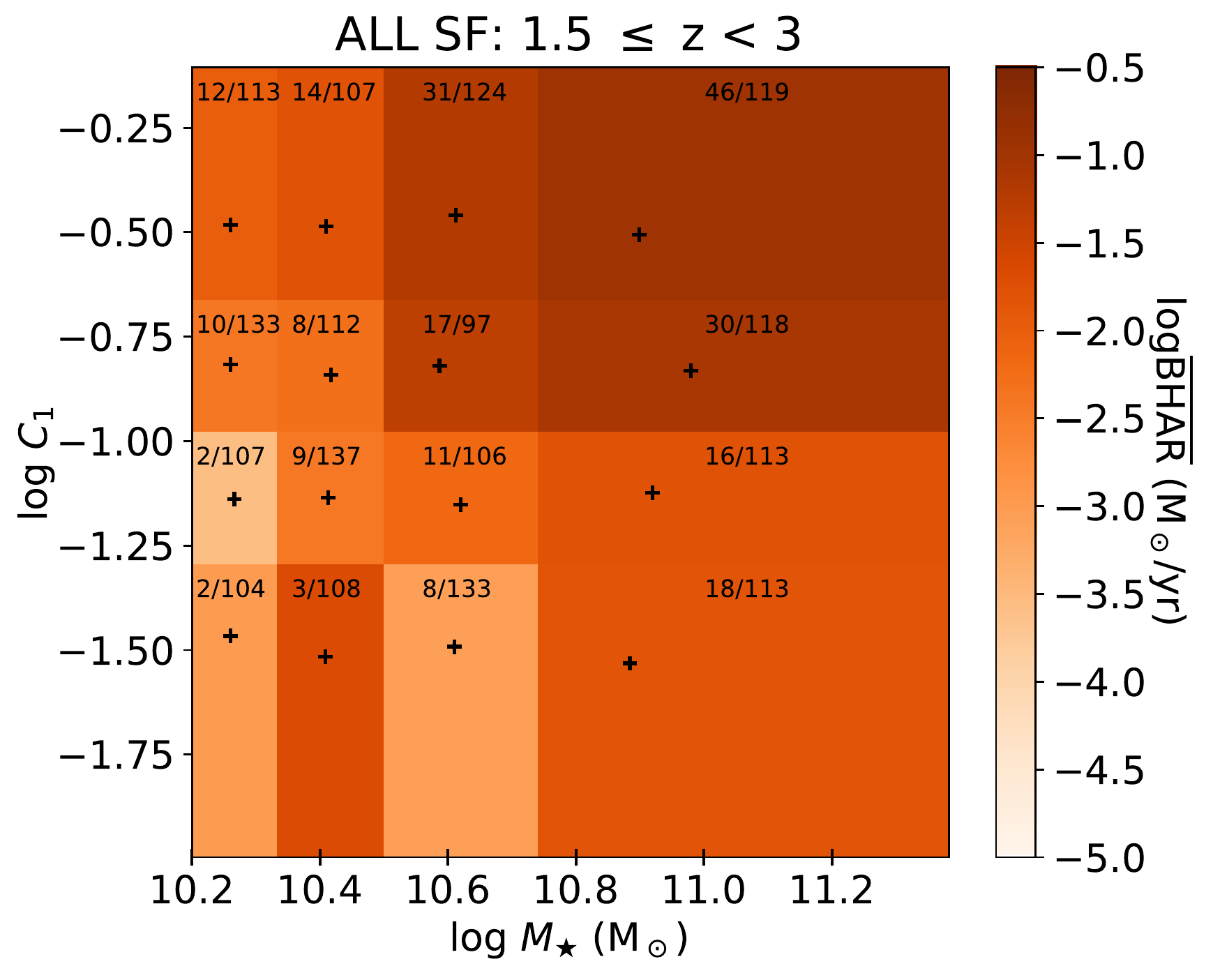}
\caption{Color-coded $\rm \overline{BHAR}$ in different bins of $M_\star$ and \cone\ for all the SF galaxies in the sample. The black plus sign indicates the median $M_\star$ and \cone\ of the sources in each bin. The median log $M_\star$, median log \cone, and log $\rm \overline{BHAR}$ are the inputs to our PCOR analyses. For each bin, the number of X-ray detected galaxies and the total number of galaxies are listed.
For bins where \bhar\ does not have a lower limit $>0$ from bootstrapping, `N/A' is shown instead.
Both the \bhar-\mstar\ and the \bhar-\cone\ relations are noticeable.
}
\label{all_pcor_sf_c1}
\end{center}
\end{figure*}

We bin all SF galaxies based on both \mstar\ and \sigmaone\ (see Figure~\ref{all_pcor_sf_s1}) to assess if the \bhar-\sigmaone\ relation is more fundamental than the \bhar-\mstar\ relation when considering all SF galaxies together.
We perform PCOR analyses with the median log\mstar, median log \sigmaone, and log \bhar\ of bins, and the results are summarized in Table~\ref{pcortableallsf}. 
We can see that for all SF galaxies at $z = 0.5$--1.5, the \bhar-\sigmaone\ relation is significant when controlling for \mstar, and the \bhar-\mstar\ relation is not significant when controlling for \sigmaone. We note that when the bin numbers are reduced from $5 \times 5$, neither the \bhar-\sigmaone\ nor \bhar-\mstar\ relations are significant at a 3$\sigma$ level, but the \bhar-\sigmaone\ relation remains more significant than the \bhar-\mstar\ relation. 
For all SF galaxies at \hbox{$z = 1.5$--3}, neither the \bhar-\sigmaone\ nor \bhar-\mstar\ relations are significant. 
Similar to the approach in Section \ref{sssec-nonbulge-s1}, we bin sources based on \mstar\ and $C_1$ (see Figure~\ref{all_pcor_sf_c1}) to test if the \bhar-\cone\ relation is significant when controlling for \mstar, thus assessing if the \bhar-\sigmaone\ relation can be explained as a secondary manifestation of the primary \bhar-\mstar\ relation. The median log\mstar, median log~\cone, and log \bhar\ of bins are the inputs to the PCOR analyses, and the results are also presented in Table~\ref{pcortableallsf}.
We found that for all SF galaxies at $z = 0.5$--1.5, the \bhar-\mstar\ relation is significant when controlling for $C_1$, and the \bhar-$C_1$ relation is also significant when controlling for \mstar. 
For all SF galaxies at $z = 1.5$--3, the \bhar-\mstar\ relation is significant when controlling for \cone, and the \bhar-\cone\ relation is not significant when controlling for \mstar.
In Section~\ref{sssec-c}, we mentioned that our results in Section \ref{sec-ar} do not change when limiting the analyses to $H < 23.5$ objects in the sample, where the \sersic\ index $n$ can be measured at the same level of accuracy as \re\ among galaxies with $H \sim$ 24.5 \citep{vanderwel2012}. 
However, if we confine our sample to $H < 23.5$ SF galaxies at $z = 1.5$--3 here ($\approx 77\%$ of all the SF galaxies at $z = 1.5$--3), the \bhar-\mstar\ and \bhar-\cone\ relations are both significant (see Table~\ref{pcortableallsf}).

Overall, the results above indicate that the \bhar-\sigmaone\ relation among all SF galaxies is not likely to be a secondary manifestation of the primary \bhar-\mstar\ relation at $z = 0.5$--3, and it is possible that the \bhar-\mstar\ relation is indeed not fundamental, but a manifestation of the link between BH growth and the central gas density, which can be reflected more effectively by the \bhar-\sigmaone\ relation among SF galaxies.

\begin{table}
\begin{center}
\caption{$p$-values (significances) of partial correlation analyses for the \bhar-\sigmaone\ relation among SF galaxies}
\label{pcortableallsf}
\begin{tabular}{ccccc}\hline\hline
Relation &   Pearson & Spearman \\\hline\hline
\multicolumn{3}{c}{All SF Galaxies: $0.5 \leqslant z < 1.5$ ($5 \times 5$ bins)} \\ \hline
\bhar-$M_\star$              &  $0.06~(1.9\sigma)$  & $0.11~(1.6\sigma)$ \\
\bhar-\sigmaone\ & $\boldsymbol {4\times 10^{-4}~(3.5\sigma)}$  & $\boldsymbol {2\times 10^{-3}~(3.1\sigma)}$ &  \\ \hline
\bhar-$M_\star$              &  $\boldsymbol {9\times 10^{-7}~(4.9\sigma)}$  & $\boldsymbol {4\times 10^{-6}~(4.6\sigma)}$ \\
\bhar-$C_1$ & $\boldsymbol {9\times 10^{-5}~(3.9\sigma)}$  & $\boldsymbol {6\times 10^{-4}~(3.4\sigma)}$ &  \\ \hline\hline
\multicolumn{3}{c}{All SF Galaxies: $1.5 \leqslant z < 3$ ($4 \times 4$ bins)} \\ \hline
\bhar-$M_\star$              &  $0.42~(0.8\sigma)$  & $0.31~(1.0\sigma)$ \\
\bhar-\sigmaone\            & $9\times 10^{-3}~(2.6\sigma)$  & $0.05~(2.0\sigma)$ &  \\ \hline
\bhar-$M_\star$              &  $\boldsymbol {2\times 10^{-3}~(3.1\sigma)}$  & $6\times 10^{-3}~(2.8\sigma)$ \\
\bhar-$C_1$                    & $7\times 10^{-3}~(2.7\sigma)$  & $0.02~(2.3\sigma)$ &  \\ \hline\hline
\multicolumn{3}{c}{All SF Galaxies: $1.5 \leqslant z < 3$ and $H < 23.5$ ($4 \times 4$ bins)} \\ \hline
\bhar-$M_\star$              &  $0.25~(1.1\sigma)$  & $0.09~(1.7\sigma)$ \\
\bhar-\sigmaone\ & $0.02~(2.3\sigma)$  & $0.02~(2.3\sigma)$ &  \\ \hline
\bhar-$M_\star$              &  $\boldsymbol {1\times 10^{-4}~(3.9\sigma)}$  & $\boldsymbol {7\times 10^{-5}~(4.0\sigma)}$ \\
\bhar-$C_1$ & $\boldsymbol {2\times 10^{-4}~(3.7\sigma)}$  & $\boldsymbol{1\times 10^{-4}~(3.8\sigma)}$ &  \\ \hline\hline 
\end{tabular}                                         
\end{center}
\end{table}

\subsubsection{The \bhar-\sigmaone\ relation among SF galaxies when $M_{\rm halo} \gtrsim 10^{12} M_\odot$} \label{sssec-103}

There is suggestive evidence in Section \ref{allsf-sigma1} for the \bhar-\mstar\ relation being a manifestation of a link between BH growth and central gas density that can be reflected more effectively by the \bhar-\sigmaone\ relation among SF galaxies.
However, we still cannot demonstrate this result confidently since the only place where the \bhar-\sigmaone\ relation ``beats'' the \bhar-\mstar\ relation in the PCOR analyses is for all SF galaxies at $z = 0.5$--1.5, and the relation cannot maintain a 3$\sigma$ significance level when the bin numbers are reduced. 
It is possible that with a larger sample size, we could draw a solid conclusion that the \bhar-\sigmaone\ relation is more fundamental than the \bhar-\mstar\ relation among SF galaxies; it is also possible that even with a larger sample, we still could not obtain significant results, as \sigmaone\ may only serve as a useful indicator of the central gas density within certain mass ranges according to theoretical proposals \citep[e.g.][]{DS1986,Dekel2019}.
As mentioned in Section \ref{wce}, these theoretical ideas argue that when $M_{\rm halo} \lesssim 10^{12} M_\odot$, supernova feedback is effective at evacuating the gas around the central BH, and thus we may not expect \sigmaone\ to serve as a good indicator of the amount of central gas. For SF galaxies at $z = 1.5$--3, our limiting \mstar\ of $10^{10.2} M_\odot$ already exceeds the \mstar\ value corresponding to $M_{\rm halo} \sim 10^{12} M_\odot$ at $z \approx 2$ \citep[e.g.][]{Legrand2018}. However, for $M_{\rm halo} \sim 10^{12}~M_\odot$ at \hbox{$z = 0.5$--1.5}, the corresponding \mstar~is $\sim 10^{10.3-10.5} M_\odot$, which is above our limiting \mstar\ of $10^{9.7} M_\odot$ at $z = 0.5$--1.5.
These theoretical ideas are consistent with our findings in the left panels of Figures \ref{comp_pcor_sf_s1}, \ref{all_pcor_sf_s1} and \ref{all_pcor_sf_c1}, where the \bhar-\sigmaone/\cone\ relation is only apparent among massive SF galaxies at $z = 0.5$--1.5.

We thus perform PCOR analyses for all log~\mstar\ $> 10.3$ (that corresponds to $M_{\rm halo} \gtrsim 10^{12} M_\odot$ at  $z \approx 1.5$) SF galaxies and SF Non-BD galaxies at $z = 0.5$--1.5, where the central gas is not expected to be evacuated by supernova feedback, and thus our assumption of a constant \mstar-to-gas ratio may roughly hold. 
The results are summarized in Table~\ref{103}.
We found that the \bhar-\sigmaone\ relation is significant when controlling for \mstar, while the \bhar-\mstar\ relation is not significant when controlling for \sigmaone, for both SF galaxies and SF Non-BD galaxies.
This clearly suggests that, at least for log \mstar\ $> 10.3$ SF galaxies/SF Non-BD galaxies at $z = 0.5$--1.5, the \bhar-\mstar\ relation is a secondary manifestation of the \bhar-\sigmaone\ relation that may reflect a link between BH growth and central gas density. 

At the same time, for log \mstar\ $\leqslant 10.3$ SF galaxies/SF Non-BD galaxies at $z = 0.5$--1.5, testing shows that neither the \bhar-\sigmaone\ nor \bhar-\mstar\ relations are significant, which is not a surprise given the limited amount of BH growth at log \mstar\ $\leqslant 10.3$ as can be seen in Figures \ref{comp_pcor_sf_s1}, \ref{all_pcor_sf_s1} and \ref{all_pcor_sf_c1}.
With a larger sample of galaxies/AGNs, we can probe if the \bhar-\sigmaone\ relation is more fundamental than the \bhar-\mstar\ relation among SF Non-BD galaxies in general, or if the \bhar-\sigmaone\ relation is only more fundamental than the \bhar-\mstar\ relation when $M_{\rm halo} \gtrsim 10^{12} M_\odot$.

\begin{table}
\begin{center}
\caption{$p$-values (significances) of partial correlation analyses for the \bhar-\sigmaone\ relation among SF galaxies with log\mstar > 10.3 at $z = 0.5$--1.5}
\label{103}
\begin{tabular}{ccccc}\hline\hline
Relation &   Pearson & Spearman \\\hline\hline
\multicolumn{3}{c}{All SF Galaxies: $0.5 \leqslant z < 1.5$, log \mstar\ $> 10.3$ ($4 \times 4$ bins)} \\ \hline
\bhar-$M_\star$              &  $0.74~(0.3\sigma)$  & $0.58~(0.6\sigma)$ \\
\bhar-\sigmaone\            & $\boldsymbol {2\times 10^{-5}~(4.3\sigma)}$  & $\boldsymbol {3\times 10^{-6}~(4.6\sigma)}$ &  \\ \hline \hline
\multicolumn{3}{c}{SF Non-BD: $0.5 \leqslant z < 1.5$, log \mstar\ $> 10.3$ ($4 \times 4$ bins)} \\ \hline
\bhar-$M_\star$              &  $0.13~(1.5\sigma)$  & $0.14~(1.5\sigma)$ \\
\bhar-\sigmaone\            & $\boldsymbol {9\times 10^{-5}~(3.9\sigma)}$  & $\boldsymbol {2\times 10^{-4}~(3.7\sigma)}$ &  \\ \hline\hline
\end{tabular}                                         
\end{center}
\end{table}

\section{Summary and Future Work} \label{sec-sum}
We have systematically studied the dependence of BH growth on host-galaxy compactness based on multiwavelength observations of the CANDELS fields. The main points from this paper are the following:
\begin{enumerate}
\item
We have built an \mstar-complete sample of CANDELS galaxies with $H <$ 24.5 and reliable structural measurements at $z = 0.5$--3 (see Section~\ref{sssec-ss} and Table~1).
We compiled galaxy redshifts, \mstar, SFR, \re, and $n$ from the CANDELS catalogs  (see Sections~\ref{ssec-morph} and \ref{ssec-msfr}), and calculated \sigmae, \sigmaone, as well as \cone\ (that is independent of \mstar) to measure the compactness of galaxies (see Section~\ref{sssec-c}).
Based on machine-learning morphological measurements, we construct a bulge-dominated sample (the BD sample) and a sample of galaxies that are not dominated by bulges (the Non-BD sample) from the \mstar-complete sample. We also select SF galaxies in these samples (see Section~\ref{ssec-sd}).
We utilized deep X-ray observations from \textit{Chandra} to calculate \bhar~for relevant subsamples of galaxies, thereby estimating the long-term average BH growth (see Section \ref{ssec-bhar}).
\item
We found that \bhar\ does not fundamentally depend on \sigmae\ in general (see Section~\ref{ssec-sigmae} and Table~\ref{pcortable}). For galaxies in the BD sample, \bhar\ does not significantly depend on \sigmae~when controlling for SFR (see Section \ref{sssec-bulge}). For galaxies in the Non-BD sample, \bhar\ does not significantly depend on \sigmae\ when controlling for \mstar\ (see Section~\ref{sssec-nonbulge}); when testing is confined to SF Non-BD galaxies, the \bhar-\sigmae\ relation is also not significant (see Table~\ref{pcortable}).
Our results indicate that the apparent \bhar-\sigmae\ relation is not fundamental, even for the overall SF galaxy population (see Appendix~\ref{sssec-allsigmae}). 
We relate our results to other results in the literature claiming elevated BH growth associated with \sigmae\ in Section~\ref{highsigmae}.

\item
We found that the current samples do not reveal a significant \bhar-\sigmaone\ relation among galaxies in the BD/Non-BD sample when controlling for SFR or \mstar\ (see Section \ref{ssec-sigma1} and Table~\ref{pcortablecore}). 
However, when testing is confined to SF Non-BD galaxies, we found a just significant (3.0$\sigma$) \hbox{\bhar-\cone} relation when controlling for \mstar\ at $z = 0.5$--1.5. This indicates that the \bhar-\sigmaone\ relation is not simply a secondary manifestation of the primary \bhar-\mstar\ relation, at least in this redshift range (see Section \ref{sssec-nonbulge-s1}).
For the overall SF galaxy population, we found not only a significant \bhar-\cone\ relation when controlling for \mstar\ at $z = 0.5$--1.5, but also suggestive evidence of a significant \bhar-\cone\ relation when controlling for \mstar\ at \hbox{$z = 1.5$--3} (see Section \ref{allsf-sigma1} and Table~\ref{pcortableallsf}), implying the existence of the \bhar-\sigmaone\ relation at high redshift as well.
The \bhar-\sigmaone\ relation may indicate a link between BH growth and the central gas density of galaxies. It is possible that the \bhar-\mstar\ relation among SF Non-BD galaxies is simply reflecting this link, which needs to be tested with a larger sample. 
The current SF Non-BD sample suggests that, at least for massive galaxies with log \mstar\ $>$ 10.3 at \hbox{$z = 0.5$--1.5}, the \bhar-\sigmaone\ relation is more fundamental than the \bhar-\mstar\ relation (see Section \ref{sssec-103} and Table~\ref{103}).
Also, a larger SF BD sample is needed to reveal if, for SF bulges, the \bhar-SFR relation in \hbox{\citet{Yang2019}} is a manifestation of this link as well, and SFR alone cannot fully indicate the central gas density.

\end{enumerate}

In the future, we plan to measure \sigmaone\ values for a larger galaxy/AGN sample utilizing the $HST$ observations in the COSMOS region, to investigate further the role of \sigmaone\ in long-term average BH growth at $z = 0.5$--1.5.
At the same time, future accumulation of ALMA pointings will enable us to probe the link between BH growth and central gas density directly: the $HST$-like resolution of ALMA can resolve the central regions of galaxies, and the gas mass can be estimated from CO lines or from the dust mass assuming a typical dust-to-gas ratio. We can also compare the central gas density obtained from ALMA with \sigmaone\ to test if \sigmaone\ among SF galaxies indeed serves as a good indicator of the central gas density.
In addition, future deep \textit{JWST} and \textit{WFIRST} imaging combined with deep X-ray observations can help us probe further the relation between BH growth and \sigmaone\ at $z \approx 1.5-3$ with a much larger sample size and much smaller $M_{\rm lim}$.

\section*{Acknowledgements}
We thank Guillermo Barro for sharing \sigmaone\ values and Dale Kocevski for helpful advice.
QN, GY, and WNB acknowledge support from \textit{Chandra} X-ray Center
grant GO8-19076X, NASA ADP grant 80NSSC18K0878,
the V.M. Willaman Endowment, and Penn State ACIS Instrument Team Contract SV4-74018 (issued by the \textit{Chandra} X-ray Center, which is operated by the Smithsonian Astrophysical Observatory for and on behalf of NASA under contract NAS8-03060).
DMA thanks the Science and Technology Facilities Council (STFC) for support from grant ST/L00075X/1.
BL acknowledges financial support from the National Key R\&D Program of China grant 2016YFA0400702 and 
National Natural Science Foundation of China grant 11673010. 
FV acknowledges financial support from CONICYT and CASSACA through the Fourth call for tenders of the CAS-CONICYT Fund.
YQX acknowledges support from the 973 Program (2015CB857004), NSFC-11890693, NSFC-11421303, the CAS Frontier Science Key Research Program (QYZDJ-SSW-SLH006), and K.C. Wong Education Foundation.
The \textit{Chandra} Guaranteed Time Observations (GTO) for the GOODS-N were selected by the ACIS Instrument
Principal Investigator, Gordon P. Garmire, currently of the Huntingdon
Institute for X-ray Astronomy, LLC, which is under contract to the
Smithsonian Astrophysical Observatory via Contract SV2-82024. 

\appendix
\section{Adding Galaxies with \texttt{GALFIT\_FLAG = 1} into the Sample} \label{append-flag1}
As explained in \citet{vanderwel2012}, \texttt{GALFIT\_FLAG = 1} does not necessarily indicate a bad fit and those results can be used after assessment on an object-by-object basis.
The properties of galaxies with \texttt{GALFIT\_FLAG = 1} are listed in Table \ref{flagfrac} with those of galaxies with \texttt{GALFIT\_FLAG = 0}.
We can see that there is no significant bias toward the X-ray detected objects. However, we note that the presence of irregularity is very high among those less-certain fits, which is expected since irregularity can lead to deviations from S$\rm \acute{e}$rsic profiles. Thus, we examined if removing galaxies with \texttt{GALFIT\_FLAG = 1} may bias our results.

\begin{comment}
\begin{table*}
\begin{center}
\caption{Properties of galaxies with \texttt{GALFIT\_FLAG = 0} and \texttt{GALFIT\_FLAG = 1} at $z = 0.5-1.5/1.5-3$.
(1) Sample name. (2) Number of galaxies in the sample. (3) The fraction of bulge-dominated galaxies in the sample. (4) The fraction of galaxies with the presence of irregularity in the sample, defined as galaxies with $f_{\rm irr} \geqslant 1/10 $. (5) The fraction of X-ray detected galaxies in the sample.}
\label{flagfrac}
\begin{tabular}{cccccccccc}\hline\hline
Sample & $\rm N_{Galaxies}$ & $f_{\rm bulge}$  & $f_{\rm irregularity}$  & $f_{\rm X-ray~Detected}$ \\ 
(1) & (2) & (3) & (4)  & (5)  \\ \hline
\multicolumn{5}{c}{$0.5 \leqslant z < 1.5$} \\ \hline
\texttt{GALFIT\_FLAG = 0} ($J$-band) & 6247 & $24.6\%^{+0.5\%}_{-0.5\%}$    & $28.3\%^{+0.6\%}_{-0.6\%}$  & $7.4\%^{+0.3\%}_{-0.4\%}$\\
\texttt{GALFIT\_FLAG = 1} ($J$-band) & 530 &$25.1\%^{+1.2\%}_{-1.2\%}$    & $43.2\%^{+2.1\%}_{-1.9\%}$ & $10.6\%^{+1.3\%}_{-1.3\%}$\\ \hline
\multicolumn{5}{c}{$1.5 \leqslant z < 3$} \\ \hline
\texttt{GALFIT\_FLAG = 0} ($H$-band) & 2595 &$ 25.9\%^{+0.8\%}_{-0.8\%}$    &$54.3\%^{+1.0\%}_{-1.0\%}$& $11.3\%^{+0.6\%}_{-0.6\%}$\\
\texttt{GALFIT\_FLAG = 1} ($H$-band) & 360 &$13.6\%^{+1.7\%}_{-1.7\%}$.  &$72.5\%^{+2.5\%}_{-2.2\%}$&$11.4\%^{+1.7\%}_{-1.7\%}$\\
\hline\hline
\end{tabular}                                         
\end{center}
\end{table*}
\end{comment}

\begin{table*}
\begin{center}
\caption{Properties of galaxies with \texttt{GALFIT\_FLAG = 0} and \texttt{GALFIT\_FLAG = 1} at $z = 0.5-1.5/1.5-3$.
(1) Sample name. (2) Number of galaxies in the sample. (3) The fraction of galaxies with the presence of irregularity in the sample, defined as galaxies with $f_{\rm irr} \geqslant 1/10 $. (4) The fraction of X-ray detected galaxies in the sample.}
\label{flagfrac}
\begin{tabular}{cccccccccc}\hline\hline
Sample & $\rm N_{Galaxies}$& $f_{\rm irregularity}$  & $f_{\rm X-ray~Detected}$ \\ 
(1) & (2) & (3) & (4)  \\ \hline
\multicolumn{4}{c}{$0.5 \leqslant z < 1.5$} \\ \hline
\texttt{GALFIT\_FLAG = 0} ($J$-band) & 6247   & $28.3\%^{+0.6\%}_{-0.6\%}$  & $7.4\%^{+0.3\%}_{-0.4\%}$\\
\texttt{GALFIT\_FLAG = 1} ($J$-band) & 530    & $43.2\%^{+2.1\%}_{-1.9\%}$ & $10.6\%^{+1.3\%}_{-1.3\%}$\\ \hline
\multicolumn{4}{c}{$1.5 \leqslant z < 3$} \\ \hline
\texttt{GALFIT\_FLAG = 0} ($H$-band) & 2595     &$54.3\%^{+1.0\%}_{-1.0\%}$& $11.3\%^{+0.6\%}_{-0.6\%}$\\
\texttt{GALFIT\_FLAG = 1} ($H$-band) & 360  &$72.5\%^{+2.5\%}_{-2.2\%}$&$11.4\%^{+1.7\%}_{-1.7\%}$\\
\hline\hline
\end{tabular}                                         
\end{center}
\end{table*}

We visually examined 890 objects in our sample with \texttt{GALFIT\_FLAG = 1} and removed $\approx 11\%$ of them that have obvious failures in structural measurements.
Then, using a sample of 9637 objects ($\approx 94\%$ of the objects in the \mstar-complete sample), we confirmed that the results throughout the paper do not change qualitatively when this alternative sample is used.

\begin{comment}
\begin{figure*}
\begin{center}
\includegraphics[scale=0.6]{fig/msigmam_galfit1.pdf}
\caption{Galaxies with \texttt{GALFIT\_FLAG = 1} plotted in the \sigmae~vs. \mstar~plane for galaxies in the low-redshift bin (left) and the high-redshift bin (right) as gray open circles.
The orange dashed contours encircle 68\%, 90\%, and 95\% of galaxies in the BD sample, and the blue solid contours encircle 68\%, 90\%, and 95\% of galaxies in the Non-BD sample.
Those \texttt{GALFIT\_FLAG = 1} galaxies removed during the visual examination are marked with black crosses, which include most of the significant outliers in the \sigmae~vs. \mstar~plane.
}
\label{msigma_outlier}
\end{center}
\end{figure*}
\end{comment}

\section{The relation between BH growth and \sigmae\ for all SF galaxies} \label{sssec-allsigmae}

In this appendix, we study the \bhar-\sigmae~relation among all the SF galaxies in the sample regardless of their morphologies.
Similar to the approach in Section \ref{ssec-dis1}, we bin sources based on both $M_\star$ and \sigmae, and calculate $\rm \overline{BHAR}$ for each bin to perform PCOR analyses.
We input the median log$M_\star$, median log \sigmae, and log $\rm \overline{BHAR}$ of bins into \texttt{PCOR.R} to calculate the significance level of the $\rm \overline{BHAR}$-$M_\star$ relation when controlling for \sigmae~and the significance level of the $\rm \overline{BHAR}$-\sigmae~relation when controlling for \mstar. The results are shown in Table \ref{pcortableallsigmae}.

For the overall SF galaxy population, we found that \bhar\ significantly depends on \mstar\ when controlling for \sigmae, and $\rm \overline{BHAR}$ does not significantly depend on \sigmae\ when controlling for \mstar, indicating that the \bhar-\sigmae\ relation is not fundamental.
We test if BH growth has any additional dependence on \re\ when controlling for \mstar\ as well, and the results are also shown in Table~\ref{pcortableallsigmae}. The \bhar-\re\ relation is not significant when controlling for \mstar\ in any case, suggesting that \re\ is not as closely related to BH growth as \cone, which combines both \re\ and $n$ to indicate the central morphology of galaxies.

%\clearpage

\begin{table}
\begin{center}
\caption{$p$-values (significances) of partial correlation analyses for the \bhar-\sigmae\ relation among SF galaxies}
\label{pcortableallsigmae}
\begin{tabular}{ccccc}\hline\hline
Relation &   Pearson & Spearman \\\hline\hline
\multicolumn{3}{c}{All SF Galaxies: $0.5 \leqslant z < 1.5$  ($5 \times 5$ bins)} \\ \hline
\bhar-$M_\star$              &  $\boldsymbol {1\times 10^{-4}~(3.9\sigma)}$  & $\boldsymbol {5\times 10^{-4}~(3.5\sigma)}$ \\
\bhar-$\Sigma_{\rm_e}$ & $9\times 10^{-3}~(2.6\sigma)$  & $0.03~(2.2\sigma)$ &  \\ \hline
\bhar-$M_\star$              &  $\boldsymbol {2\times 10^{-7}~(5.2\sigma)}$  & $\boldsymbol {2\times 10^{-7}~(5.2\sigma)}$ \\
\bhar-\re\ & $0.02~(2.4\sigma)$  & $0.02~(2.3\sigma)$ &  \\ \hline
\multicolumn{3}{c}{All SF Galaxies: $1.5 \leqslant z < 3$ ($4 \times 4$ bins)} \\ \hline
\bhar-$M_\star$             &  $\boldsymbol {3 \times 10^{-3}~(3.0 \sigma)}$  & $\boldsymbol {3 \times 10^{-3}~(3.0\sigma)}$ \\
\bhar-$\Sigma_{\rm_e}$ &  $0.14~(1.5 \sigma)$  & $0.42~(0.8\sigma)$ \\ \hline
\bhar-$M_\star$              &  $\boldsymbol {9\times 10^{-4}~(3.3 \sigma)}$  & $\boldsymbol {3\times 10^{-4}~(3.6\sigma)}$ \\
\bhar-\re\ & $0.18~(1.3\sigma)$  & $0.04~(2.0\sigma)$ &  \\ \hline\hline
\end{tabular}                                         
\end{center}
\end{table}

\bibliography{compact.bib}
\bibliographystyle{mnras}

\bsp	% typesetting comment
\label{lastpage}
\end{document}